\documentclass[english,11pt,a4paper,twoside]{article}
\usepackage[german,english]{babel}
\usepackage{amsmath}
\usepackage{latexsym,amssymb}
\usepackage{epsfig,changebar}
\usepackage{amscd}
\usepackage{diagrams}		
\usepackage{curves}		
\usepackage{axodraw}		
\numberwithin{equation}{subsection}  

\newcommand{\p}{\partial}
\newcommand{\df}{\delta}
\newcommand{\n}{\nabla}
\newcommand{\ti}[1]{\ensuremath{\tilde#1}}
\newcommand{\trm}[1]{\textrm{#1}}
\newarrow{Identic} =====						
\newarrow{Fromout} |...>						

\begin{document}
\title{\vspace*{-15mm}\Huge{\sc{Gauge Invariance and Gauge--Factor Group in Causal Yang--Mills Theory}}
\\[15mm]}
\author{{\bf Dissertation} \\[7mm]
{\bf zur}\\[2mm]{\bf Erlangung der naturwissenschaftlichen Doktorw"urde} 
\\[2mm]
{\bf (Dr. sc. nat.)}\\[2mm]
{\bf vorgelegt der}\\[2mm]{\bf Mathematisch-naturwissenschaftlichen Fakult"at}
 \\[2mm]{\bf der} \\[8mm]
{\bf Universit"at Z"urich} \\[8mm]{\bf von} \\[2mm]
{\bf Niklaus Josef Emmenegger} \\[2mm]{\bf von}\\[2mm]{\bf Hasle LU} \\
[8mm]
{\bf Begutachtet von} \\[7mm]
{\bf Prof. Dr. G"unter Scharf}
\\[10mm]}
\date{Z"urich 2001}  

\maketitle
\thispagestyle{empty}

\newpage
\thispagestyle{empty} 
\small \noindent
Die vorliegende Arbeit wurde von der Mathematisch-naturwissenschaftlichen 
Fakult"at der Universit"at Z"urich auf Antrag von Prof. Dr. G"unter Scharf  
und Prof. Dr. Daniel Wyler als Dissertation angenommen.
\normalsize
\newpage
\selectlanguage{german}
\vspace{15cm}
\setcounter{page}{1}
\pagenumbering{roman}
\centerline{\Large{Zusammenfassung}}
\vspace{3 cm}

In der vorliegenden Arbeit wird die Eichinvarianz der kausalen Yang--Mills Theorie bewiesen. Dazu muss gezeigt werden, dass die in der kausalen Konstruktion der \textbf{S}--Matrix auftretenden operatorwertigen Distributionen $T_n$ bzw. $D_n\big|_{\trm{ret}}$ unter der Eichvariation $d_Q$ eine Divergenzform aufweisen. Weil aber nur lokale Terme zu einer Eichverletzung f"uhren k"onnen, werden einzig lokale Ausdr"ucke genauer untersucht.\\
Im ersten Teil der Arbeit wird in Anlehnung an die Eich--Cohomologie Theorie die lokale Eich--Faktorgruppe definiert und gezeigt, dass s"amtliche Elemente aus dieser Gruppe unter der Eichtransformation $d_Q$ unmittelbar zu einer Divergenzform der gesamten operatorwertigen Distribution $d_Q\,T_n$ f"uhren.\\
Im zweiten Teil der Arbeit werden alle in der kausalen Yang--Mills Theorie m"oglichen lokalen Terme systematisch untersucht. Es wird gezeigt, dass jede lokale operatorwertige Distribution Element der Eich--Faktorgruppe oder identisch zu Null ist, und somit die kausale Yang--Mills Theorie ohne zus"atzliche Einschr"ankungen eichinvariant geschrieben werden kann.
\newpage
\selectlanguage{english}
\vspace{15cm}
\centerline{\Large{Summary}}
\vspace{3 cm}

In the present work the gauge invariance of causal Yang--Mills theory will be proven with the aid of the gauge--factor group. For that purpose it must be shown, that the operator valued distributions $T_n$ and $D_n\big|_{\trm{ret}}$ occurring in the causal \textbf{S}--matrix construction can be written, after applying the gauge variation $d_Q$, as a divergence. Since merely local terms lead to gauge destroying expressions, one has to focus on them exclusively.\\
In the first part of the work the local gauge--factor group will be defined in the style of the concept of gauge cohomology theory. It will be shown, that every element out of the so defined factor group under the transformation $d_Q$ leads to a divergence of the entire operator valued distribution $d_Q\,T_n$.\\
In the second part all local terms arising in causal Yang--Mills theory are systematically investigated. Without further restrictions there can be proven, that every local operator valued distribution is an element of the gauge factor group or equal to zero. This concludes the demonstration of gauge invariance of causal Yang--Mills theory.
\newpage
\tableofcontents
\newpage
\setcounter{page}{1}
\pagenumbering{arabic}
\section{Introduction}	\label{INTRO}

The purpose of this work is to introduce and apply a simple and powerful technique for demonstrating the gauge invariance of causal Yang--Mills theory. The technique uses the free field structure of the \textbf{S}--matrix in the causal perturbation theory. This is in sharp contrast to the BRST invariance, which deals with interacting fields.

After introducing first the principals of causal quantum field theory followed by the pointing out of the gauge variation for Yang--Mills theory, the next part is devoted to the in--depth analysis of gauge invariant construction of the \textbf{S}--matrix. We will see that the determination of the gauge transformed $T_n$'s can be executed by two different  constructions whose results differ in local terms only. Furthermore one realizes, that the general set of all local terms arising under the splitting procedure $(d_Q\, D_n) \rightarrow (d_Q\, D_n)\big|_{\trm{ret}}$ builds a superset over all local terms coming from $d_Q (T_n\big|_{\trm{ret}})$. Thus gauge invariance is proven, if adding each single local term out of the superset to the not local part $d_Q\, D_n$ also results in a gauge invariant form of the entire operator valued distribution. The introduction of the gauge--factor group in the subsequent chapter specifies this construction idea in a clearly mathematical formulation.

In the last part of this work the explicit determination of the gauge factor group elements shows gauge invariance for all $T_n$'s which are plagued by any local term. In this connection there must be a careful distinction between local $3$ and $4$--leg terms describing graphs with totally disjunct arguments and those with a pair or pairs of equal arguments.
%
\section{Causal Approach to Quantum Field Theory}	\label{KQFT}

The causal QFT, also known as finite QFT (confer G.~Scharf~\cite{scharf}) is an alternative formalism for describing quantum field theories. The goal of the theory is to solve problems, which occur in the path--integral method. In the 1950's E.~C.~G.~St\"uckelberger and N.~N.~Bogoliubov (see e.g. N.~N.~Bogoliubov, D.~V.~Shirkov~\cite{bogshi}) described the causal restriction in the construction of the \textbf{S}--matrix.

In 1973 H.~Epstein and V.~Glaser elaborated on a compact treatment of the topic in the Ann. Inst. Henry Poincar\'e \cite{epstein}. Herein they introduced the distributional splitting of causal terms which lead to the construction of the causal $T_n$'s order by order.
In the spirit of Epstein and Glaser, G.~Scharf~\cite{scharf} published an extended working out of Finite Quantum Electrodynamics in 1989. In the latter, the causal \textbf{S}--matrix theory is shown in a modern formulation with many examples executed in mathematical strength. The notations used in G.~Scharf~\cite{scharf} will also serve for this work.

First, a short insight into the principals of causal QFT will help to introduce techniques which are used in the later chapters of this work. Next step is devoted to give a short summary about causal \textbf{S}--matrix construction and causal distribution splitting. In the following chapter, the description of gauge invariance for causal \textbf{S}--matrices will be discussed.

\subsection{The Causal QFT}	\label{KQFT1}

As mentioned above, the causal QFT is a \textbf{S}--matrix theory. As shown by \cite{epstein},\cite{scharf}, under the restrictions of causality and Poincar\'e--invariance the \textbf{S}--matrix is free of (Ultraviolet) \textbf{UV}--divergences. The determination of the \textbf{S}--matrix is done inductively order by order. But in contrast to the Feynman method one always keeps in mind, that the multiplication of $\Theta$--functions with a propagator must not necessarily be well defined in the sense of distributional multiplication.

The tool which helps to preserve mathematically well defined expressions is called distributional splitting, which will be explained in detail in a subsequent chapter. But not only the so--called distributional splitting distinguishes the causal method from Feynman calculations, but also the fact, that the considered operator valued distributions are free field operators in Fock space $\mathcal{F}$.
The starting point for the calculations serve the free (Fock space) field operators $A , u , \ti{u}$ with their (anti)commutation relations and the first order $T_1$ of the \textbf{S}--matrix. Whereby the last expression $T_1$ corresponds to the first interacting Lagrange term in path integral method (confer e.g. \cite{ryder})

Since the \textbf{S}--matrix consists in operator valued distributions, according to the considerations above, one has to smear out the resulting $T_n$'s with test functions $g$ in order to obtain valuable physical expressions. The test functions are in Schwarz space $g \in \mathcal{S} (\mathbb{M})$\footnote{$\mathbb{M}$ stands for Minkowski space}, in order to apply Fourier transformation within the theory. Additionally, the test functions cut off the long reaching interaction terms (e.g. Coulomb interaction) so the smeared out \textbf{S}--matrix elements are free of (Infrared) \textbf{IR}--divergences. Thus the \textbf{S}--matrix is free of both, \textbf{UV}--divergences and \textbf{IR}--divergences. However \textbf{IR}--divergences can arise in the adiabatic limit $g \to 1$ later on, and so have to be considered separately in that respect.

\subsection{The \textbf{S}--Matrix in Causal Perturbation Theory} \label{KQFT2}

The following short introduction to the construction of the causal \textbf{S}--matrix only touches on the main points. For a detailed explanation refer to \cite{epstein},\cite{scharf}.
In quantum field theory there is no dynamical equation which describes the time evolution of states. Therefore one must use perturbation series for describing time depending states. The Dyson series which occurs in the \textbf{S}--matrix, can be formally written as an exponential series in $T_n$'s
\begin{equation} 	\label{cqft1}
	\textbf{S}(g) = 1 + \sum_{n=1}^{\infty}\frac{1}{n!}\int d^4 \! x_1 \dots  			d^4 \! x_n T_n(x_1,\ldots,x_n)g(x_1)\cdot \ldots \cdot g(x_n)
\end{equation}
whereas the $T_n$'s are well defined operator valued distributions, and the smeared out expression $\textbf{S}(g)$ describes an operator in Fock space.
More specifically, the $T_n$ in $\eqref{cqft1}$ consist in a distributional part $t_n^l$ and an operator valued part $\mathcal{O}_l$. Additionally, the distributions $t_n^l$ depend only on relative coordinates
\begin{equation}		\label{cqft2}
	T_n(x_1,\ldots,x_n) = \sum_l t_n^l(x_1 - x_n,\ldots,x_{n-1} - x_n): 				\mathcal{O}_l (x_1,\ldots,x_n):
\end{equation}
since the $\textbf{S}(g)$ and consequently the $T_n$'s as well, depend only in translation invariant coordinates. Furthermore, the numerical distribution $t_n^l$ can have the form of a tensor. The Wick monomial $:\mathcal{O}_l:$ itself consists of a product of \emph{free} field operators.
Moreover one can invert the exponential series $\eqref{cqft1}$ formally and achieves
\begin{align}
	\textbf{S}(g)^{-1} 
		&= 1 + \sum_{n=1}^{\infty}\frac{1}{n!}\int d^4 \! x_1 \dots d^4 \! x_n   			\ti{T}_n(x_1,\ldots,x_n)g(x_1)\cdot \ldots \cdot g(x_n)																			\label{cqft3}\\
		&=\left( 1 + T \right)^{-1} = 1 + \sum_{r=1}^{\infty}\left( -T \right)^r 														\label{cqft4}
\end{align}
The second identity shows, that the $\ti{T}_n$ clearly can be written in terms of $T_n$'s \footnote{For more details see \cite{epstein},\cite{scharf}}. Taking into consideration the symmetry in the arguments in $\eqref{cqft3}$ and $\eqref{cqft4}$ one is lead to $\ti{T}_n$ in $\eqref{cqft3}$ as a sum over all permutations in $x_1,\ldots,x_n$. The latter explicitly yields for the $\ti{T}_n$ with the aid of $\eqref{cqft4}$ to
\begin{equation}		\label{cqft5}
	\ti{T}_n(X) = \sum_{r=1}^{\infty} (-1)^r \sum_{\trm{$P_{r}$}} T_{n_{1}}(X_1) \cdot 			\ldots \cdot T_{n_{r}}(X_r)
\end{equation}
Wherein $X$ stands for the unordered set of n points in $\mathbb{M}^{\otimes n}$ \begin{equation}		\label{cqft6}
	X:= \left\{ X= \left( x_{\pi(1)},\ldots,x_{\pi(n)} \right) \mid \forall x_i 		\in \mathbb{M} \text{ and } \forall \text{ permutations } \pi \text{ of } 		\{ 1,\ldots,n \} \right\}
\end{equation}
and the second sum runs over all partitions $P_r$ out of $X$ into $r$ disjoint subsets $X_k$. Consequently the combination over all subsets $X_k$ leads to
\begin{equation}		\label{cqft7}
	X_1 \cup X_2 \cup \ldots \cup X_r = X \text{ whereas } X_k \ne \emptyset \quad 	\forall k \in \left\{ 1, \ldots , r\right\}
\end{equation}
The next topic is causality. If one considers two test functions $g_1, g_2 \in \mathcal{S}(\mathbb{M})$ with $supp(g_1) < supp(g_2)$\footnote{$supp(g_i)< supp(g_j)$ or simply $g_i < g_j$ means, that in any Lorenz system the time coordinate of $x_i \in supp(g_i)$ is smaller then the time component of $x_j \in supp(g_j)$}, the causality condition for the \textbf{S}--matrix can be written as
\begin{align}	\label{cqft8}
	\textbf{S}(g_1 + g_2) &= \textbf{S}(g_2)\textbf{S}(g_1) &&\forall g_1, g_2
	\in \mathcal{S}(\mathbb{M}), g_1 < g_2
\end{align}
Furthermore as shown in \cite{scharf}, even in the adiabatic limit the defined perturbation series is physically valuable.
In addition one can show with the aid of the causality condition, that the n--point function occurring in the $T_n$'s can be constructed inductively order by order from the initial $T_1$.
The interaction Lagrangian $T_1$ therefore  must be given, as mentioned above (confer \cite{fabcs}).

After these important preliminary remarks one can proceed to the causal construction of the \textbf{S}--matrix and thus to the inductive determination of the $T_n$'s.
Combining $\eqref{cqft8}$ and $\eqref{cqft1}$ with $\eqref{cqft3}$ one gets after permutation of all arguments $x_1,\ldots,x_n$ (confer again 3.1 in \cite{scharf}) the following expressions
\begin{align}	\label{cqft9}
	T_n(x_1,\ldots,x_n) 
	&= T_m(x_1,\ldots,x_m) \cdot T_{n-m}(x_{m+1},\ldots,x_n)
	&& \text{ with } \left\{x_1,\ldots,x_m \right\} > \left\{x_{m+1},\ldots,x_n 			\right\}
\end{align}
and
\begin{align}		\label{cqft10}
	\ti{T}_n(x_1,\ldots,x_n) 
	&= \ti{T}_m(x_1,\ldots,x_m) \cdot \ti{T}_{n-m}(x_{m+1},\ldots,x_n)
	&& \text{ with } \left\{x_1,\ldots,x_m \right\} > \left\{x_{m+1},\ldots,x_n 			\right\}
\end{align}
whereas the $T_n$'s and $\ti{T}_n$'s obviously are time ordered products. There is still an advanced possible factorization, since the time support of the n--point function can be written to fulfil the time ordering relation $x_1^0 > x_2^0 > x_3^0 > \ldots > x_n^0$ \footnote{If there are not two arguments which have the same time argument}. So one is lead to the simple expression
\begin{equation}		\label{cqft11}
	T_n(x_1,\ldots,x_n) 
	= T_1(x_1) \cdot  T_1(x_2) \cdot \ldots \cdot T_1(x_n)
\end{equation}
The same considerations obviously can be applied to the $\ti{T}_n(x_1,\ldots,x_n)$.

Next we effect the general construction step from $T_{n-1} \to T_n$ under the assumption, that all $T_m$'s with $1 \leqslant m < n$ are already known under the above restrictions.
The three n--point functions $A_n' ,R_n'$ and $D_n$ below constitute the principle terms in the construction of $T_n$'s
\begin{align}
	A_n'(x_1,\ldots,x_n) 
		&:= \sum_{\trm{$P_2$}}\ti{T}_{n_1}(X)T_{n-n_1}(Y,x_n)	\label{cqft12}\\
	R_n'(x_1,\ldots,x_n) 
		&:= \sum_{\trm{$P_2$}}T_{n-n_1}(Y,x_n)\ti{T}_{n_1}(X)	\label{cqft13}\\
 	D_n(x_1,\ldots,x_n) 
		&:= R_n'(x_1,\ldots,x_n) - A_n'(x_1,\ldots,x_n)			\label{cqft14}
\end{align}
whereas in the expression above the summation runs over all disjoint subsets of the form
\begin{align}		\label{cqft15}
	P_2 &: \left\{ x_1, x_2 , \ldots , x_{n-1} \right\} = X \cup Y
		&& X \ne \emptyset
\end{align}
If one extends the summation in $\eqref{cqft12}$ and $\eqref{cqft13}$ over the additionally possible subset $X = \emptyset$, one is lead to the new terms
\begin{align}
	A_n(x_1,\ldots,x_n) 
		&:= \sum_{\trm{$P_2^0$}} \ti{T}_{n_1}(X)T_{n-n_1}(Y,x_n)	
		= A_n'(x_1,\ldots,x_n) + T_n(x_1,\ldots,x_n)			\label{cqft16}
\intertext{and}
	R_n(x_1,\ldots,x_n) 
		&:= \sum_{\trm{$P_2^0$}} T_{n-n_1}(Y,x_n)\ti{T}_{n_1}(X)	
		= R_n'(x_1,\ldots,x_n) + T_n(x_1,\ldots,x_n)			\label{cqft17}
\end{align}
Obviously the difference $D_n$, which is defined above, is identical to \begin{equation}		\label{cqft18}
	D_n := R_n' - A_n' = R_n - A_n
\end{equation}
since the two $T_n$'s in the right hand expression add up to zero.
If one knows one of the two terms $A_n$ or $R_n$, it is easy to rearrange the equations $\eqref{cqft16}$ or $\eqref{cqft17}$ and extract $T_n$ as follows
\begin{equation}		\label{cqft19}
	T_n = A_n - A_n' = R_n - R_n'
\end{equation}
Since for the algebraic calculations in the present work only the definitions and the notions of the causal construction of $T_n$'s are used, the technical proofs of the statements above are skipped. The interested reader finds the extended treatment in chapter 3.1 of \cite{scharf}. Nevertheless the following considerations in the elaboration of the proofs are of special interest and are  briefly mentioned.

The supports of the above defined operator valued distributions $A_n$ and $R_n$ lie inside the light cone. Furthermore it can be shown, that the support of $A_n$ ($R_n$) only consists of the forward (backward) light cone
\begin{align}
	supp(A_n) &\subseteq \Gamma^-_n			\label{cqft20}\\
	supp(R_n) &\subseteq \Gamma^+_n			\label{cqft21}
\end{align}
Taking into account, that after the construction rule the $A_n'$ and $R_n'$ both have causal support, the $D_n$ (as a difference of $A_n'$ and $R_n'$) also has causal support.
By splitting the numerical distribution $d_n$ into the two disjointed parts $a_n$ or $r_n$ respectively, the operator valued distribution $D_n$ is separated into two parts according to the sets $\eqref{cqft20}$ and $\eqref{cqft21}$.
Altogether the $T_n$ itself also has causal support, and the problem of constructing the latter is reduced to the splitting of a distribution $d_n$.
In the next part we will get acquainted with distribution splitting.

\subsection{Causal Distribution Splitting}	\label{KQFT3}

According to the construction rule and the above considerations the operator valued distributions  $D_n$ have the following form
\begin{equation}			\label{cqft22}
	D_n(x_1,\ldots,x_n) = \sum_l d_n^l(x_1 - x_n,\ldots,x_{n-1} - x_n): 				\mathcal{O}_l (x_1,\ldots,x_n):
\end{equation}
where the (eventually tensorial) factors $d_n^l \in \mathcal{S} (\mathbb{M}^{\otimes n})$ have causal support. Furthermore, the free field operators of the Wick monomial $:\mathcal{O}_l (x_1,\ldots,x_n):$ do not cause further restrictions of the support. So one has (according to the above) only to split the tempered distribution $d_n$ into an advanced and a retarded part.
The problem which has to be solved can now be reformulated as follows:
\\
Given a tempered distribution $d_n \in \mathcal{S} (\mathbb{M}^{\otimes n})$ (one chooses $d_n \in \mathcal{S}(\mathbb{M}^{\otimes n})$ in order to get well defined\footnote{confer e.g. \cite{blanchard}, \cite{gelfand}} Fourier transformed $d_n$'s) with causal support
\begin{equation}		\label{cqft23}	
	supp (d_n) \subset \Gamma^+_n(0) \cup \Gamma^-_n(0)
\end{equation}
It is possible to split the distribution $d_n$ into two disjoint parts $r_n$ and $a_n$ with the following properties
\begin{equation}		\label{cqft24}
	\begin{aligned}
		r_n , a_n 	& \in \mathcal{S} (\mathbb{M}^{\otimes n})\\
		supp(r_n)	& \subset \Gamma^+_n(0)\\
		supp(a_n)	& \subset \Gamma^-_n(0)
	\end{aligned}
\end{equation}
whereas $d_n$ can be written as the difference
\begin{equation} 	\label{cqft25}
	d_n = r_n - a_n
\end{equation}
As it is shown in \cite{malgrange}, the searched distributions are regularly separated and thus always can be split into two separate parts\footnote{Confer \cite{malgrange}}.
The general form of the latter parts $a_n$ and $r_n$ emerge from the considerations above, whereas the supports $supp(r_n)$ and $supp(a_n)$ both have a common point --- the origin. This shows, that the distributions $r_n$, $a_n$ are not uniquely in the latter point and thereby do not define the splitted distributions definitely.
Only the difference $r_n - a_n$ is clearly defined.
But the ambiguousness of the expressions $r_n$, $a_n$ can be easily determined, if one takes into consideration, that the support of the difference of two different decompositions (splitted $d_n$'s)
\begin{equation}		\label{cqft26}
	supp(r_n - \ti{r}_n) =supp(a_n - \ti{a}_n) = \{ 0 \}
\end{equation}
has point nature. With the aid of theorem 3.9 in \cite{blanchard} it follows immediately, that $r_n$ as well as $a_n$ must be extended by the sum
\begin{equation} 	\label{cqft27}
	\sum_{|q| \leqslant p} c_{q} \, \mathcal{D}^{q} \df (x_1 - x_n, 						\ldots,x_{n-1} - x_n)
\end{equation}
Whereas $c_{q}$ stands for numbers out of $\mathbb{C}$, and $\mathcal{D}^{q}$ for the multi-derivation
\begin{align}	\label{cqft28}
	\mathcal{D}^{q}&:= \frac{\p^{q_1 + q_2 + \ldots}} 										{\p^{q_1}\p^{q_2} \ldots}	&&\text{with the multi-index }q
\end{align}
The summation index in $\eqref{cqft27}$ has taken values which can be clearly found by analyzing the distribution $d_n$.
As it is shown in the outline of the construction of distributions with point support $supp(T)= \left\{ x_0 \right\}$\footnote{Confer e.g. Theorem 3.9 in \cite{blanchard} or \text{chapter II}, \S 4.5 in \cite{gelfand}}, the summation over derivated $\df$--distributions is limited by the singular order $p$ of the distribution $T(x_0)$ itself
\begin{align}	\label{cqft29}
	T	&= \sum_{|q| \leqslant p} c_q \mathcal{D}^q \df(x - x_0)
		&& supp(T) = x_0
\end{align}
Altogether the general solutions of the splitting are found to have the form
\begin{align}
	r_n &= r_n^0 + \sum_{|q| \leqslant p} c_q \mathcal{D}^q \df(0)																			\label{cqft30}\\
	a_n &= a_n^0 + \sum_{|q| \leqslant p} c_q \mathcal{D}^q \df(0)
														\label{cqft31}
\end{align}
whereas $r_n^0 , a_n^0$ are special splitting solutions. More about the latter topic follows shortly.
To find the explicit summation index $p$, one has to determine the singular order of the distribution $d_n$ in the splitting point. This is best executed with the aid of quasi asymptotics of $d_n$ at the relevant point. For that purpose one needs the subsequent definitions, where one finds, without loss of arbitrariness, for the special splitting point $x_0 = 0$:\\
A distribution $d_n \in \mathcal{S}'(\mathbb{R}^m)$ has quasi asymptotics $d_0 \in \mathcal{S}'(\mathbb{R}^m)$ at the point $x=0$, if the limes
\begin{equation}	\label{cqft32}
	\lim_{\df \searrow 0} \rho(\df) \df^m d(\df x) = d_0(x) \ne 0
\end{equation}
under consideration of any function $\rho(\df)$, $\rho \in \mathcal{C}^0[0,\infty)$, exists\footnote{Confer \cite{epstein},\cite{scharf}}.
With the help of Fourier transformation one gets the related expression in momentum space as
\begin{equation}	\label{cqft33}
	\lim_{\df \searrow 0} \rho(\df) \hat{d}(\frac{p}{\df}) = 							\hat{d}_0(p) \ne 0
\end{equation}
which shows the behaviour of the distribution $\hat{d}(\frac{p}{\df})$ for $p \to \infty$\footnote{Obviously, here for all $\check{\varphi} \in \mathcal{S}(\mathbb{R}^m)$ exists the limit $\lim_{\df \searrow 0} \langle \rho(\df) \hat{d}(\frac{p}{\df}), \check{\varphi}(p) \rangle = \langle\hat{d}_0(p), \check{\varphi}(p) \rangle$}. To determine the power series characteristics of $d_0 (x)$, $\hat{d_0}(p)$ respectively, one must carry out a scale transformation\footnote{We only consider the transformation in momentum space, calculations in $x$--space are similar.} $p \to ap$ in $\mathcal{S}(\mathbb{R}^m)$. Building the following quotient
\begin{equation}	\label{cqft34}
	\lim_{\df \searrow 0} \frac{\rho(a \df)}{\rho(\df)}= a^{\omega}
\end{equation}
one gets the power expression $a^{\omega}$. This shows, that for all $a \in (0,\infty)$ there exists a $\omega \in \mathbb{R}$ in such a way, that the quasi asymptotic $d_0$ of $d$ has the form
\begin{equation}	\label{cqft35}
	d_0 = \rho(\df) = \df^{\omega}
\end{equation}
Obviously the terms of the Taylor expanding vanish for all $p > \omega$ at $x=0$ and at $p= \infty$ respectively. This shows, that in $\eqref{cqft27}$ to $\eqref{cqft31}$ the summation index must be chosen as $p=\omega$.

As the last step in this chapter, it shall be resumed, how the special splitting solutions $r_n^0 , a_n^0$ can be found. The two cases each have to be treated in a different manner:\\
\textbf{Case $\omega < 0$}:\\
One sees, that in this case the sum over the local distributions in $\eqref{cqft30}\text{ and }\eqref{cqft31}$ contains no terms. Furthermore one can show \cite{epstein}, \cite{scharf}, that $d_n$ can be splitted trivially. The $r_n , a_n$ are simply given by multiplying $d_n$ with a $\Theta$--function
\begin{align}	\label{cqft36}
	r_n(x_1 - x_n, \ldots , x_{n-1} - x_n) 
	&= \Theta 
	\Big( \sum_j v_j ( x_j - x_n ) \Big) d_n(x_1 - x_n,\ldots, x_{n-1}- x_n)
	&&\forall \omega(d_n) <0
\end{align}
where $j$ runs over all elements $j \in \{ 1, \ldots , n\}$ and $v_j \in V^+$.
\\
\textbf{Case $\omega \geqslant 0$}:\nopagebreak[4]\\
Here the sums over the $\df$--distributions are not of a trivial type. The special splitting solutions $r_n^0 , a_n^0$ are given from the dispersion integral (for mass $m \ne 0$) as
\begin{equation}	\label{cqft37}
	\hat{r}_n^0 (p) = \frac{i}{2 \pi} \int dt 										\frac{\hat{d}(tp)}{(t-i0)^{\omega +1}(1-t+i0)}
\end{equation}
The latter is the so called central splitting solution, since for $r_n^0 (p)$ the origin $p=0$ carries the normalization condition
\begin{align}	\label{cqft38}
	\mathcal{D}^q \, \hat{r}_n^0 (0) &= 0 	&&|q| \leqslant \omega
\end{align}
The attentive reader surely already realized, that the central splitting mentioned above is not in correspondence with the assumption of mass-free field expressions, which are the only terms considered in this work.
But one can show \cite{YMII},\cite{scharf}, that in the massless case the redefined dispersion integral with a normalization point $p \ne 0$ is well defined too. So one gets for the normalization point $\ti{q}$ the special solution
\begin{align}	\label{cqft39}
	\hat{r}_q^0 (p) 
	&= \frac{i}{2 \pi} \int dt 															\frac{\hat{d}(tp + \ti{q})}{(t-i0)^{\omega +1}(1-t+i0)}
	&&\mbox{whereas $\mathcal{D}^q \, r_q^0(\ti{q})=0$ ,} \quad|q| \leqslant \omega
\end{align}
Taking all the above points into consideration, one is well supplied to go over to the construction of gauge invariant causal \textbf{S}--matrices.

\subsection{Summary of Chapter \mbox{\boldmath $\ref{KQFT}$}}	\label{KQFT4}

Altogether what has been summarized are the following important properties of causal QFT. The \textbf{S}--matrix is after construction \textbf{UV}--divergence and \textbf{IR}--divergence free. The action is obtained by smearing out the operator valued distribution series $T_n$ in each power. The $T_n$'s can be calculated order by order from all already known $T_m$'s $(m<n)$. As the last step the numerical distribution part in $D_n = D_n(T_1, \ldots , T_{n-1})$ must be split. The latter is done by calculating the dispersion integral for $d_n$'s with a singular order $\omega(d_n(q)) \geqslant 0$ and adding the sum $\sum_{|q| \leqslant \omega} c_q \mathcal{D}^q \df(x-q)$ to the result. In contrast to the last point, a simple multiplication with a $\Theta$--function leads to the splitted terms for distributions with $\omega(d_n(q)) < 0$.
\section{Gauge Invariance of the \textbf{S}--Matrix}	\label{EICH}

The aim of the present chapter is to get acquainted with the gauge invariant construction of causal \textbf{S}--matrices. Since the \textbf{S}--matrix is constructed order by order, no strict statement about the theory can be given except those, which are proven inductively.

In the first step the notion of gauge invariance in QFT will be introduced. 
Similar to classical field theory, gauge invariance in QFT is defined with the help of the action\footnote{confer e.g. \cite{bogshi} ,\cite{kugo}}.
Then based on the Noether theorem one knows, that every symmetry of the action leads to a current $j$, which is preserved up to a total divergence of an anti-symmetrical tensor. The spatial integral of the current
\begin{equation}	\label{eich1}
	Q=\int d^3 \!x \, j_0(x)
\end{equation}
is called the according charge. The latter is an invariant of time.
Additionally one can show, that the three terms
\begin{center}
	\emph{charge , field operator , infinitesimal symmetry transformation}
\end{center} 
are not unrelated expressions. Rather the three terms are connected to each other by the following (anti)commutation relations
\begin{align}	\label{eich2}
	& \forall F \in \mathcal{F} \quad \exists \quad G \in \mathcal{F} 
	&&\mbox{with $[ Q , F ]_{\pm} = i G$}
\end{align}
where the above defined quantity $Q$ (=gauge charge) takes the r\^ole of an infinitesimal representative of the symmetry transformation
\begin{equation}	\label{eich3}
	F_i(x) \to F_i'(x) = F_i(x) + \epsilon \, G_i(F(x))
\end{equation}
in Fock space $\mathcal{F}$.
In the calculation of the infinitesimal field transformation $G$ in $\eqref{eich2}$ one has to distinguish between the two subsequent cases.
$F$ acts as a substitute for a Fermi field. Then $i\,G$ in $\eqref{eich2}$ is represented by the commutation relation $[ Q , F ]_-$. In contrast, if $F$ takes the place of a Bose field, $[ Q , F ]_+$ has to be taken as a anti commutator.

In the following we will restrict ourselves to gauge transformations of \textbf{S}--matrices. Here again it is obvious, that the action is invariant, if $G_i$ in $\eqref{eich2}$, $\eqref{eich3}$ has the structure of a divergence.
Taking into account, that the \textbf{S}--matrix is evaluated by linear calculation (integration) and with respect to the considerations above, gauge invariance of the latter can be represented simply in any order $n$ of the expansion by
\begin{equation}	\label{eich3a}
	[Q,T_n] = \p^{\mu} \sum_j T^{\mu}_{n,j}
\end{equation}
The special form of the divergence term on the righthand side is clearly given by the lower order gauge transformed expressions $d_Q T_m$'s $(m < n)$ \footnote{Confer \cite{scharf}, \cite{kugo}.} except for local terms.
After this basic statements about symmetry transformations one can turn to the special problems which occur in the description of gauge invariance for \textbf{S}--matrices in causal Yang--Mills theory.

\subsection{Gauge Invariance in Yang--Mills theory}	\label{EICH1}

As in path integral method \cite{kugo} as well as in causal Yang--Mills theory, the gauge charge (infinitesimal representative) represents the gauge transformation in the best way. Analogous to causal QED \cite{scharf} one defines the $Q$--operator in causal Yang--Mills as follows\footnote{A good overview of gauge theory in causal QFT is given by \cite{dutsch}.}
\begin{equation}	\label{eich4}
	Q=\int d^3 \!x \p_{\nu} A_a^{\nu} \overset{\leftrightarrow}{\p_0} u_a
\end{equation}
where $A,u$ are free field operators.

\subsubsection{The Algebra of Free Field Operators and Gauge Invariance of \textbf{S}(g) in First Order}	\label{EICH1.1}

As mentioned above, the term $T_1$ must be given as the starting point of the causal theory. In the first step a reasonable $T_1$ is defined by the gluon interacting term $T_1^A$
\begin{equation}	\label{eich6}
	T_1^A := i g f_{abc} :A^{\mu}_a A^{\nu}_b F^{\mu\nu}_c:
\end{equation}		
which is also known as the first order (in $g$) gauge fixing term in path integral method \cite{ryder}, \cite{kugo}, \cite{feynman}, whereas $F^{\mu\nu}_a$ stands as usual for\footnote{In this work only the algebraic structures of the field operators are used, therefore the Lorentz indices can be written as superscripts without any loss of information. This helps to make for a more readable text.}
\begin{equation}	\label{eich7}
	F^{\mu\nu}_a := \p^{\mu}A^{\nu}_a - \p^{\nu}A^{\mu}_a
\end{equation}	
In the second step one can show \cite{YMI}, that the action in first order
\begin{equation}	\label{eich5}
	\textbf{S}_1(g) = 1 + \int d^4 \! x T_1^A(x)g_1(x)
\end{equation}
can only be written gauge invariantly, if one also introduces the free scalar Fermi field $\ti{u}_a$.
A detailed calculation of the gauge transformed of $T_1^A$ shows that $\textbf{S}_1(g)$ in the above written form does not represent a gauge invariant action. But if one extends $T_1^A$ with the ghost term
\begin{equation}	\label{eich8}
	T_1^u := i g f_{abc} :A^{\mu}_a u_b \p^{\mu}\ti{u}_c:
\end{equation}
to the new expression $T_1 := T_1^A + T_1^u$ it results in a gauge invariant $\textbf{S}_1(g)$, since the transformed $T_1$ now writes as
\begin{equation}	\label{eich9}
\begin{aligned}
	\left[ Q , T_1 \right] 	&= g f_{abc} \left[ \p^{\mu}  : A^{\mu}_a 					u_bF^{\mu\nu}_c: + 
	\frac{1}{2} \p^{\nu}  : u_a u_b \p^{\nu}\ti{u}_c: \right]			\\
							&=: \p^{\nu} T_{1,1}^{\nu}
\end{aligned}
\end{equation}
To calculate the above results in detail, it is necessary to know the algebra of the introduced field operators. Therefore the $Q$--transformed $T_1$ can easily be found by applying algebraic identities.
According to the definitions and with the help of earlier remarks, the following identities result for the field operators
\begin{subequations}
\begin{align}
	\Box A_a^{\mu} &= 0									\label{eich10a}	\\
	[A_a^{\mu}(x),A_b^{\nu}(y)] &= i \df_{ab} g^{\mu\nu} D_0 (x-y)
														\label{eich10b}	\\
	[\p_{\mu}A_a^{\mu}(x),\p_{\nu}A_b^{\nu}(y)]=0 										\qquad&\qquad	\phantom{\p_{\nu}}
	[\p_{\mu}A_a^{\mu}(x),F_b^{\lambda\kappa}(y)]=0		\label{eich10c}
\end{align}
\begin{align}
	\Box u_a = 0 	\qquad	\quad&\qquad\qquad	\phantom{u_a}\phantom{u} 
	\Box \ti{u}_a = 0									\label{eich10d}	\\
	\{u_a,u_b\} = 0	\qquad\quad&\qquad\qquad	
	\{\ti{u}_a,\ti{u}_b\} = 0							\label{eich10e}	\\				\{u_a(x),\ti{u}_b(y)\} &= -i \df_{ab} D_0 (x-y)		\label{eich10f}
\end{align}
\end{subequations}
where $D_0$ represents the massless Jordan--Pauli distribution.

But the most important property for this work is held by the charge gauge operator $Q$. 
With the aid of Leibnitz's rule, the product of two such operators lead to
\begin{equation}	\label{eich11}
\begin{aligned}
	Q^2	 	&= \frac{1}{2} \{ Q,Q\} \\
			&= \frac{1}{2} \int d^3 \!x \int d^3 \!y
			 \left\{ \left[ \p_{\mu}A_a^{\mu}(x),\p_{\nu}A_b^{\nu}(y) \right] 				\overset{\leftrightarrow}{\p_{x^0}}
		\overset{\leftrightarrow}{\p_{y^0}}
			 (u_a(x)u_b(y))	\right.										\\
			&\left. \phantom{= \frac{1}{2} \{ Q,Q\} 
			 = \frac{1}{2} \int d^3 \!x \int d^3 \!y 
			 \left\{ [\p_{\mu}\right.}
			 + (\p_{\mu}A_a^{\mu}(x)\p_{\nu}A_b^{\nu}(y))
		\overset{\leftrightarrow}{\p_{x^0}}
		\overset{\leftrightarrow}{\p_{y^0}}
			\{  u_a(x),u_b(y) \} \right\}								\\
			&= 0
\end{aligned}
\end{equation}
where $\eqref{eich10c}$ and $\eqref{eich10e}$ are used.
Obviously $Q$ is a nilpotent operator of second rank. Furthermore, one realizes immediately, that $Q$ defines in the sense of homological algebra (confer e.g. \cite{greub}) a graded differential operator in Fock space $\mathcal{F}$. This is true, since the involution $\omega$ in $\mathcal{F}$
\begin{align}	\label{eich12}
	\omega(F) := (-1)^{Q_g(F)} \cdot F	&&	\forall F \in \mathcal{F}
\end{align}
where $Q_g$ is the ghost charge operator \cite{krahe}
\begin{equation}	\label{eich13}
	Q_g := \int d^3 \! x \, \ti{u}_a(x) \overset{\leftrightarrow}{\p_0}u_a(x)
\end{equation}
and $Q_g(F)$ reads as
\begin{align}	\label{eich14}
	Q_g(F) := [Q_g,F]_{\pm} = zF	&& z \in \mathbb{Z}
\end{align}
builds the quadruplet $\{ \mathcal{F}, Q_g, \omega, Q \}$ which satisfies the conditions for a graded differential algebra \cite{greub}.
According to the usual style in differential form writing, one can write, with all that in mind, the operators $Q$ and $Q_g$ as follows
\begin{align}	
	Q 	&:= d_Q		\label{eich15}	\\
	Q_g	&:= \df_g	\label{eich16}
\end{align}
With this definition the gauge transformation of a field operator $F\in \mathcal{F}$ and equation $\eqref{eich11}$ are writable in the more elegant form
\begin{align}
	d_QF   :=& 	[Q,F]_{\pm} = QF - \omega(F)Q		\label{eich18}	\\
	d_Q^2 	=&	0									\label{eich17}
\end{align}
and the whole set of gauge transformed free field operators is found to satisfy the identities \cite{dutsch},\cite{krahe},\cite{YMI}
\begin{subequations}
\begin{align}
	d_Q A_a^{\mu}(x)			&=	i\p^{\mu}u_a(x)			\label{eich19a}\\
	d_Q \p_{\mu}A_a^{\mu}(x)	&=	0						\label{eich19b}\\
	d_Q F_a^{\mu\nu}(x)			&=	0						\label{eich19c}\\
	d_Q u_a(x)					&=	0						\label{eich19d}\\
	d_Q	\ti{u}_a(x)				&=	-i\p_{\nu}A_a^{\nu}(x)	\label{eich19e}\\
	d_Q	\p^{\mu}\ti{u}_a(x)		&=	-i\p_{\nu}F_a^{\mu\nu}(x)\label{eich19f}
\end{align}
\end{subequations}
As the differential operator $d_Q$ acts linearly in the (graded) Fock space, the product rule leads to
\begin{equation}	\label{eich20}
	d_Q(FG) = d_Q(F)G + \omega(F)d_Q(G)
\end{equation}
With this remark the overview about gauge invariance of the \textbf{S}--matrix in the first order is finished and the algebraic tools, which are needed in the following, are noted.
The subsequent section is devoted to gauge invariance of $T_n$'s in higher order.

\subsubsection{Gauge Invariance in Higher Order}	\label{EICH1.2}

In contrast to the treatment of gauge invariance in the last section, for $T_n$'s $(n \geqslant 2)$ additional difficulties arise. The reason for these problems lies in the occurrence of local terms in the construction of $T_n$'s with singular order $\omega(T_n) \geqslant 0$, which leads to the below given enhanced form of $\eqref{cqft19}$
\begin{align}
	T_n	&=R_n - R_n'														\\
		&= \sum_j r_n^j \mathcal{O}_n^j (\bar{T}_n) - \sum_j {r'}_n^j 														\mathcal{O}_n^j (\bar{T}_n)\notag	\\
		&= \sum_j \left( r^0 + \sum_{q\leqslant \omega} c_q \mathcal{D}^q \df 					\right)_n^j \mathcal{O}_n^j (\bar{T}_n) - {r'}_n^j 																\mathcal{O}_n^j (\bar{T}_n)\notag	\\
		&= \sum_j \left( r^0 - r' \right)_n^j \mathcal{O}_n^j (\bar{T}_n) + 
			\sum_{q\leqslant \omega} c_q \mathcal{D}^q \,\df \;																\mathcal{O}_n^j (\bar{T}_n) 	\label{eich21}
\end{align}
Herein stands $\mathcal{O}(\bar{T}_n)$ for all possible Wick ordered products of $n$ field operators out of $T_n$\footnote{Thus the expression $\eqref{eich21}$ reads as the sum of all possible Wick monomials $\bar{T}_n$ (multiplied with the belonging numerical distribution). Written extendedly, $\eqref{eich21}$ writes as\begin{equation*}
	T_n	= \sum_{\bar{T}_n} (r_n^0 - r_n')\mathcal{O}(\bar{T}_n) +
		\sum_{\bar{T}_n}\sum_{q\leqslant \omega} c_q \mathcal{D}^q \df(0) \mathcal{O}(\bar{T}_n)
\end{equation*}	}.
In contrast, a $T_n$ with $\omega(T_n)< 0 $ is simply given by
\begin{align}
	T_n	&= R_n - R_n'										\notag	\\
		&= \left(r_n^0 - r_n'\right) \mathcal{O}(\bar{T}_n) 	\label{eich22}
\end{align}
Obviously, for gauge invariant treatment of $T_n$'s with $n\geqslant 2$ the additional local terms in $\eqref{eich21}$ and $\eqref{eich22}$ must be taken into account. But based on the inductive construction of the $T_n$'s, the gauge invariant form of the operator valued distributions can be newly calculated in each order with respect to the already determined lower order expressions. Then clearly the operator $Q$ does not act on the numerical distributions, and thus one can consider $[Q,D_n]$ instead of $[Q,T_n]$. The latter is true, since $T_n$ and $D_n$ only differ in their distributive parts.
Executing the gauge transformation on $D_n$ (it will be shown as an example for \mbox{$n=2$})\footnote{\label{eichfoot1}For the cases with $n>2$ one has to take into account a sum over all partitions of two subsets of the form $(X,y)$, with $X$ a subset of $n-1$ points out of $\mathbb{M}^n$ instead of a sum over two terms in $\eqref{eich23}$.} one acquires
\begin{align}
	\left[ Q,D_2 \right]
		&= \left[ Q,\left[ T_1(x),T_1(y) \right] \right]	\notag	\\
		&= \left[\left[ Q, T_1(x)\right],T_1(y) \right] -
			\left[\left[ Q, T_1(y)\right],T_1(x) \right]	\label{eich23}
\intertext{where the inner brackets in equation $\eqref{eich23}$ can be written as divergences according to the presuppositions. This means, that the above treated transformation $d_Q$ of $T_2$, since $T_1$ is already found as gauge invariant, writes as}
		&= [ \p^{\mu}_x T^{\mu}_{1,1}(x),T_1(y) ] -
			[ \p^{\mu}_y T^{\mu}_{1,1}(y),T_1(x) ]	\notag	\\
		&=  \p^{\mu}_x  [T^{\mu}_{1,1}(x),T_1(y) ] -
			 \p^{\mu}_y [ T^{\mu}_{1,1}(y),T_1(x) ] \label{eich24}
\end{align}
The latter shows, that the whole expression is a divergence\footnote{For $n>2$
the terms $T_{n-1}(X)$ in footnote $\ref{eichfoot1}$ are writable as divergences. Therefore, the sum over all partitions also describes a divergence.}.
Altogether $D_n$ trivially is gauge invariant!
But in the above considerations it was neglected, that in the transition from $D_n$ to $T_n$
\begin{equation*}										
	D_n \longrightarrow T_n
\end{equation*}
the form of the numerical distribution
\begin{equation*}
	d_n = r_n' - a_n'\longrightarrow t_n =
		\begin{cases}
			r_n^0 -r_n'					& \omega < 0			\\
			r_n^0 -r_n' + \sum \df		& \omega \geqslant 0
		\end{cases}
\end{equation*}
does not keep its structure. To include the possibly neglected local terms into the calculations, one had used until now the following method for the different $T_n$'s:
\\[2ex]
\textbf{The case $T_2$:} Here every single local term is treated on its own \cite{YMI}. The free variables $c_q$ in $\eqref{cqft30}$ or $\eqref{cqft31}$ were chosen in such a way, that all local terms emerging from the splitting procedure and the local terms coming from $[Q,T_n]$ figure up to zero \cite{YMI}. Doing so, one got the $4$--gluon interacting term similar to the one occurring in path integral method.
\\[2ex]
\textbf{The case $T_n \; (n>2)$:} One had determined the divergence of local terms origin from the splitting procedure with the aid of $C_g$--identities \cite{YMII}, \cite{YMIII}.
Similar to the treatment above, the free variable $c_q$ were again so defined, that all local terms added up to zero \cite{YMII}, \cite{YMIII}, \cite{YMIV}.
\\[2ex]
With all the above results taken into account, one was lead to the gauge invariant form $\eqref{eich3a}$ of the $T_n$'s.
With this result the overview of the current calculation method for gauge invariance calculations of massless \textbf{S}--matrices in causal Yang--Mills theory is finished. In the next chapter we will get familiar with the new method of algebraic determination of invariant terms $T_n$.

\subsection{Summary of Chapter \mbox{\boldmath $\ref{EICH}$}} 	\label{EICH2}

It was shown, that the \textbf{S}--matrix is exactly gauge invariant, if the transformed $T_n$ takes the form
\begin{equation*}
	d_Q T_n = [Q,T_n] =\p^{\mu} \sum_j T_{n,j}^{\mu}
\end{equation*}
where $d_Q$ is a nilpotent differential operator of second rank in Fock space $\mathcal{F}$ and $\{ \mathcal{F}, Q_g, \omega, Q \}$ represents a graded differential algebra. Furthermore gauge invariance of $T_1$ was executed explicitely. Additionally it has been resumed, that gauge invariance treatment for $T_n$'s with $n \geqslant 2$ can be split into two separate calculations. The first step comprises the trivial invariance of $D_n$--terms. The second one separately considers the local terms occurring in the splitting procedure.
\section{Algebraic Determination of Gauge Invariant \textbf{S}--Matrix in Causal Yang--Mills Theory} 		\label{ALGEBRA}

In the preceding chapter it was shown, that the quadruple $\{ \mathcal{F}, Q_g, \omega, Q \}$ builds a graded differential algebra. Obviously, one should use this additional structure of Fock space $\mathcal{F}$ to simplify the calculations of gauge invariance.
A strong algebraic tool for analyzing a differential algebra is the cohomology theory. As in path integral method \cite{dixon1}, \cite{dixon2}, \cite{piguet}, one can adapt the definition of the cohomology group to the specific problem at hand\footnote{The linear part \textbf{$\df_{\trm{BRS}}$} of the BRS--operator~\textbf{$s$} modulo divergence $d$ \cite{piguet}, which is used to define the cohomology in path integral methods, does not fulfil the restrictions for our problem. Confer for this topic the following considerations.}.
In the subsidiary sections the algebraic determination of gauge invariant $T_n$'s will be systematically introduced.
The sections are comprised of the following problems: part $\ref{ALGEBRA1}$ compares the different gauge invariant \textbf{S}--matrix constructions. In $\ref{ALGEBRA2}$ it will be shown, which graphs can contain gauge destroying local terms and thus must be considered further. In the last section $\ref{ALGEBRA3}$ the gauge factor group for causal Yang--Mills theory will be introduced.

\subsection{The Gauge Invariant \textbf{S}--Matrix Constructions} 																				\label{ALGEBRA1}

As already mentioned in chapter $\ref{EICH1.2}$, the inductive gauge invariance proof can be split into two separate problems. In this work we does not follow the ideas used in \cite{YMI} -- \cite{YMIV} (e.g. the way via $C_g$--identities) but rather prove gauge invariance with the aid of the gauge--factor group which will be defined below.
Finding gauge invariant $T_n$'s is best described by the following diagram~3.1
\scriptsize
\begin{diagram}
\sum_{\trm{\tiny{Part}}} \ti{T}_{n-l} \cdot T_l 						&																					&
\rMapsto																&																					&
\sum_{\trm{\tiny{Part}}} \ti{T}_{n-l}\cdot T_l \biggr|_{\text{ret}} + 						\sum^{\omega}_{i=0} \chi_i \mathcal{D}^i \df				\\																					&
D_n																		&			\rTo^{splitting}														&
T_n \bigr|_{\text{ret}}													&
																		\\
\dMapsto																&
\dTo^{d_Q}																&
																		&
\dTo_{d_Q}																&
\dMapsto																\\
																		&
																		&																					&
d_Q \left(T_n\bigr|_{\text{ret}} \right)								&
d_Q \left( \sum_{\trm{\tiny{Part}}}\ti{T}_{n-l}\cdot T_l 
			\biggr|_{\text{ret}} \right) + 
					\sum^{\omega}_{i=0} d_Q \chi_i \mathcal{D}^i \df	\\
																		&
d_Q D_n																	&
\rTo^{splitting}														&
(d_Q D_n )\bigr|_{\text{ret}}											&
																		\\
d_Q \sum_{\trm{\tiny{Part}}} \ti{T}_{n-l} \cdot T_l					&
																		&			\rMapsto^{splitting}													&
d_Q\left(\sum_{\trm{\tiny{Part}}}\ti{T}_{n-l}\cdot T_l \right) 
\Biggr|_{\text{ret}} + \sum^{\omega '}_{i=0} \chi'_i \mathcal{D}^i \df	&
																		\\
\dIdentic_{ind. hyp.}													&
																		&
																		&
																		&
																		\\
\p_{\mu} \sum_j\left( \sum_{\trm{\tiny{Part}}} \ti{T}_{n-l} 
									\cdot T_l \right)^{\mu}_{,j}		&
																		&
\rMapsto^{splitting}													&
\p_{\mu} \sum_j\left( \sum_{\trm{\tiny{Part}}}\ti{T}_{n-l}\cdot 
				T_l \right)^{\mu}_{,j} \Biggr|_{\text{ret}} + 
					\sum^{\omega '}_{i=0} \bar{\chi}'_i \mathcal{D}^i \df	&
																		\\
&&&&\\
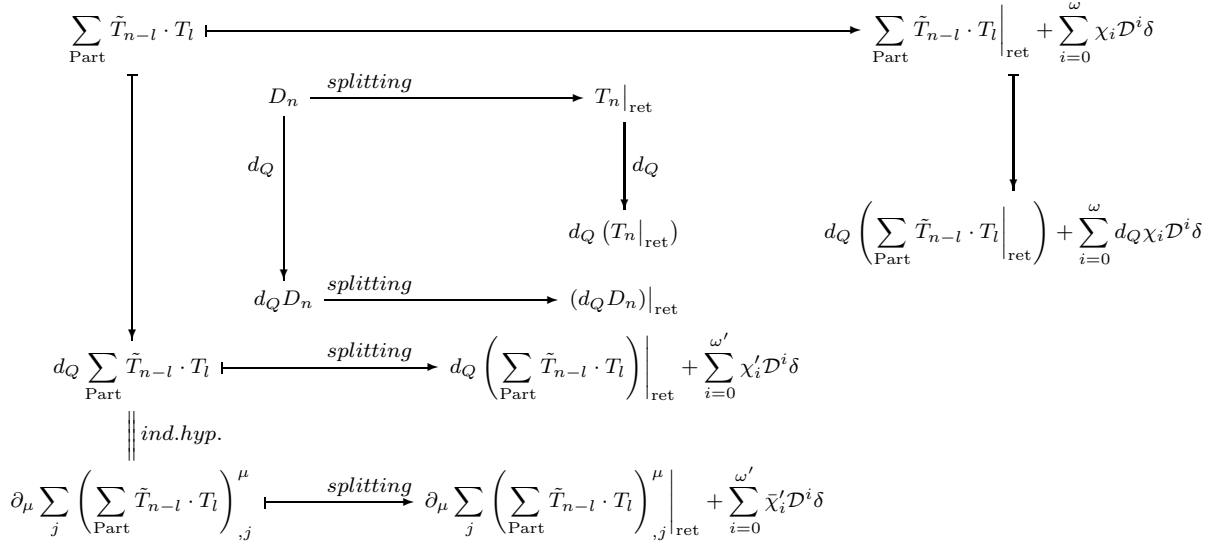
&&&\text{\normalsize diagram 3.1}&										\\
\end{diagram}
\normalsize
\\[.3cm]
where the subsequent remarks help to understand the above representation
\begin{itemize}
	\item	all $T_m$'s with $(m<n)$ are, according to inductive presumption, 					already expressible as gauge invariant terms
			\begin{align}		\label{algeb1}
				d_Q T_m = \sum_j \p^{\mu} T^{\mu}_{m,j}		&& m<n
			\end{align}
	
	\item	the expressions $(\ldots)\bigr|_{\text{ret}}$ represent the retarded 			operator valued distributions which are given by the central 						splitting solution of the embraced parts.
	
	\item	for a better overview about the mappings, the outer part of the 					diagram also indicates the terms contained in the considered sets.
\end{itemize}	
In the following steps the different mappings from $D_n$ to $d_Q (T_n \bigr|_{\text{ret}})$ and $D_n$ to $(d_Q D_n)\bigr|_{\text{ret}}$ in the diagram above will be explored respectively.
Since the diagram~3.1 does not commute
\begin{equation}	\label{algeb2}
	d_Q (T_n \bigr|_{\text{ret}}) \ne (d_Q D_n)\bigr|_{\text{ret}}
\end{equation}
one seems to be forced to calculate gauge invariant $T_n$'s clockwise, e.g. by the sequence of the mappings
\begin{equation}	\label{algeb3}
	d_Q \circ splitting \,(D_n) \overset{?}{=} \sum_j \p^{\mu} T^{\mu}_{n,j}
\end{equation}
But this construction rule does not take into account the already determined (by induction) gauge invariant $T_m$'s, and so one should try to adapt the alternative
\begin{equation}	\label{algeb4}
	splitting \circ d_Q \,(D_n) \overset{?}{=} \sum_j \p^{\mu} T^{\mu}_{n,j}
\end{equation}
counter clockwise mapping in the diagram above, which respects inductive results. The latter is correct, since the linearity of $d_Q$ allows the possibility of writing the interesting terms in form $II$ (in diagram 3.2 below) which is well suited to the induction hypothesis.
At first sight, there seems to be a high price paid for the embedding of recursive information, since diagram 3.2 does not commute. So we have to focus ourselves on the difference between the considered terms $d_Q (T_n \bigr|_{\text{ret}})$ and $(d_Q D_n)\bigr|_{\text{ret}}$. With this knowledge, we are led to a well defined construction rule for the mapping $\eqref{algeb4}$ instead of $\eqref{algeb3}$.
The difference in $\eqref{algeb2}$ is best found by simply comparing the images 
of any term
\begin{align}	\label{algeb5}
	\ti{t}_{n-l}\ti{\chi}_{n-l}\cdot t_l \chi_{l} \in D_n 
	&& t_k \in \mathcal{S} (\mathbb{M}^{\otimes k}),\,
	 \chi_{k} \in \mathcal{O}(\bar{T}_k)
\end{align}
of $D_n$ in diagram 3.1
\scriptsize
\begin{diagram}
\ti{t}_{n-l}\ti{\chi}_{n-l}\cdot t_l \chi_{l}							&		\rMapsto^{splitting}													&			\ti{t}_{n-l} t_l \Bigr|_{\text{ret}} \ti{\chi}_{n-l} \cdot \chi_{l}+ 											\sum^{\omega}_{i=0} \chi_i \mathcal{D}^i \df\\																					&																					&
\dMapsto^{d_Q}															\\
\dMapsto^{d_Q}															&
																		&
\underbrace{\ti{t}_{n-l} t_l \Bigr|_{\text{ret}}(d_Q \ti{\chi}_{n-l}\cdot\chi_{l} + 
\ti{\chi}_{n-l}\cdot d_Q \chi_{l})}_{\trm{I}} + 
\sum^{\omega}_{i=0} d_Q\chi_i\mathcal{D}^i \df							\\
																		&
																		&																					\\
\begin{minipage}[b][7ex]{35ex}
	\begin{multline*}
		\ti{t}_{n-l} d_Q \ti{\chi}_{n-l}\cdot t_l \chi_{l}  	\\  
		 + \ti{t}_{n-l} \ti{\chi}_{n-l}\cdot t_l d_Q \chi_{l}
	\end{multline*}
\end{minipage}															&
\rMapsto^{splitting}													&
\underbrace{\ti{t}_{n-l}t_l \Bigr|_{\text{ret}} (d_Q \ti{\chi}_{n-l}\cdot\chi_{l} + 
	\ti{\chi}_{n-l}\cdot d_Q \chi_{l})}_{\trm{II}} +
	 	\sum^{\omega '}_{i=0} \chi'_i \mathcal{D}^i\df 					\\
\dIdentic_{ind. hyp.}													&
																		&
																		\\
\begin{minipage}[c][7ex]{39ex}
	\begin{multline*}
		\p_{\mu}^{n-l}\sum_j[(\ti{t}_{n-l}\ti{\chi}_{n-l})^{\mu}_{,j} 
										\cdot t_l\chi_{l}] \\  
		+ \p_{\mu}^{l}\sum_j[\ti{t}_{n-l}\ti{\chi}_{n-l}\cdot 																	(t_l\chi_{l})^{\mu}_{,j}]
	\end{multline*}
\end{minipage}															&
\rMapsto^{splitting}													&
\underbrace{\p_{\mu}^{n-l}\sum_j [(\ti{t}_{n-l}\ti{\chi}_{n-l})^{\mu}_{,j} \cdot t_l \chi_{l}] \Bigr|_{\text{ret}} + \p_{\mu}^{l}\sum_j[\ti{t}_{n-l}\ti{\chi}_{n-l}\cdot 
(t_l\chi_{l})^{\mu}_{,j}] \Bigr|_{\text{ret}}}_{\trm{III}} + 
\sum^{\omega '}_{i=0} \bar{\chi}'_i \mathcal{D}^i \df					\\
&&\\
&&\makebox[4in][l]{\normalsize diagram: 3.2}							\\
\end{diagram}
\normalsize
\\[.3cm]
Whereas the same notation was used as in diagram~3.1.
Additionally, $d_Q$ acts only on the Wick part of the terms. Immediately one realizes by comparison of the expressions $\trm{I}$--$\trm{III}$, that the three images
\begin{subequations}
 	\begin{align}
 		&d_Q \circ splitting \, (\ti{t}_{n-l}\ti{\chi}_{n-l}\cdot t_l 						\chi_{l})								\label{algeb6a}	\\
		&splitting \circ d_Q \, (\ti{t}_{n-l}\ti{\chi}_{n-l}\cdot t_l 						\chi_{l})								\label{algeb6b}	\\
		&splitting \circ ind.hyp \circ d_Q \, (\ti{t}_{n-l} \ti{\chi}_{n-l} 					\cdot t_l \chi_{l})					\label{algeb6c}
	\end{align}
\end{subequations}
does not differ in their not--local parts, since based on the inductive construction the expressions $\trm{II}$ and $\trm{III}$ are equivalent and $\trm{I}$ and $\trm{II}$ obviously does not differ!
Combining all the above results one finds, that gauge invariance of $T_n$'s, by causal inductive construction, can be destroyed only by local terms. This leads to the question, in which graphs can such local terms arise at all? 
The answer is given in the next section.

\subsection{Determination of Non Gauge Invariant Graphs}	\label{ALGEBRA2}

As shown above, only local expressions can destroy gauge invariance. But not all $T_n$ of $\mathbb{M}^{\otimes n}$ can lead to local terms.
Remembering equation$\eqref{cqft35}$ and the above remarks, one realizes, that only $T_n$'s with a singular order $\omega$
\begin{equation} 	\label{algeb7}
	\omega(\trm{graph}) \geqslant 0
\end{equation}
result, by causal construction, in gauge destroying parts. From \cite{YMII} the estimation of the local order of a graph in Yang--Mills is already known. It is shown there, that the range of the singular order of an arbitrary operator valued distribution is limited by
\begin{align}	\label{algeb8}
	\omega \leqslant 4 - b - g_u -g_{\tilde{u}} - d
\end{align}
where the following abbreviations were used
\begin{align*} 
	b 			&\text{ for the number of external gluons}	\\
	g_u 		&\text{ for the number of external ghosts}	\\
	g_{\ti{u}} 	&\text{ for the number of external anti--ghosts}	\\
	d 			&\text{ for the number of derivations on external field 																			operators}
\end{align*}
With the aid of the mapping $d_Q$ of free field operators in $\eqref{eich19a}$ --$\eqref{eich19f}$ one immediately finds with respect of $\eqref{algeb8}$, that the gauge transformation raises the singular order of a considered term by one
\begin{align}	\label{algeb9}
	\omega(d_Q \mathit{f}) = \omega (\mathit{f}) + 1 && \forall \mathit{f} \in \mathcal{F}
\end{align}
The latter is obvious, because $d_Q$ leads to an additional derivation in the field operator part of the terms.
\begin{quote}
	\emph{Altogether one finds, that only those operator valued distributions 					$T_n$ in diagram~3.1 and 3.2 contain local terms, which have at most 			five external legs in their accompanying graphs.}
\end{quote}
This is best seen with the aid of $\eqref{algeb8}$ and $\eqref{algeb9}$:
Since the interesting local expressions occur either in $d_Q \left( T_n \bigr|_{\text{ret}} \right)$ or $(d_Q D_n )\bigr|_{\text{ret}}$, the maximal singular order for both cases is given by
\begin{align}										
	\omega \left[ (d_Q D_n )\bigr|_{\text{ret}} \right]
		&= \omega (d_Q D_n )							\notag	\\
		&= \omega(D_n) + 1								\notag	\\
		&\leqslant 4 - b - g_u -g_{\tilde{u}} - d + 1	\label{algeb10} 
\end{align}
with respect to the fact, that the singular order is not changed by the splitting procedure\footnote{Confer for this considerations \cite{epstein}.}.
Furthermore $\eqref{algeb9}$ was used for the second identity and in the last step $\omega$ was substituted by $\eqref{algeb8}$\footnote{Obviously the same holds for the expression $d_Q \left( T_n \bigr|_{\text{ret}} \right)$.}. 
Thus one finds immediately with the aid of $\eqref{algeb10}$ under the restriction $d=0$, that only graphs with a maximum of five external legs~$j:= b + g_u + g_{\tilde{u}}$ have singular order $\omega \geqslant 0$, since
\begin{align}	\label{algeb11}
	0 \leqslant \omega \left[ (d_Q D_n )\bigr|_{\text{ret}} \right] \leqslant
		4 +1 - j_{\trm{max}} && \Longleftrightarrow && j_{\trm{max}}=5
\end{align}
Up until now it was disregarded, that in the set of all local terms of $\eqref{algeb6b}$ and $\eqref{algeb6c}$ (the lower two rows in diagram~3.2) only local expressions occurring in $\eqref{algeb6a}$ must be considered. This is obvious, since after the construction rule each $d_Q$--transformed $T_n$ must be calculated according to $\eqref{algeb6a}$, whereas the expressions $(I)$ -- $(III)$ in diagram~3.2 (as shown above) remain untouched. Consequently only the local expressions of $\eqref{algeb6a}$ must be taken into consideration for gauge invariance determination.

This leads to the additional restriction, that only terms with a maximum limit of four external legs have to be taken into account, because under the mapping $d_Q$ the mathematical structure of the local expression part in the set $T_n \bigr|_{\text{ret}}$ will not be affected, and the latter set has according to relation $\eqref{algeb8}$ a maximum singular order of four.
So, we can indeed restrict ourselves to local expressions with up to four external legs.
With this statement all the preliminary facts are listed, and one can advance in the following section to the definition of an appropriate gauge--factor group.

\subsection{Definition of the Gauge--Factor Group}	\label{ALGEBRA3}

In order to simplify the notation in the subsequent calculations, the following definitions will be introduced:
The set of all local terms in Fock space $\mathcal{F}$ is defined as
\begin{align}	
	\mathcal{F}_{\trm{loc}} &:= \left\{ t_i \chi_i \mid t_i \in \mathcal{D}^q 			\df\;,\;\chi_i \in :\mathcal{O}_i: \right\}			\label{algeb12}
\intertext{Furthermore the sets of all local expressions with ghost number $n_u = 0$, $n_{u} = 1$ are signed by  $\mathcal{M}$ and $\mathcal{L}$ respectively}	
	\mathcal{M}
		&:=\left\{ \mathit{f} \in \mathcal{F}_{\trm{loc}} \mid n_u(f)= 0\right\}															\label{algeb13}	\\	
	\mathcal{L} 
		&:=\left\{ \mathit{f} \in \mathcal{F}_{\trm{loc}} \mid n_u(f)= 1\right\}															\label{algeb14}
\end{align}
As shown in section~$\ref{ALGEBRA1}$ the occurring terms $\eqref{algeb6a}$ --$\eqref{algeb6c}$	(in diagram~3.1 and diagram~3.2) only differ in their local expressions. Because of the different compositions of the mappings one has no control over whether the latter terms are the same in both cases or not.
The sums can essentially vary, and so it is not wise to treat the entire expression. It is better to examine the behaviour of each local term of the sum under the transformation $d_Q$ on its own.
For if one can show, that every single local term can be written in a gauge invariant way\footnote{More about gauge invariance follows below.}, the same applies to all finite linear combinations\footnote{The singular order of the operator valued distributions in Yang--Mills is limited by the relation $\eqref{algeb11}$ in chapter~$\ref{ALGEBRA2}$, and so the sum must be finite.\label{eichfuss2}}, since the mappings $d_Q, \p$ which act on the local terms are linear.
\begin{quote}
	Thus, to be able to express gauge invariance of the sum of local terms in 			$\eqref{algeb6a}$, one can show invariance for \emph{each single term} 				$\mathit{l}_i \in \mathcal{L}$ \emph{alone}, and rebuild the wanted sum 			afterwards as a linear combination.
\end{quote}
The latter obviously is true, since the set $\mathcal{L}\bigl|_{\eqref{algeb10}}$ contains all possible local terms occurring in $\eqref{algeb6a}-\eqref{algeb6c}$
\begin{equation}	\label{algeb15}
	\mathcal{L}\bigl|_{\eqref{algeb10}} \supset \left(\sum^{\omega '}_{i=0} 			\bar{\chi}'_i \mathcal{D}^i \df +
	\sum^{\omega '}_{i=0} \chi'_i \mathcal{D}^i \df	\right)
\end{equation}
But in which way does the above mentioned invariance of local terms manifest itself? To see this in the base construction, it is best to rewrite diagram~3.1 with all the above founded restrictions and identities inserted
\scriptsize
\begin{diagram}
\dFromout&&&&															\\
&&&&																	\\		
d_Q \sum_{\trm{\tiny{Part}}} \ti{T}_{n-l} \cdot T_l					&
																		&			\rMapsto^{splitting}													&
d_Q\left(\sum_{\trm{\tiny{Part}}}\ti{T}_{n-l}\cdot T_l \right) 
\Biggr|_{\text{ret}}													&
																		\\
\dIdentic_{ind. hyp.}													&
																		&
																		&
																		&
																		\\
\p_{\mu} \sum_j\left( \sum_{\trm{\tiny{Part}}} \ti{T}_{n-l} 
									\cdot T_l \right)^{\mu}_{,j}		&
																		&
\rMapsto^{splitting}													&
\p_{\mu} \sum_j\left( \sum_{\trm{\tiny{Part}}}\ti{T}_{n-l}\cdot 
T_l \right)^{\mu}_{,j} \Biggr|_{\text{ret}} + \sum^{\omega '}_{i=0} \bar{\chi}'_i \mathcal{D}^i \df											&
																		\\
&&&&																	\\
&&\text{\normalsize diagram 3.3}&&										\\
\end{diagram}
\normalsize
\\[.3cm]
A special remark should be made about the expression on the lower right--hand side. Wherein all local terms which can occur in $\eqref{algeb6b}$ and $\eqref{algeb6c}$ are added. This is possible according to the identity ''$ind. hyp.$'' in diagram~3.3, which lets commute the inquired sub-diagram. Obviously this leads to a local term--free expression shown in the top right line of diagram~3.3 \footnote{Confer the definite consideration in splitting procedure e.g. in \cite{scharf}\label{eichfuss3}}.
With these remarks one clearly finds, that gauge invariance of $d_Q D_n\bigr|_{\text{ret}} + \mathit{l}_i$ is demonstrated, if for \emph{every single local term} $\mathit{l}_i \in \mathcal{L}$ exists a $\mathit{m}_i \in \mathcal{M}$ and a $\tilde{\mathit{l}}^{\mu}_i \in \mathcal{L}$ in such a way, that
\begin{align}	\label{algeb16}
	\mathit{l}_i = d_Q\mathit{m}_i + \p^{\mu}\tilde{\mathit{l}}^{\mu}_i
	&&\forall \, \mathit{l}_i \in \mathcal{L} \quad \exists \quad \mathit{m}_i 			\in \mathcal{M} \,,\, \tilde{\mathit{l}}^{\mu}_i \in \mathcal{L}
\end{align}
holds. Then all linear combinations of local terms in diagram~3.3 can be rewritten using the identity above in the following way
\scriptsize
\begin{diagram}
\dFromout&&&&															\\
&&&&																	\\
d_Q \sum_{\trm{\tiny{Part}}} \ti{T}_{n-l} \cdot T_l					&
																		&			\rMapsto^{splitting}													&
d_Q\left(\sum_{\trm{\tiny{Part}}}\ti{T}_{n-l}\cdot T_l \right) 
\Biggr|_{\text{ret}}													&
																		\\
\dIdentic_{ind. hyp.}													&
																		&
																		&
																		&
																		\\
\p_{\mu} \sum_j\left( \sum_{\trm{\tiny{Part}}} \ti{T}_{n-l} 
									\cdot T_l \right)^{\mu}_{,j}		&
																		&
\rMapsto^{splitting}													&
\p_{\mu} \sum_j\left( \sum_{\trm{\tiny{Part}}}\ti{T}_{n-l}\cdot 
T_l \right)^{\mu}_{,j} \Biggr|_{\text{ret}} + \sum _i d_Q\mathit{m}_i + \p^{\mu}\tilde{\mathit{l}}^{\mu}_i										&
																		\\
&&&&																	\\
&&\text{\normalsize diagram 3.4}&&										\\
\end{diagram}
\normalsize
\\[.3cm]
and since the diagrams~3.3 and 3.4 commute, as mentioned above, the $d_Q$--term can simply be taken to the left hand side in $\eqref{algeb17}$ below
\begin{equation}	\label{algeb17}
	d_Q\Big(\sum_{\trm{\tiny{Part}}}\ti{T}_{n-l}\cdot T_l \Big) 
	\biggr|_{\text{ret}} - \sum _i d_Q\mathit{m}_i = 
	\p_{\mu} \sum_j \Big( \sum_{\trm{\tiny{Part}}}\ti{T}_{n-l}\cdot 
	T_l \Big)^{\mu}_{,j} \biggr|_{\text{ret}} + 										\sum _i \p^{\mu}\tilde{\mathit{l}}^{\mu}_i
\end{equation}
With the linearity of the operators $d_Q$ and $\p^{\mu}$ the equation shows gauge invariance in the required way
\begin{equation}	\label{algeb18}
	d_Q\bigg(\sum_{\trm{\tiny{Part}}}\ti{T}_{n-l}\cdot T_l 
	\biggr|_{\text{ret}} - \sum _i \mathit{m}_i \bigg) = 
	\p_{\mu} \bigg(\sum_j \Big( \sum_{\trm{\tiny{Part}}}\ti{T}_{n-l}\cdot 
	T_l \Big)^{\mu}_{,j} \biggr|_{\text{ret}} + 										\sum _i \tilde{\mathit{l}}^{\mu}_i \bigg)
\end{equation}
The two identities $\eqref{algeb16}$ and $\eqref{algeb18}$ furthermore lead to the observation, that if two elements $\mathit{l}_i,\mathit{l}_j \in \mathcal{L}$ (or $\mathit{m}_i,\mathit{m}_j \in \mathcal{M}$) are equivalent up to a divergence
\begin{align}	\label{algeb19}
	&\mathit{l}_i  = \mathit{l}_j + \p^{\mu}\tilde{\mathit{l}}^{\mu}_i	
	&&\mathit{m}_i = \mathit{m}_j + \p^{\mu}\tilde{\mathit{m}}^{\mu}_i
\end{align}
they either both result in gauge invariant expressions (which differ only in a total divergence) or neither of them do.
\\[1ex]
In mathematical terminology the latter reads as follows:
%
Not all local terms of $\mathcal{L}$ and $\mathcal{M}$ have to be taken into consideration for our calculations, instead only one representative $\mathit{l}_i$ and $\mathit{m}_i$ of each coset of $\n$ in $\mathcal{L}$ and of $\n$ in $\mathcal{M}$ need to be considered. Explicitly, only local terms $\mathit{l}_i'$, $\mathit{m}_i'$ out of the factor sets
\begin{align}
	\mathcal{L}' &:= \,\raisebox{1ex}{$\mathcal{L}$}
		\raisebox{0ex}{$\! \Big/ \!$} \raisebox{-1ex}{$\n$}	\label{algeb20}
	\\[1ex]
	\mathcal{M}' &:= \, \raisebox{1ex}{$\mathcal{M}$} 										\raisebox{0ex}{$\! \Big/ \!$}\raisebox{-1ex}{$\n$}	\label{algeb21}
\end{align}
with $\n$ representing a total divergence of a local term, must be considered in detail.
After all the preparatory explanations we can progress to the definition of an adequate gauge--factor group for our problem.

First of all the sets of modified closed and exact local forms on Fock space are introduced as follows: Let $\mathit{l}_i'$ be a \emph{single}\footnote{Confer for this restriction to footnotes $\eqref{eichfuss2} - \eqref{eichfuss3}$ and the additional remarks in this chapter.} local term of $\mathcal{L}'$, then one defines the sets
\begin{equation}
	\textbf{Z}(\mathcal{M})	\bigcup_{\mathit{l}_i' \in \mathcal{L}'} 
		\textbf{z}(\mathit{l}_i'\,,\,\mathcal{M})			\label{algeb22.1}
\end{equation}
\begin{equation}
	\textbf{B}(\mathcal{M})	:= \big\{ \textbf{b}(\mathcal{M}) \big\}																			\label{algeb22.2}
\end{equation}
%
which comprises all elements
\begin{align} 
	\textbf{z}(\mathit{l}_i'\, , \,\mathcal{M}) &:= 
	\big\{ \mathit{m}_i \in \mathcal{M} \bigm| d_{Q}\mathit{m}_i \equiv 					\mathit{l}_i'	\pmod \n  , \; \mathit{l}_i' \in \mathcal{L}' \big\}
															\label{algeb22}
	\\[1ex]
  	\textbf{b}(\mathcal{M}) &:= 
	\big\{ \mathit{m}_i \in \mathcal{M} \bigm| \mathit{m}_i \equiv 							d_{Q}(\mathit{f}_i) \pmod \n  , \; \mathit{f}_i \in 									\mathcal{F} \bigr|_{n_u=-1} \big\}				\label{algeb23}
\end{align}
respectively. Remembering that the quadruplet $\{ \mathcal{F}, Q_g, \omega, Q \}$ is a differential algebra, one immediately understands, that the set
\begin{equation}	\label{algeb24}
	\textbf{H}(\mathcal{M}) := \, 
	\raisebox{1ex}{$\textbf{Z}(\mathcal{M})$} 													\raisebox{0ex}{$\! \Big/ \!$} 
	\raisebox{-1ex}{$\textbf{B}(\mathcal{M})$}	
\end{equation}
builds a related factor group, from now on referred to as the gauge--factor group.
With all the above notations in mind, one can reformulate the problem of finding a gauge invariant form of each $\mathit{l}_i \in \mathcal{L}$ (occurring in equation $\eqref{algeb16}$) as
\begin{quote}
	A $\mathit{l}_i \in \mathcal{L}$ is exactly then gauge invariantly 					writable, if $\mathit{l}_i$ is an element of the gauge--factor group 				$\textbf{H}(\mathcal{M})$.
	For each $\mathit{h}_i \in \textbf{H}(\mathcal{M})$ has after construction 			the $d_Q$--transformed form
	\begin{align}	\label{algeb25}
		d_Q(\mathit{h}_i) = 0 + \p^{\mu} \mathit{f}_i^{\mu} + \mathit{l}_i
		&& \text{with } \mathit{l}_i \in \mathcal{L}_i, \; \mathit{f}_i^{\mu} 				\in \mathcal{L}
	\end{align}
	with $\mathit{l}_i$ again a \emph{single} local term out of $\mathcal{L}_i$.
\end{quote}
A simple rearranging of the last equation leads immediately to the insight, that $\eqref{algeb16}$ and $\eqref{algeb25}$ are identical, and thus $\eqref{algeb25}$ states nothing else than gauge invariance of 
$d_Q D_n\bigr|_{\text{ret}} + \mathit{l}_i'$.
Altogether we have found, that the set of all gauge invariantly writable local terms $\mathit{l}_i' \in \mathcal{L}'$ can be written as the gauge transformed elements $\mathit{h}_i$ of $\textbf{H}(\mathcal{M})$ modulo a divergence. Furthermore, the subsets $\textbf{B}(\mathcal{M})$ and $\textbf{Z}(\mathcal{M})$ build an additional structure on $\mathcal{M}$. Visualizing the above results is best done by the following figure 1,
\begin{figure}[h]
\footnotesize
\setlength{\unitlength}{.7mm}
\linethickness{0.2mm}
\begin{center}
	\begin{picture}(200,80)(-15,-10)
		\put(0,0){\begin{picture}(80,60)\closecurve(0,30, 40,20, 80,30, 40,40) 								\end{picture}}
		\put(20,18){\begin{picture}(60,40)\closecurve(-14,12, 16,6, 48,12,16,18)
						\end{picture}}
		\put(26,18){\begin{picture}(30,20)\closecurve(0,12, 16,8, 32,12, 16,16)
						\end{picture}}
 		\put(120,0){\begin{picture}(40,60)\closecurve(0,30, 13,0, 26,30, 13,60)
						\end{picture}}
 		\put(122,14){\begin{picture}(80,60)\closecurve(0,19.5, 9.5,0 																		,19,19.5, 9.5,39)
						\end{picture}}
	\scriptsize
		\put(85,30){\linethickness{0.3mm}\vector(1,0){30}}
		\put(70,30){$\mathcal{M}$}
 		\put(13,31){\textbf{B}($\mathcal{M}$)}
 		\put(43,28.5){\textbf{Z}($\mathcal{M}$)}
 		\put(97.5,32){\text{$d_Q$}}
 		\put(123,30){$d_Q \textbf{H}(\mathcal{M}$)}
 		\put(133,3){$\mathcal{L}'$}
	\end{picture}
\end{center}
\normalsize	
\caption{The mapping $d_Q$}		
\label{picture1}
\end{figure}
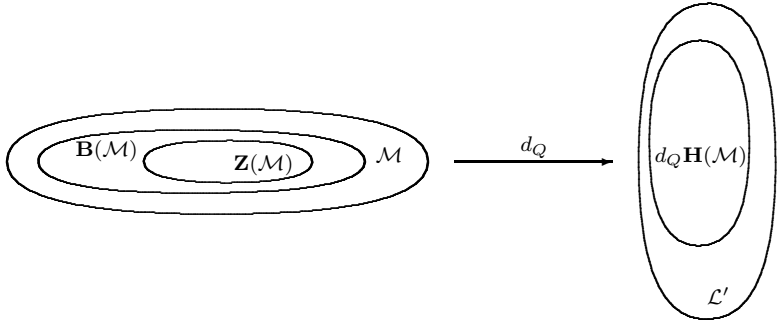
which shows the mapping of $\textbf{H}(\mathcal{M})$ under $d_Q$.
So, to find whether $\mathcal{L}'$ has gauge invariant form or not, one has to execute the following tasks:
\begin{enumerate}
	\item	determination of the elements of $\mathcal{L}'$ (confer the 						definition $\eqref{algeb20}$)
	\item	determination of modified closed and exact local forms of 							$\mathcal{F}_{\trm{loc}}$ according to the definitions 								$\eqref{algeb22}$ and $\eqref{algeb23}$
	\item	determine the gauge--factor group $\textbf{H}(\mathcal{M}) := \, 
			\raisebox{1ex}{$\textbf{Z}(\mathcal{M})$} \raisebox{0ex}{$\! \Big/ 					\!$}\raisebox{-1ex}{$\textbf{B}(\mathcal{M})$}$
	\item	$d_Q$--mapping of $\textbf{H}(\mathcal{M})$ onto $\mathcal{L}'$
	\item	demonstrating, that all elements of $\mathcal{L}'$ are contained in 				$d_Q(\textbf{H}(\mathcal{M}))$.
\end{enumerate}			
The last step results in the fact, that the sets $d_Q(\textbf{H}(\mathcal{M}))$ and $\mathcal{L}'$ are equivalent\footnote{This conclusion clearly holds, since $d_Q(\textbf{H}(\mathcal{M}))$ is after construction a subset of $\mathcal{L}'$.}.
With these preliminary results one is well suited to proceed to the calculations of gauge invariance for the different cases of external leg numbers.

\subsection{Summary of Chapter \mbox{\boldmath $\ref{ALGEBRA}$}}\label{ALGEBRA4} 
In this chapter it was shown, that gauge invariance of causal constructed \textbf{S}--matrices in each order $T_n$ can only be destroyed by local terms with at most four external legs. It was also found, that the image (under $d_Q$) of all single elements of the factor group $\textbf{H}(\mathcal{M})$ lead to a gauge invariant form of the $T_n$'s.
And so, finally, all local expressions occurring in $\eqref{algeb6a}$ have a gauge invariant form then and only then if the set $\mathcal{L}'$ is a subset of $d_Q(\textbf{H}(\mathcal{M}))$.
\newpage
\section{Calculations}	\label{BERECH}

In this chapter the gauge invariance calculations will be explicitely executed with the aid of the basic concepts developed in the preceeding chapters. As shown in section $\ref{ALGEBRA2}$ only graphs with $2,3$ or $4$ external legs can lead to local terms and so only these cases must be considered in detail.
The following chapter is separated into four parts: in the first section principal remarks about the  calculations will be introduced, in the three subsequent parts the local terms with $2,3$ and $4$ external legs respectively will be examined.

\subsection{Basic Aspects of the Calculations}				\label{BERECH1}

In this section it will be explained, how gauge destroying local terms emerge from graphs of $T_n$'s and the way they must be treated to show gauge invariance.
For that purpose one has first to consider in detail the distributional part of the local terms of $\mathcal{L},\mathcal{M} \in \mathcal{F}_{\trm{loc}}$\footnote{The definitions of $\mathcal{L},\mathcal{M}$ and $\mathcal{F}_{\trm{loc}}$ are found in section \ref{ALGEBRA3}.}. Because of the translation invariance of the $D_n(x_1,\ldots, x_n)$'s, described in $\eqref{cqft27}$, the singular terms of $\mathcal{L},\mathcal{M}$ must have the same property. So the $\df$--distributions occurring in chapter $\ref{ALGEBRA}$ have the general form\footnote{As usual, a $\df$--distribution with more than one argument has to be read as a product of $\df$--distributions with single arguments.$\label{footnote-delta}$}
\begin{align}		\label{berech1.1}
	\df(x_1,\ldots, x_n) 
	&= \delta(x_1 - x_n,x_2 - x_n, \ldots ,x_{n-1} - x_n)			\notag	\\
	&= \df(x_1 - x_n) \df(x_2 - x_n) \cdot \ldots \cdot \df(x_{n-1} - x_n)			\end{align}
The translation invariant form shown above obviously leads to the important distributional identity
\begin{equation}		\label{berech1.2}
	\sum_i \p^{\mu}_{x_i} \df = \p_{x_1}^{\mu} \df + 									\p_{x_2}^{\mu} \df + \ldots + \p_{x_n}^{\mu}\df = 0
\end{equation}
which plays an important r\^ole in the following calculations. More about that topic will be discussed below. But first, the explicit form and the origin of the local graphs
with $2,3$ and $4$ external legs will be investigated.

The distributive parts $\eqref{berech1.1}$ of the terms under consideration lead to the following local terms after smearing out with the test functions $g_1(x_1), \ldots , g_n(x_n)$
\\[2ex]
\begin{equation}		\label{berech1.3}
\begin{array}{ccc}
\begin{picture}(100,60)(-50,-30)
\Line(-40,0)(0,0)
\Line(0,0)(40,0)
\Vertex(0,0){2}
\end{picture}
&
\begin{picture}(100,60)(-50,-30)
\Line(-40,-25)(0,0)
\Line(-40,25)(0,0)
\Line(0,0)(40,0)
\Vertex(0,0){2}
\end{picture}
&
\begin{picture}(100,60)(-50,-30)
\Line(-40,-25)(0,0)
\Line(-40,25)(0,0)
\Line(0,0)(40,-25)
\Line(0,0)(40,25)
\Vertex(0,0){2}
\end{picture}
\\
2\text{--leg graph} & 3\text{--leg graph} & 4\text{--leg graph}
\end{array}
\end{equation}
\\[2ex]
It is obvious, that these singular graphs originate from different not local $2,3$ and $4$ leg terms of $T_n$\footnote{In the splitting procedure with singular order $\omega \geqslant 0$}. The following source graphs below lead to the latter
\\[2ex]
\begin{subequations}
\begin{equation}			\label{berech1.4a}
\begin{array}{ccc}
\footnotesize\begin{picture}(100,60)(-50,-30)
\Line(-40,0)(0,0)
\Line(0,0)(40,0)
\GCirc(0,0){12}{.8}
\end{picture}
&
\begin{picture}(100,60)(-50,-30)
\Line(-40,-25)(0,0)
\Line(-40,25)(0,0)
\Line(0,0)(40,0)
\GCirc(0,0){12}{.8}
\end{picture}
&
\begin{picture}(100,60)(-50,-30)
\Line(-40,-25)(0,0)
\Line(-40,25)(0,0)
\Line(0,0)(40,-25)
\Line(0,0)(40,25)
\GCirc(0,0){12}{.8}
\end{picture}
\end{array}
\end{equation}
\begin{equation}			\label{berech1.4b}
\begin{array}{ccc}
\phantom{\begin{picture}(100,60)(-50,-30)
\Line(-40,0)(0,0)
\Line(0,0)(40,0)
\GCirc(0,0){12}{.3}
\end{picture}}
&
\begin{picture}(100,60)(-50,-30)
\Line(-40,-25)(-20,0)
\Line(-40,25)(-20,0)
\Line(-20,0)(0,0)
\Line(0,0)(40,0)
\GCirc(0,0){12}{.8}
\end{picture}
&
\begin{picture}(100,60)(-50,-30)
\Line(-40,-25)(-20,0)
\Line(-40,25)(-20,0)
\Line(-20,0)(0,0)
\Line(0,0)(40,-25)
\Line(0,0)(40,25)
\GCirc(0,0){12}{.8}
\end{picture}
\\
&&
\begin{picture}(100,60)(-50,-30)
\Line(-40,-25)(-20,0)
\Line(-40,25)(-20,0)
\Line(-20,0)(0,0)
\Line(20,0)(40,-25)
\Line(20,0)(40,25)
\Line(20,0)(0,0)
\GCirc(0,0){12}{.8}
\end{picture}\normalsize
\end{array}
\end{equation}
\end{subequations}
With these intermediate remarks in mind we can go back to the local terms themselves.
At first, all terms with disjunct arguments in their field operator part and
a $\df$--distribution of the form $\eqref{berech1.1}$ describe local terms of $T_n$ emerging from graphs of the form $\eqref{berech1.4a}$ above. The gauge invariance proofs for these most common cases (with disjunct arguments for the field operators) are given in the first parts of the three subsequent sections. There remains the gauge invariance proofs for the reducible three graphs of $\eqref{berech1.4b}$ which must be treated separately.
This will be done following the proofs for the $3$--leg and $4$--leg graphs with disjunct arguments respectively. It will be shown next, that the executed calculations for disjunct arguments can be adapted by restriction to identical arguments to the cases of reducible graphs. This can be simply achieved, if one takes into account, that by the transitions
\begin{equation}		\label{berech1.4.1}
\begin{aligned}
	\mathit{l}_{i,j}\big|_{\trm{P}} 
		&\longrightarrow 
			\mathit{l}_{i,j}\big|_{\trm{P},x_l = x_k}						\\
	\mathit{m}_{i,j}\big|_{\trm{P}} 
		&\longrightarrow 
			\mathit{m}_{i,j}\big|_{\trm{P},x_l = x_k}						\\
	\textbf{Z}(\mathcal{M}') \,,\,			
  	\textbf{B}(\mathcal{M}') \,,\,
	\textbf{H}(\mathcal{M}') 
		&\longrightarrow 
			\textbf{Z}(\mathcal{M}')\big|_{x_l = x_k} \,,\,	
  			\textbf{B}(\mathcal{M}')\big|_{x_l = x_k} \,,\,
			\textbf{H}(\mathcal{M}')\big|_{x_l = x_k} 
\end{aligned}
\end{equation}
only the equivalence proofs which make use of identity $\eqref{berech1.2}$ or the Leibnitz' rule will be destroyed and thus must be recalculated. The latter is true, since in these two cases the Leibnitz' rule for distributions\footnote{It will be shown, that all elements of $\mathcal{L}'$ and $\mathcal{M}'$ in both $3$--leg graphs as well as $4$--leg graphs, are equivalent to the elements which have derivatives on external field operators only. To then make use of identity $\eqref{berech1.2}$ one requires Leibnitz' rule to shift the derivation onto the $\df$--distribution.}
\begin{align}		\label{berech1.4.2}
	\p_{x_i}(T_1 \cdot T_2) &= \p_{x_i}T_1 \cdot T_2 + T_1 \cdot \p_{x_i}T_2
		&& \forall \; T_1 \,,\, T_2 \in \mathcal{S}
\end{align}
do not lead to identical results for terms with identical arguments compared with those with disjunct arguments, as the following example for the term $\mathit{l}_{0,1}$ on page $\pageref{berech3.29}$ in equation $\eqref{berech3.29}$ shows\footnote{For simplicity, the whole permutation invariant sum will not be written, but only the first term $\eqref{berech3.27a}$}:
\begin{equation}		\label{berech1.4.3}
	\begin{aligned}
	\mathit{l}_{0,1}\big|_{\p^{\nu}_{x_1}\p^{\nu}_{x_3}} 
		&= :\p^{\nu}A_a^{\mu}(x_1) A^{\mu}_b(x_2) \p^{\nu}u_c(x_3): \df		\\
		&= \n_{x_1} - :A_a^{\mu}(x_1) A^{\mu}_b(x_2)\p^{\nu}u_c(x_3): 							\p^{\nu}_{x_1}\df												
\\[3ex]
	\mathit{l}_{0,1}\big|_{\p^{\nu}_{x_1}\p^{\nu}_{x_3}\,,\, x_1 = x_2}
		&= :\p^{\nu}A_a^{\mu}(x_1) A^{\mu}_b(x_1) \p^{\nu}u_c(x_3): \df		\\
		&= \n_{x_1} - :A_a^{\mu}(x_1) \p^{\nu}A^{\mu}_b(x_1)\p^{\nu}u_c(x_3): 					\df - :A_a^{\mu}(x_1)A^{\mu}_b(x_1)\p^{\nu}u_c(x_3): 							\p^{\nu}_{x_1}\df
	\end{aligned}
\end{equation}
It will be demonstrated, that simple recalculation of the equivalences in question (equivalences which rely on $\eqref{berech1.2}$ or the Leibnitz' rule) lead for both, the reducible $3$ and $4$--leg terms, to the same results as for terms with disjunct arguments.
Since all other equivalences trivially remain correct under the transition to identical arguments, the gauge invariance proofs for reducible graphs simply comprise of the recalculations of the identities including Leibnitz' rule or identity $\eqref{berech1.2}$.
With this results one is well suited for the explicit calculations for all local terms in the next chapters.
\subsection{\boldmath $2$--Leg Calculations}		\label{BERECH2}

The gauge invariance proofs depend on the different color structure for distinct external leg numbers.
In the case of two external legs, the color structure is very simple due to the adjoint representation of $SU(n)$. This immediately shows, that the color connection coefficient with two arguments can only occur in a diagonal form (i.e. with two identical color arguments). Explicitely, the local terms always have the form
\begin{equation}	\label{berech2.1}
	\mathrm{D}^{\beta}:\varLambda_a(x_1) \varGamma_a(x_2):\mathrm{D}^{\alpha}\df
\end{equation}
Obviously, all terms with a particular color structure can be examined separately. Thus, for the gauge calculations in this section, one can neglect the color structure.
Additionally, equation $\eqref{berech1.2}$ has for two leg terms the simple form
\begin{equation}	\label{berech2.2}
	\sum_i \p^{\mu}_{x_i} \df = \p_{x_1}^{\mu} \df + 									\p_{x_2}^{\mu} \df + \sum_{\trm{inner}} \p^{\mu}\df = 0
\end{equation}
whereas all the inner derivatives of the $\df$--distribution simply lead to divergence expressions, since these derivatives do not affect the field operator part and thus can be considered before the whole term.
With these preliminary remarks in mind one can start with the determination of the factor group according to the procedure given at the end of section $\ref{ALGEBRA3}$.

\subsubsection{Determination of Elements of \boldmath$\mathcal{L}$} 																		\label{BERECH2.1}

To determine $\mathcal{L}$ in a systematic way, the whole set will be divided into subsets $\mathcal{L}_d$ with the same number of $d$ derivatives on the external field operators. Equation $\eqref{algeb10}$ reads for the $2$--leg case as follows
\begin{equation}	\label{berech2.3}
	\omega(2\text{--leg}) \leqslant 4 - 2 - d + 1 = 3 - d
\end{equation}
which states, that $2$--leg terms with $d \leqslant 3$ can have singular order $\omega\geqslant 0$. Consequently $\mathcal{L}$ can be written as the union of the subsets
\begin{equation}	\label{berech2.4}
	\mathcal{L} = \mathcal{L}_0 \cup \mathcal{L}_1 \cup\mathcal{L}_2 									\cup\mathcal{L}_3
\end{equation}
With respect to the ghost number of the elements out of $\mathcal{L}$, all $\mathcal{L}_i$ must satisfy the restriction $n_u(\mathcal{L}_i) = 1$. The only way to meet the latter is given by the field operator form $A^{\mu}u$.
In the following subsections the subsets $\mathcal{L}_0,\mathcal{L}_1, \mathcal{L}_2, \mathcal{L}_3$ will be determined systematically.

\paragraph{The Subset \boldmath$\mathcal{L}_0$}		\label{BERECH2.1.0}

The subset $\mathcal{L}_0$ has singular order
\begin{equation}	\label{berech2.5}
	\omega(\mathcal{L}_0) \leqslant 4 - 2 - 0 + 1 = 3
\end{equation}
and thus is given by the terms
\begin{equation}	\label{berech2.6}
	\mathcal{L}_0 = \big\{ :A^{\mu}(x_1)u(x_2):\mathcal{D}^{\beta} \df \bigm| 
			|\beta| \leqslant 3 \; , \; x_1 \longleftrightarrow x_2 \big\}
\end{equation}
whereas in the entire set of $\mathcal{L}$ the arguments of $A^{\mu}$ and $u$
do not occur just in the written way, but also (as indicated by the arrow in $\eqref{berech2.6}$) in there exchanged position.
Thus, one can suppress, without loss of generality, in the following lists of elements of $\mathcal{L}_i$ the terms with exchanged arguments. One only has to 
keep in mind, that if necessary, the arguments can be exchanged without problems\footnote{This is true, since the permutation invariant sums over all elements remains unchanged under these exchange of arguments.}.
Since the terms with $|\beta| = 0$ and $|\beta| = 2$ cannot be Lorentz scalars, the set of $\mathcal{L}_0$ only consists of the following expressions\\
$\mathit{l}_{0,i} \in \mathcal{L}_0 \bigr|_{|\beta| = 1}$
\begin{align}	\label{berech2.7}
	&\mathit{l}_{0,1}\overset{def}{=}\, 																		:A^{\mu}(x_1)u(x_2):\p^{\mu}_{x_1}\delta\,&
	&\mathit{l}_{0,2}\overset{def}{=}\, 																		:A^{\mu}(x_1)u(x_2):\p^{\mu}_{x_2}\delta\,&
	&\mathit{l}_{0,3}\overset{def}{=}\, 
			:A^{\mu}(x_1)u(x_2):\sum_{\trm{inner}}\p^{\mu}\delta\,&
\end{align}
and one single element of\\
$\mathcal{L}_0 \bigr|_{|\beta| = 3}$
\begin{equation}	\label{berech2.8}
	\mathit{l}_{0,4}\overset{def}{=}\,:A^{\mu}(x_1)u(x_2):\p^{\mu}
	\p^{\nu} \p^{\nu}\df:
\end{equation}
with the derivation $\p^{\mu}\p^{\nu}\p^{\nu}$ in $\mathit{l}_{0,4}$ as the below defined expression:
Any multi--derivation $\p^{\mu}\p^{\nu}\cdot \ldots \cdot \p^{\xi}$ on a $\df$--distribution represents the set
\begin{multline}	\label{berech2.9}
	\p^{\mu}\p^{\nu}\cdot \ldots \cdot \p^{\xi} :=
	\bigg\{ \p^{\mu}_{x_i}\p^{\nu}_{x_j}\cdot \ldots \cdot\p^{\xi}_{x_m} \df 			\biggm| 
	\p^{\kappa}_{x_p} \in \big\{ \p^{\kappa}_{x_1}, \p^{\kappa}_{x_2},\ldots, 			\p^{\kappa}_{x_{\trm{inner}-1}}, \sum_{\trm{inner}} \p^{\kappa} \big\} \;,\;
	\\ \kappa \in \{ \mu , \nu , \ldots , \xi \} \; , \;	x_p \in \{ x_i , x_j 	, \ldots , x_n \} \bigg\}
\end{multline}
And so, for example, the expression $\mathit{l}_{0,4}$ represents the $18$--element--set
\begin{multline}	\label{berech2.10}
 	\mathit{l}_{0,4} = 
	\bigg\{
	:A^{\mu}(x_1)u(x_2): \p^{\mu}_{x_1}\p^{\nu}_{x_1}\p^{\nu}_{x_1} \df \;,\;
	:A^{\mu}(x_1)u(x_2): \p^{\mu}_{x_1}\p^{\nu}_{x_1}\p^{\nu}_{x_2} \df \;,\;
	\ldots \;,\;	\\
	:A^{\mu}(x_1)u(x_2): \sum_{\trm{inner}}\p^{\mu}\sum_{\trm{inner}} 					\p^{\nu}\sum_{\trm{inner}}\p^{\nu} \df
	\bigg\}
\end{multline}
Now we can go over to the remaining subsets $\mathcal{L}_i$, which will be noted without extended explanatory remarks.

\paragraph{The Subset \boldmath$\mathcal{L}_1$}		\label{BERECH2.1.1}

The singular order is given by 
\begin{equation}	\label{berech2.11}
	\omega(\mathcal{L}_1) \leqslant 4 - 2 - 1 + 1 = 2
\end{equation}
and thus $\mathcal{L}_1$ consists of the elements
\begin{equation}	\label{berech2.12}
	\mathcal{L}_1 = \big\{ 																\mathcal{D}^{\alpha}\!\!:A^{\mu}(x_1)u(x_2):\mathcal{D}^{\beta} \df \bigm| 
	|\alpha|=1\;,\;|\beta| \leqslant 2 \; , \; x_1 \longleftrightarrow x_2\big\}
\end{equation}
Herein the terms with $|\beta| =1$ cannot be Lorentz scalars, and so only the restricted subsets\\
$\mathit{l}_{1,i} \in \mathcal{L}_1 \bigr|_{|\beta| = 0}$
\begin{align}	\label{berech2.13}
	&\mathit{l}_{1,1}\overset{def}{=}\,:\p^{\mu}_{x_1}A^{\mu}(x_1)u(x_2):\df&\,
	&\mathit{l}_{1,2}\overset{def}{=}\,:A^{\mu}(x_1)\p^{\mu}_{x_2}u(x_2):\df
\end{align}
and $\mathit{l}_{1,i} \in \mathcal{L}_1 \bigr|_{|\beta| = 2}$
\begin{equation}	\label{berech2.14}
\begin{aligned}
	\mathit{l}_{1,3}&\overset{def}{=}\, :\p^{\mu}_{x_1}A^{\mu}(x_1)u(x_2): 					\p^{\nu}\p^{\nu} \df					\qquad\qquad				&			\mathit{l}_{1,4}&\overset{def}{=}\, :\p^{\nu}_{x_1}A^{\mu}(x_1)u(x_2):					\p^{\mu}\p^{\nu} \df												\\
	\mathit{l}_{1,5}&\overset{def}{=}\, :A^{\mu}(x_1)\p^{\mu}_{x_2}u(x_2):					\p^{\nu}\p^{\nu} \df												&
	\mathit{l}_{1,6}&\overset{def}{=}\, :A^{\mu}(x_1)\p^{\nu}_{x_2}u(x_2):					\p^{\mu}\p^{\nu} \df
\end{aligned}
\end{equation}
lead to permitted expressions.
Again, the derivations $\p^{\xi}\p^{\kappa}$ represent each single derivation of the whole set according to definition $\eqref{berech2.9}$.

\paragraph{The Subset \boldmath$\mathcal{L}_2$}		\label{BERECH2.1.2}

This subset has singular order 
\begin{equation}	\label{berech2.15}
	\omega(\mathcal{L}_2) \leqslant 4 - 2 - 2 + 1 = 1
\end{equation}
which leads to the set
\begin{equation}	\label{berech2.16}
	\mathcal{L}_2 = \big\{ 																\mathcal{D}^{\alpha}\!\!:A^{\mu}(x_1)u(x_2):\mathcal{D}^{\beta} \df \bigm| 
	|\alpha|=2\;,\;|\beta| \leqslant 1 \; , \; x_1 \longleftrightarrow x_2\big\}
\end{equation}
But herein the terms with $|\beta| =1$ respect Lorentz covariance only, and so $\mathcal{L}_2$ consists of\\
$\mathit{l}_{2,i} \in \mathcal{L}_2\bigr|_{|\beta| = 1}$
\begin{equation}	\label{berech2.17}
\begin{aligned}
	\mathit{l}_{2,1}&\overset{def}{=}\, :\p^{\mu}_{x_1}\p^{\nu}_{x_1} 						A^{\mu}(x_1)u(x_2):\p^{\nu} \df 									\\
	\mathit{l}_{2,2}&\overset{def}{=}\, :\p^{\mu}_{x_1}A^{\mu}(x_1) 						\p^{\nu}_{x_2} u(x_2):\p^{\nu} \df									\\
	\mathit{l}_{2,3}&\overset{def}{=}\, :\p^{\mu}_{x_1}A^{\nu}(x_1) 						\p^{\mu}_{x_2} u(x_2):\p^{\nu} \df									\\
	\mathit{l}_{2,4}&\overset{def}{=}\, :\p^{\mu}_{x_1}A^{\nu}(x_1) 						\p^{\nu}_{x_2}u(x_2): \p^{\mu} \df\equiv \mathit{l}_{2,3} \pmod{\n}	\\
	\mathit{l}_{2,5}&\overset{def}{=}\, :A^{\mu}(x_1)\p^{\mu}_{x_2} 						\p^{\nu}_{x_2}u(x_2):\p^{\nu} \df
\end{aligned}
\end{equation}
whereas $\p^{\kappa}\df$ represents again each single element of the set $\{ \p^{\kappa}_{x_1}\df, \p^{\kappa}_{x_2}\df, \sum_{\trm{inner}}\p^{\kappa}\df \}$.

\paragraph{The Subset \boldmath$\mathcal{L}_3$}		\label{BERECH2.1.3}

The singular order has the value
\begin{equation}	\label{berech2.18}
	\omega(\mathcal{L}_3) \leqslant 4 - 2 - 3 + 1 = 0
\end{equation}
which leads to
\begin{equation}	\label{berech2.19}
	\mathcal{L}_3 = \big\{ 																\mathcal{D}^{\alpha}\!\!:A^{\mu}(x_1)u(x_2):\mathcal{D}^{\beta} \df \bigm| 
	|\alpha|=3\;,\;|\beta| = 0 \; , \; x_1 \longleftrightarrow x_2\big\}
\end{equation}
Obviously the set $\mathcal{L}_3$ only include the two elements\\
$\mathit{l}_{3,i} \in \mathcal{L}_3 \bigr|_{|\beta| = 0}$
\begin{align}	\label{berech2.20}
	&\mathit{l}_{3,1}\overset{def}{=}\, :\p^{\mu}_{x_1}\p^{\nu}_{x_1} 						A^{\mu}(x_1)\p^{\nu}_{x_2}u(x_2): \df							&	
	&\mathit{l}_{3,2}\overset{def}{=}\, :\p^{\mu}_{x_1}A^{\nu}(x_1) 						\p^{\nu}_{x_2}\p^{\mu}_{x_2}u(x_2): \df
\end{align}
since all other terms vanish due to the wave equations $\eqref{eich10a}$ and $\eqref{eich10d}$.
The following subset is devoted to the determination of the factor set $\mathcal{L}' := \,\raisebox{1ex}{$\mathcal{L}$}\raisebox{0ex}
{$\! \Big/\!$}\raisebox{-1ex}{$\n$}$.

\subsubsection{Determination of Equivalent Elements in \boldmath$\mathcal{L}$} 																	\label{BERECH2.2}

To show the equivalence of different terms, one rewrites the expressions with the help of Leibnitz' rule. The technique can be best demonstrated by explicitely writing down the successive steps as an example. This will be done in full length for the term $\mathit{l}_{0,1}$ only. To do so, one finds after three steps
\begin{equation}	\label{berech2.21}
\begin{aligned}
	\mathit{l}_{0,1} 	
		&= :A^{\mu}(x_1)u(x_2): \p^{\mu}_{x_1} \df 						\\
		&= \p^{\mu}_{x_1} \left( :A^{\mu}(x_1)u(x_2): \df \right) - 							:\p^{\mu}_{x_1}A^{\mu}(x_1)u(x_2): \df 
									&&\qquad \bigm|\text{Leibnitz rule}	\\
		&= \nabla - :\p^{\mu}_{x_1} A^{\mu}(x_1)u(x_2): \df				\\
			\mathit{l}_{0,1} & \equiv \mathit{l}_{1,1} \pmod{\n}
\end{aligned}
\end{equation}
the equivalence of $\mathit{l}_{0,1}$ and $\mathit{l}_{1,1}$. Whereas in the last two steps the abbreviation $\n$ was used to indicate any divergence term. Here one has to point to the fact, that, as stated in section $\ref{ALGEBRA3}$ from $\eqref{algeb19}$ to $\eqref{algeb24}$, no information is lost within this 	simplified notation.
In the following the terms of the subsets $\mathcal{L}_d$ will be investigated for different derivations $\p^{\mu}_{x_i}$
\begin{equation}	\label{berech2.22}
\begin{aligned}
	\mathit{l}_{0,2} 	&= :A^{\mu}(x_1)u(x_2):\p^{\mu}_{x_2}\df			&			\mathit{l}_{0,3}	&= :A^{\mu}(x_1)u(x_2): \sum \partial^{\mu}\delta	\\
						&= \n - :A^{\mu}(x_1)\p^{\mu}_{x_2}u(x_2):\df		&			\qquad\qquad		&= \sum_{\trm{inner}}\partial^{\mu} 																(:A^{\mu}(x_1)u(x_2): \df )			\\
	\mathit{l}_{0,2}	&  \equiv \mathit{l}_{1,2}\pmod{\n}					&			\mathit{l}_{0,3}	&= \n
\end{aligned}
\end{equation}
Obviously the right term $\mathit{l}_{0,3}$ has (as all terms with inner derivatives on $\df$'s) a divergent structure\footnote{Confer the remark given in the following of $\eqref{berech2.2}$}. Thus, from now on one does not have to take into consideration expressions of this type.
For each expression in the next set of calculations one derivative $\partial^{\nu}$ or $\partial^{\mu}$ will be restricted to a specific argument, whereas the remaining derivations still represent (according to definition $\eqref{berech2.9}$) all possible combinations of $\partial^{\nu}_{x_i}$'s and $\partial^{\mu}_{x_j}$'s.
We start with the term $\mathit{l}_{0,4}$ under the restriction $\partial^{\mu}_{x_1}$
\begin{equation}	\label{berech2.23}
\begin{aligned}
	\mathit{l}_{0,4}\bigr|_{\p^{\mu}_{x_1}}
			&= :A^{\mu}(x_1)u(x_2):\p^{\mu}_{x_1}\p^{\nu} \p^{\nu}\df		&	
	\mathit{l}_{0,4}\bigr|_{\p^{\mu}_{x_2}}
			&= :A^{\mu}(x_1)u(x_2):\p^{\mu}_{x_2}\p^{\nu} \p^{\nu}\df		\\
			&= \n - :\p^{\mu}_{x_1}A^{\mu}(x_1)u(x_2):\p^{\nu} \p^{\nu}\df	&
	\qquad\qquad	
			&= :A^{\mu}(x_1)\p^{\mu}_{x_2}u(x_2):\p^{\nu} \p^{\nu}\df		\\
			&	\equiv \mathit{l}_{1,3}\pmod{\n}							&
			&	\equiv \mathit{l}_{1,5}\pmod{\n}
\end{aligned}
\end{equation}
For $\partial^{\nu}_{x_1}$ one finds
\begin{equation}	\label{berech2.24}
\begin{aligned}
	\mathit{l}_{0,4}\bigr|_{\p^{\nu}_{x_1}}
			&= :A^{\mu}(x_1)u(x_2):\p^{\mu}\p^{\nu}_{x_1} \p^{\nu}\df		&
	\mathit{l}_{0,4}\bigr|_{\p^{\nu}_{x_2}}
			&= :A^{\mu}(x_1)u(x_2):\p^{\mu}\p^{\nu}_{x_2} \p^{\nu}\df		\\
			&= \n - :\p^{\nu}_{x_1}A^{\mu}(x_1)u(x_2):\p^{\mu} \p^{\nu}\df	&
	\qquad\qquad
			&= :A^{\mu}(x_1)\p^{\nu}_{x_2}u(x_2):\p^{\mu} \p^{\nu}\df		\\
			&	\equiv \mathit{l}_{1,4}\pmod{\n}&
			&	\equiv \mathit{l}_{1,6}\pmod{\n}
\end{aligned}
\end{equation}
By the same procedure the terms $\mathit{l}_{1,i} \in \mathcal{L}_1$ lead to the equivalences (since equivalent elements for $\mathit{l}_{1,1},\mathit{l}_{1,2}$ are already found, the considerations begin with $\mathit{l}_{1,3}$)
\begin{equation}	\label{berech2.25}
\begin{aligned}
	\mathit{l}_{1,3}\bigr|_{\p^{\nu}_{x_1}}
			&= :\p^{\mu}_{x_1}A^{\mu}(x_1)u(x_2):\p^{\nu}_{x_1}\p^{\nu}\df	&
	\mathit{l}_{1,3}\bigr|_{\p^{\nu}_{x_2}}
			&= :\p^{\mu}_{x_1}A^{\mu}(x_1)u(x_2):\p^{\nu}_{x_2}\p^{\nu}\df	\\
			&= \n - :\p^{\nu}_{x_1}\p^{\mu}_{x_1}A^{\mu}(x_1)u(x_2):
				\p^{\nu}\df
	\qquad																	&
			&= \n - :\p^{\mu}_{x_1}A^{\mu}(x_1)\p^{\nu}_{x_2}u(x_2):
				\p^{\nu}\df													\\
			& \equiv \mathit{l}_{2,1} \pmod{\n}								&
			& \equiv \mathit{l}_{2,2} \pmod{\n}
\end{aligned}
\end{equation}
\\
\begin{equation}	\label{berech2.26}
\begin{aligned}
	\mathit{l}_{1,4}\bigr|_{\p^{\mu}_{x_1}}
			&= :\p^{\nu}_{x_1}A^{\mu}(x_1)u(x_2):\p^{\mu}_{x_1}\p^{\nu}\df	&
	\mathit{l}_{1,4}\bigr|_{\p^{\mu}_{x_2}}
			&= :\p^{\nu}_{x_1}A^{\mu}(x_1)u(x_2):\p^{\mu}_{x_2}\p^{\nu}\df	\\
			&= \n - :\p^{\nu}_{x_1}\p^{\mu}_{x_1}A^{\mu}(x_1)u(x_2): 								\p^{\nu}\df	
	\qquad																	&					&= \n - :\p^{\nu}_{x_1}A^{\mu}(x_1)\p^{\mu}_{x_2}u(x_2):
				\p^{\nu}\df													\\
			& \equiv \mathit{l}_{2,1} \pmod{\n}															&
			& \equiv \mathit{l}_{2,4} \pmod{\n}
\\[.6cm]
	\mathit{l}_{1,4}\bigr|_{\p^{\nu}_{x_1}}
			&= :\p^{\nu}_{x_1}A^{\mu}(x_1)u(x_2):\p^{\mu}\p^{\nu}_{x_1}\df	&
	\mathit{l}_{1,4}\bigr|_{\p^{\nu}_{x_2}}
			&= :\p^{\nu}_{x_1}A^{\mu}(x_1)u(x_2):\p^{\mu}\p^{\nu}_{x_2}\df	\\
			&= \n - :\p^{\nu}_{x_1}\p^{\nu}_{x_1}A^{\mu}(x_1)u(x_2):
				\p^{\mu}\df																\qquad																	&
			&= \n - :\p^{\nu}_{x_1}A^{\mu}(x_1)\p^{\nu}_{x_2}u(x_2):
				\p^{\mu}\df													\\
			&= 0								&
			& \equiv \mathit{l}_{2,3} \pmod{\n}
\end{aligned}
\end{equation}
\\
\begin{equation}	\label{berech2.27}
\begin{aligned}
	\mathit{l}_{1,5}\bigr|_{\p^{\nu}_{x_1}}
			&= :A^{\mu}(x_1)\p^{\mu}_{x_2}u(x_2):\p^{\nu}_{x_1}\p^{\nu}\df	&
	\mathit{l}_{1,5}\bigr|_{\p^{\nu}_{x_2}}
			&= :A^{\mu}(x_1)\p^{\mu}_{x_2}u(x_2):\p^{\nu}_{x_2}\p^{\nu}\df	\\
			&= \n - :\p^{\nu}_{x_1}A^{\mu}(x_1)\p^{\mu}_{x_2}u(x_2):
			\p^{\nu}\df																	\qquad																	&
			&= \n - :A^{\mu}(x_1)\p^{\nu}_{x_2}\p^{\mu}_{x_2}u(x_2):\p^{\nu} 						\df															\\
			& \equiv \mathit{l}_{2,4} \pmod{\n}								&
			& \equiv \mathit{l}_{2,5} \pmod{\n}
\end{aligned}
\end{equation}
\\
\begin{equation}	\label{berech2.28}
\begin{aligned}
	\mathit{l}_{1,6}\bigr|_{\p^{\mu}_{x_1}}
			&= :A^{\mu}(x_1)\p^{\nu}_{x_2}u(x_2):\p^{\mu}_{x_1}\p^{\nu}\df	&
	\mathit{l}_{1,6}\bigr|_{\p^{\mu}_{x_2}}
			&= :A^{\mu}(x_1)\p^{\nu}_{x_2}u(x_2):\p^{\mu}_{x_2}\p^{\nu}\df	\\
			&= \n - :\p^{\mu}_{x_1}A^{\mu}(x_1)\p^{\nu}_{x_2}u(x_2):
			\p^{\nu}\df																	\qquad																	&
			&= \n - :A^{\mu}(x_1)\p^{\mu}_{x_2}\p^{\nu}_{x_2}u(x_2):\p^{\nu} 						\df															\\
			& \equiv \mathit{l}_{2,2} \pmod{\n}								&
			& \equiv \mathit{l}_{2,5} \pmod{\n}
\\[.6cm]
	\mathit{l}_{1,6}\bigr|_{\p^{\nu}_{x_1}}
			&= :A^{\mu}(x_1)\p^{\nu}_{x_2}u(x_2):\p^{\mu}\p^{\nu}_{x_1}\df	&
	\mathit{l}_{1,6}\bigr|_{\p^{\nu}_{x_2}}
			&= :A^{\mu}(x_1)\p^{\nu}_{x_2}u(x_2):\p^{\mu}\p^{\nu}_{x_2}\df	\\
			&= \n - :\p^{\nu}_{x_1}A^{\mu}(x_1)\p^{\nu}_{x_2}u(x_2):
			\p^{\mu}\df																	\qquad																	&
			&= \n - :A^{\mu}(x_1)\p^{\nu}_{x_2}\p^{\nu}_{x_2}u(x_2):\p^{\mu} 						\df															\\
			& \equiv \mathit{l}_{2,3} \pmod{\n}								&
			& = 0
\end{aligned}
\end{equation}
The above calculations show the equivalence of all elements of $\mathcal{L}_0$ with those of $\mathcal{L}_1$. Moreover, all elements of $\mathcal{L}_2$ are equivalent to terms of $\mathcal{L}_1$.
It only remains to show, if further equivalences concerning the elements of $\mathcal{L}_3$ exist or not.
The subsequent transformation of the elements $\mathit{l}_{2,1}$ and $\mathit{l}_{2,5}$ show such equivalences for both elements of $\mathcal{L}_3$:
\begin{equation}	\label{berech2.29}
\begin{aligned}
	\mathit{l}_{2,1}\bigr|_{\p^{\nu}_{x_1}}
		&= :\p^{\mu}_{x_1}\p^{\nu}_{x_1} A^{\mu}(x_1)u(x_2):\p^{\nu}\df		\\
		&= \n - :\p^{\mu}_{x_1}\p^{\nu}_{x_1} A^{\mu}(x_1)\p^{\nu}u(x_2):\df\\
		& \equiv \mathit{l}_{3,1} \pmod{\n}
\end{aligned}
\end{equation}
and
\begin{equation}	\label{berech2.30}
\begin{aligned}
	\mathit{l}_{2,5}\bigr|_{\p^{\nu}_{x_1}}
		&= :A^{\mu}(x_1)\p^{\mu}_{x_2} \p^{\nu}_{x_2}u(x_2):\p^{\nu}\df		\\
		&= \n - :\p^{\nu}A^{\mu}(x_1)\p^{\mu}_{x_2} \p^{\nu}_{x_2}u(x_2):\df\\
		& \equiv \mathit{l}_{3,2} \pmod{\n}
\end{aligned}
\end{equation}
Up to now we have found, that all elements of $\mathcal{L}$ can be represented by the single subset $\mathcal{L}_1$ alone. But this is not the final result, since internal symmetries in $\mathcal{L}_1$ lead to an even smaller factor group.
The following calculations uncover the internal symmetries of $\mathit{l}_{1,1}$ and $\mathit{l}_{1,2}$
\begin{equation}	\label{berech2.31}
\begin{aligned}
	\mathit{l}_{1,1} 	&= :\p^{\mu}_{x_1}A^{\mu}(x_1)u(x_2):\df			\\
						&= \n - :A^{\mu}(x_1)u(x_2):\p^{\mu}_{x_1}\df		\\
						&= \n + :A^{\mu}(x_1)u(x_2): (\p^{\mu}_{x_2} + 											\sum_{\trm{inner}})\df
						&&\qquad\qquad \big|\text{with }\eqref{berech2.2}	\\
						&= \n + :A^{\mu}(x_1)u(x_2):\p^{\mu}_{x_2}\df		\\
						&\equiv \mathit{l}_{1,2} \pmod{\n}	
\end{aligned}
\end{equation}
of $\mathit{l}_{1,3}$ and $\mathit{l}_{1,5}$
\begin{equation}	\label{berech2.32}
\begin{aligned}
	\mathit{l}_{1,3}
		&= :\p^{\mu}_{x_1}A^{\mu}(x_1)u(x_2):\p^{\nu}\p^{\nu}\df			\\
		&= \p^{\mu}_{x_1}\big(:A^{\mu}(x_1)u(x_2):\p^{\nu}\p^{\nu}\df\big)	
			- :A^{\mu}(x_1)u(x_2):\p^{\nu}\p^{\nu}\p^{\mu}_{x_1}\df			\\
		&= \n^{\mu}_{x_1} + :A^{\mu}(x_1)u(x_2):\p^{\nu}\p^{\nu} 								(\p^{\mu}_{x_2} + \sum_{\trm{inner}})\df
					&&\quad \big|\text{with }			\eqref{berech2.2}	\\
		&= \n^{\mu}_{x_1} + \n^{\mu}_{\sum} + 													:A^{\mu}(x_1)u(x_2):\p^{\nu}\p^{\nu}\p^{\mu}_{x_2} \df			\\
		&= :A^{\mu}(x_1)u(x_2):\p^{\nu}\p^{\nu}\p^{\mu}_{x_2} \df \pmod{\n}	\\
		& \equiv \mathit{l}_{1,5} \pmod{\n}											\end{aligned}
\end{equation}
of $\mathit{l}_{1,4}$ and $\mathit{l}_{1,6}$
\begin{equation}	\label{berech2.33}
\begin{aligned}
	\mathit{l}_{1,4}
		&= :\p^{\nu}_{x_1}A^{\mu}(x_1)u(x_2):\p^{\mu}\p^{\nu}\df			\\
		&= \p^{\nu}_{x_1}\big(:A^{\mu}(x_1)u(x_2):\p^{\mu}\p^{\nu}\df\big)	
				- :A^{\mu}(x_1)u(x_2):\p^{\mu}\p^{\nu}\p^{\nu}_{x_1}\df		\\
		&= \n^{\nu}_{x_1} + :A^{\mu}(x_1)u(x_2):\p^{\mu}\p^{\nu} 								(\p^{\nu}_{x_2} + \sum_{\trm{inner}})\df
					&&\quad \big|\text{with }			\eqref{berech2.2}	\\
		&= \n^{\nu}_{x_1} + \n^{\nu}_{\sum} + 													:A^{\mu}(x_1)u(x_2):\p^{\mu}\p^{\nu}\p^{\nu}_{x_2} \df			\\
		& \equiv \mathit{l}_{1,6} \pmod{\n}
\end{aligned}
\end{equation}
and finally of $\mathit{l}_{1,3}$ and $\mathit{l}_{1,4}$
\begin{align}
	\mathit{l}_{1,3}\bigr|_{\p^{\nu}_{x_1}}
		&= :\p^{\mu}_{x_1}A^{\mu}(x_1)u(x_2):
											\p^{\nu}_{x_1}\p^{\nu}\df\notag	\\
		&= \n - :\p^{\nu}_{x_1}\p^{\mu}_{x_1}A^{\mu}(x_1)u(x_2):
											\p^{\nu}\df				\notag	\\
		&= \n - :\p^{\nu}_{x_1}A^{\mu}(x_1)u(x_2):
					\p^{\nu}\p^{\mu}_{x_1}\df
			\equiv \mathit{l}_{1,4}\bigr|_{\p^{\mu}_{x_1}}\label{berech2.34}\\
		&= \n - :\p^{\nu}_{x_1}A^{\mu}(x_1)u(x_2):\p^{\nu}
				(\p^{\mu}_{x_2} + \sum_{\trm{inner}})\df
					&&\quad \big|\text{with }\eqref{berech2.2}		\notag	\\
		&= \n - :\p^{\nu}_{x_1}A^{\mu}(x_1)u(x_2):
					\p^{\nu}\p^{\mu}_{x_2}\df
			\equiv \mathit{l}_{1,4}\bigr|_{\p^{\mu}_{x_2}}\label{berech2.35}
\end{align}
In the last transformation the equations $\eqref{berech2.34}$ and $\eqref{berech2.35}$ show the equivalence $\mathit{l}_{1,3} \sim \nolinebreak[4]\mathit{l}_{1,4}$, since according to the latter two equations $\mathit{l}_{1,3}\bigr|_{\p^{\nu}_{x_1}}$ represents both possible terms $\mathit{l}_{1,4}$ restricted either by $\p^{\mu}_{x_1}$ or $\p^{\mu}_{x_2}$.
Collecting all results we find, that the factor group $\mathcal{L}'$ comprises of the elements
\begin{align}	\label{berech2.36}
	\mathcal{L}' 
	:= \,\raisebox{1ex}{$\mathcal{L}$}\raisebox{0ex}
				{$\! \Big/\!$}\raisebox{-1ex}{$\n$}
	= \big\{ \mathit{l}_{1,1}\, , \, \mathit{l}_{1,3} \big\}	
\end{align}
only. With that result we can close this section and turn to the set $\mathcal{M}$ in the following.

\subsubsection{Determination of the Elements of \boldmath$\mathcal{M}$} 																		\label{BERECH2.3}

Completely analogous to the determination in the preceding section, we find the elements of $\mathcal{M}$ in a similar way.
First, one can profit by breaking $\mathcal{M}$ into the subsets\footnote{Similar to the subset structure of $\mathcal{L}$, the subscripts again specify the number of external derivations.}
\begin{equation}	\label{berech2.37}
	\mathcal{M} = \mathcal{M}_0 \cup \mathcal{M}_1 \cup \mathcal{M}_2 				\end{equation}
According to equation $\eqref{algeb8}$
\begin{align}	\label{berech2.38}
	\omega(2\text{--leg}) \leqslant 4 - 2 - d = 2 - d
\end{align}
only two leg terms with a maximum of two external derivatives lead to singular graphs. Furthermore the ghost number $n_u(\mathcal{M}) = 0$ must equal zero and thus restrict the field operator combinations to
\begin{align}	 \label{berech2.39}
	&A^{\kappa}(x_1)A^{\xi}(x_2) 	&u(x_1)\ti{u}(x_2)
\end{align}
Now the determination of the elements of $\mathcal{M}_i$ will be executed in detail.

\paragraph{The Subset \boldmath$\mathcal{M}_0$}		\label{BERECH2.3.0}

The singular order is given by
\begin{equation}	\label{berech2.40}
	\omega(\mathcal{M}_0) \leqslant 4 - 2 - 0 = 2
\end{equation}
which gives the general set
\begin{equation}	\label{berech2.41}
	\mathcal{M}_0 =
	\Bigl\{:\Lambda(x_1) \Gamma(x_2):\, \mathcal{D}^{\beta} \df \Bigm| \left( 			\Lambda \Gamma \right) \in \left\{ A^{\kappa}A^{\xi},u \ti{u} \right\} \,, 			\,|\beta| \leqslant 2 \, , \,\, x_1 \longleftrightarrow x_2 \Bigr\}
\end{equation}
Due to the fact that terms with $|\beta| = 1$ cannot be Lorentz scalars, only the following expressions\\
$\mathit{m}_{0,i} \in \mathcal{M}_0 \bigr|_{|\beta| = 0}$
\begin{align}	\label{berech2.42}	
	\mathit{m}_{0,1}&\overset{def}{=}\,:A^{\mu}(x_1) A^{\mu}(x_2):\df\,		&
	\mathit{m}_{0,2}&\overset{def}{=}\,:u(x_1)\ti{u}(x_2):\df\,
\end{align}
and $\mathit{m}_{0,i} \in \mathcal{M}_0 \bigr|_{|\beta| = 2}$
\begin{equation}	\label{berech2.43}
\begin{aligned}
	\mathit{m}_{0,3}&\overset{def}{=}\,:A^{\mu}(x_1)A^{\mu}(x_2): 																		\p^{\nu}\p^{\nu}\df		&
	\qquad\qquad
	\mathit{m}_{0,4}&\overset{def}{=}\,:A^{\mu}(x_1)A^{\nu}(x_2): 																		\p^{\mu}\p^{\nu}\df		\\
	\mathit{m}_{0,5}&\overset{def}{=}\,:u(x_1)\ti{u}(x_2):
													\p^{\nu}\p^{\nu}\df
\end{aligned}
\end{equation}
satisfy all restrictions.

\paragraph{The Subset \boldmath$\mathcal{M}_1$}		\label{BERECH2.3.1}

The same inequality as above leads to the singular order
\begin{equation}	\label{berech2.44}
	\omega(\mathcal{M}_1) \leqslant 4 - 2 - 1 = 1
\end{equation}
which restricts $\mathcal{M}_1$ to
\begin{equation}	\label{berech2.45}
	\mathcal{M}_1 =
	\Bigl\{\mathcal{D}^{\alpha}:\Lambda(x_1) \Gamma(x_2):\, 							\mathcal{D}^{\beta}\df \Bigm| \left( \Lambda \Gamma \right) \in \left\{ 			A^{\kappa}A^{\xi},u \ti{u} \right\} \, ,\,|\alpha| = 1 \, , \,|\beta| 				\leqslant 1 \, , \,\, x_1 \longleftrightarrow x_2 \Bigr\}
\end{equation}
Thus the list of all elements is given as\footnote{Again, terms with $|\beta| =0$ are not Lorentz scalars, and thus must be ignored.}\\
$\mathit{m}_{1,i} \in \mathcal{M}_1 \bigr|_{|\beta| = 1}$
\begin{equation}	\label{berech2.46}
\begin{aligned}
	\mathit{m}_{1,1}&\overset{def}{=}\,:\p^{\mu}_{x_1}A^{\mu}(x_1) 															A^{\nu}(x_2):\p^{\nu} \df			&
	\qquad\qquad
	\mathit{m}_{1,2}&\overset{def}{=}\,:\p^{\mu}_{x_1}A^{\nu}(x_1) 															A^{\mu}(x_2):\p^{\nu} \df			\\
	\mathit{m}_{1,3}&\overset{def}{=}\,:\p^{\mu}_{x_1}A^{\nu}(x_1) 															A^{\nu}(x_2):\p^{\mu} \df 			\\
	\mathit{m}_{1,4}&\overset{def}{=}\,:\p^{\mu}_{x_1}u(x_1) 																\ti{u}(x_2):\p^{\mu}\df				&
	\mathit{m}_{1,5}&\overset{def}{=}\,:u(x_1)\p^{\mu}_{x_2} 																\ti{u}(x_2):\p^{\mu}\df
\end{aligned}
\end{equation}

\paragraph{The Subset \boldmath$\mathcal{M}_2$}		\label{BERECH2.3.2}

Obviously the last subset has singular order zero and thus $\mathcal{M}_2$ comprises of
\begin{equation}	\label{berech2.47}
	\mathcal{M}_2 =
	\Bigl\{ \mathcal{D}^{\alpha}:\Lambda(x_1) \Gamma(x_2):\, \mathcal{D}^{\beta} 	\df \Bigm| \left( \Lambda \Gamma \right) \in \left\{ A^{\kappa}A^{\xi},u 			\ti{u} \right\} \, ,\,|\alpha| = 2 \, , \,|\beta| = 0 \, , \,\, x_1 				\longleftrightarrow x_2 \Bigr\}
\end{equation}
which leads to the explicit list\\
$\mathit{m}_{2,i} \in \mathcal{M}_2$
\begin{equation}	\label{berech2.48}
\begin{aligned}
	\mathit{m}_{2,1}&\overset{def}{=}\,:\p^{\mu}_{x_1}\p^{\nu}_{x_1} 													A^{\mu}(x_1) A^{\nu}(x_2):\df			&
 	\qquad\qquad
	\mathit{m}_{2,2}&\overset{def}{=}\,:\p^{\mu}_{x_1}A^{\mu}(x_1) 														\p^{\nu}_{x_2} A^{\nu}(x_2):\df			\\
	\mathit{m}_{2,3}&\overset{def}{=}\,:\p^{\mu}_{x_1} u(x_1) 															\p^{\mu}_{x_2}\ti{u}(x_2):\df			\\
	\mathit{m}_{2,4}&\overset{def}{=}\,:\p^{\mu}_{x_1}A^{\nu}(x_1) 														\p^{\mu}_{x_2} A^{\nu}(x_2):\df			&
	\mathit{m}_{2,5}&\overset{def}{=}\,:\p^{\mu}_{x_1}A^{\nu}(x_1) 														\p^{\nu}_{x_2}A^{\mu}(x_2): \df
\end{aligned}
\end{equation}

\subsubsection{Determination of Equivalent Elements in \boldmath$\mathcal{M}$} 																	\label{BERECH2.4}

To find the equivalent elements in $\mathcal{M}$, we will follow the same argumentation as in section $\ref{BERECH2.2}$. Thus, in the following set of calculations, we restrict ourself to a systematical listing of all equivalences without further explanation.
We begin our treatment with $\mathcal{M}_0$. Immediately one realizes, that $\mathit{m}_{0,1}$ and $\mathit{m}_{0,2}$ cannot be equivalent to other terms, since the remaining expressions contain at least one derivation (which is not true for $\mathit{m}_{0,1}\, , \, \mathit{m}_{0,2}$). Consequently we start with $\mathit{m}_{0,3}$:
\begin{equation}	\label{berech2.49}
\begin{aligned}
	\mathit{m}_{0,3}\bigr|_{\p^{\nu}_{x_1}} 
			&= :A^{\mu}(x_1)A^{\mu}(x_2):\p^{\nu}_{x_1}\p^{\nu}\df			&
	\qquad
	\mathit{m}_{0,3}\bigr|_{\p^{\nu}_{x_2}} 
			&= :A^{\mu}(x_1)A^{\mu}(x_2):\p^{\nu}_{x_2}\p^{\nu}\df			\\
			&= \n - :\p^{\nu}_{x_1}A^{\mu}(x_1)A^{\mu}(x_2):\p^{\nu}\df		&
			&= \n - :A^{\mu}(x_1)\p^{\nu}_{x_2}A^{\mu}(x_2):\p^{\nu}\df		\\
			& \equiv \mathit{m}_{1,3} \pmod{\n}								&
			& \equiv \mathit{m}_{1,3} \pmod{\n}
\end{aligned}
\end{equation}
The terms $\mathit{m}_{0,4}$ and $\mathit{m}_{0,5}$ give
\begin{equation}	\label{berech2.50}
\begin{aligned}
	\mathit{m}_{0,4}\bigr|_{\p^{\mu}_{x_1}} 
			&= :A^{\mu}(x_1)A^{\nu}(x_2):\p^{\mu}_{x_1}\p^{\nu}\df			&
	\qquad
	\mathit{m}_{0,4}\bigr|_{\p^{\mu}_{x_2}} 
			&= :A^{\mu}(x_1)A^{\nu}(x_2):\p^{\mu}_{x_2}\p^{\nu}\df			\\
			&= \n - :\p^{\mu}_{x_1}A^{\mu}(x_1)A^{\nu}(x_2):\p^{\nu}\df		&
			&= \n -  :A^{\mu}(x_1)\p^{\mu}_{x_2}A^{\nu}(x_2):\p^{\nu}\df	\\
			& \equiv \mathit{m}_{1,1} \pmod{\n}								&	
			& \equiv \mathit{m}_{1,2} \pmod{\n}
\\[.6cm]
	\mathit{m}_{0,4}\bigr|_{\p^{\nu}_{x_1}} 
			&= :A^{\mu}(x_1)A^{\nu}(x_2):\p^{\mu}\p^{\nu}_{x_1}\df			&
	\qquad
	\mathit{m}_{0,4}\bigr|_{\p^{\nu}_{x_2}} 
			&= :A^{\mu}(x_1)A^{\nu}(x_2):\p^{\mu}\p^{\nu}_{x_2}\df			\\
			&= \n - :\p^{\nu}_{x_1}A^{\mu}(x_1)A^{\nu}(x_2):\p^{\mu}\df		&
			&= \n - :A^{\mu}(x_1)\p^{\nu}_{x_2}A^{\nu}(x_2):\p^{\mu}\df		\\
			& \equiv \mathit{m}_{1,2} \pmod{\n}								&
			& \equiv \mathit{m}_{1,1} \pmod{\n}
\end{aligned}
\end{equation}
\\
\begin{equation}	\label{berech2.51}
\begin{aligned}
	\mathit{m}_{0,5}\bigr|_{\p^{\nu}_{x_1}} 
			&= :u(x_1)\ti{u}(x_2):\p^{\nu}_{x_1}\p^{\nu}\df					&
	\qquad
	\mathit{m}_{0,5}\bigr|_{\p^{\nu}_{x_2}} 
			&= :u(x_1)\ti{u}(x_2):\p^{\nu}_{x_2}\p^{\nu}\df					\\
			&= \n - :\p^{\nu}_{x_1}u(x_1)\ti{u}(x_2):\p^{\nu}\df			&
			&= \n - :u(x_1)\p^{\nu}_{x_2}\ti{u}(x_2):\p^{\nu}\df			\\
			& \equiv \mathit{m}_{1,4} \pmod{\n}								&
			& \equiv \mathit{m}_{1,5} \pmod{\n}
\end{aligned}
\end{equation}
Altogether the above calculations point out, that all elements of $\mathcal{M}_1$ can be represented by those of $\mathcal{M}_0$.
Transforming the terms of $\mathcal{M}_1$, we will find further equivalences between those and $\mathcal{M}_2$. The detailed calculation below 
\begin{equation}	\label{berech2.52}
\begin{aligned}
	\mathit{m}_{1,1}\bigr|_{\p^{\nu}_{x_1}}
	 		&=:\p^{\mu}_{x_1}A^{\mu}(x_1)A^{\nu}(x_2):\p^{\nu}_{x_1}\df		&
	\qquad
	\mathit{m}_{1,1}\bigr|_{\p^{\nu}_{x_2}} 	
			&=:\p^{\mu}_{x_1}A^{\mu}(x_1)A^{\nu}(x_2):\p^{\nu}_{x_2}\df		\\
			&= \n - :\p^{\nu}_{x_1}\p^{\mu}_{x_1}A^{\mu}(x_1)A^{\nu}(x_2):\df&
			&= \n - :\p^{\mu}_{x_1}A^{\mu}(x_1)\p^{\nu}_{x_2}A^{\nu}(x_2):\df\\
			& \equiv \mathit{m}_{2,1} \pmod{\n}								&
			& \equiv \mathit{m}_{2,2} \pmod{\n}
\end{aligned}
\end{equation}
\\
\begin{equation}	\label{berech2.53}
\begin{aligned}
	\mathit{m}_{1,2}\bigr|_{\p^{\nu}_{x_1}} 
			&=:\p^{\mu}_{x_1}A^{\nu}(x_1)A^{\mu}(x_2):\p^{\nu}_{x_1}\df		&
	\qquad
	\mathit{m}_{1,2}\bigr|_{\p^{\nu}_{x_2}} 
			&=:\p^{\mu}_{x_1}A^{\nu}(x_1)A^{\mu}(x_2):\p^{\nu}_{x_2}\df		\\
			&= \n - :\p^{\nu}_{x_1}\p^{\mu}_{x_1}A^{\nu}(x_1)A^{\mu}(x_2):\df&
			&= \n - :\p^{\mu}_{x_1}A^{\nu}(x_1)\p^{\nu}_{x_2}A^{\mu}(x_2):\df\\
			& \equiv \mathit{m}_{2,1} \pmod{\n}								&
			& \equiv \mathit{m}_{2,5} \pmod{\n}
\end{aligned}
\end{equation}
\\
\begin{equation}	\label{berech2.54}
\begin{aligned}
	\mathit{m}_{1,3}\bigr|_{\p^{\mu}_{x_1}} 
			&=:\p^{\mu}_{x_1}A^{\nu}(x_1)A^{\nu}(x_2):\p^{\mu}_{x_1}\df		&
	\qquad
	\mathit{m}_{1,3}\bigr|_{\p^{\mu}_{x_2}} 
			&=:\p^{\mu}_{x_1}A^{\nu}(x_1)A^{\nu}(x_2):\p^{\mu}_{x_2}\df		\\
			&= \n - :\p^{\mu}_{x_1}\p^{\mu}_{x_1}A^{\nu}(x_1)A^{\nu}(x_2):\df&
			&= \n - :\p^{\mu}_{x_1}A^{\nu}(x_1)\p^{\mu}_{x_2}A^{\nu}(x_2):\df\\
			& = 0															&
			& \equiv \mathit{m}_{2,4} \pmod{\n}
\end{aligned}
\end{equation}
\\
\begin{equation}	\label{berech2.55}
\begin{aligned}
	\mathit{m}_{1,4}\bigr|_{\p^{\mu}_{x_1}} 
			&=:\p^{\mu}_{x_1}u(x_1)\ti{u}(x_2):\p^{\mu}_{x_1}\df			&
	\qquad
	\mathit{m}_{1,4}\bigr|_{\p^{\mu}_{x_2}} 
			&=:\p^{\mu}_{x_1}u(x_1)\ti{u}(x_2):\p^{\mu}_{x_2}\df			\\
			&= \n - :\p^{\mu}_{x_1}\p^{\mu}_{x_1}u(x_1)\ti{u}(x_2):\df		&
			&= \n - :\p^{\mu}_{x_1}u(x_1)\p^{\mu}_{x_2}\ti{u}(x_2):\df		\\
			& = 0 															&
			& \equiv \mathit{m}_{2,3} \pmod{\n}
\end{aligned}
\end{equation}		
\\
\begin{equation}	\label{berech2.56}
\begin{aligned}
	\mathit{m}_{1,5}\bigr|_{\p^{\mu}_{x_1}} 
			&=:u(x_1)\p^{\mu}_{x_2}\tilde{u}(x_2):\p^{\mu}_{x_1}\df			&
	\qquad
	\mathit{m}_{1,5}\bigr|_{\p^{\mu}_{x_2}} 
			&=:u(x_1)\p^{\mu}_{x_2}\tilde{u}(x_2):\p^{\mu}_{x_2}\df			\\
			&= \n - :\p^{\mu}_{x_1}u(x_1)\p^{\mu}_{x_2}\tilde{u}(x_2):\df	&
			&= \n - :u(x_1)\p^{\mu}_{x_2}\p^{\mu}_{x_2}\tilde{u}(x_2):\df	\\
			& \equiv \mathit{m}_{2,3} \pmod{\n}								&
			& = 0
\end{aligned}
\end{equation}
shows, that all terms of $\mathcal{M}_2$ can be represented by those of $\mathcal{M}_1$. Thus the factor group $\mathcal{M}' := \,\raisebox{1ex} {$\mathcal{M}$}\raisebox{0ex}{$\! \Big/\!$}\raisebox{-1ex}{$\n$}$ can be written as the subset $\mathcal{M}_0$ only.
Next we consider the question, if there are inner equivalences between elements of $\mathcal{M}_0$, which would lead to a further reduction of representing elements. The answer is that there are no such equivalences, and so the final factor group reads as 
\begin{align}	\label{berech2.57}
	\mathcal{M}' 
	:= \,\raisebox{1ex}{$\mathcal{M}$}\raisebox{0ex}
				{$\! \Big/\!$}\raisebox{-1ex}{$\n$}
	= \big\{ \mathit{m}_{0,1}, \mathit{m}_{0,2}, \mathit{m}_{0,3}, 							\mathit{m}_{0,4}, \mathit{m}_{0,5} \big\}	
\end{align}

\subsubsection{The Subgroups \boldmath$\textbf{B}(\mathcal{M}')$, 						$\textbf{Z}(\mathcal{M}')$ and $\textbf{H}(\mathcal{M}')$}\label{BERECH2.5}

We determine the modified exact and closed subgroups of $\mathcal{M}'$ according to definition $\eqref{algeb22}$ and $\eqref{algeb23}$. Knowing these two subgroups, one easily realizes the factor group of $\textbf{H}(\mathcal{M}')$ and its gauge transformed set $d_Q(\textbf{H}(\mathcal{M}'))$. Therefore, to prove gauge invariance, it remains only to show that the image of $\textbf{H}(\mathcal{M}')$ under $d_Q$ is a superset of $\mathcal{L}'$.
For a detailed explanation of this connection, confer with the introductory chapter $\ref{ALGEBRA}$. According to the results showed there, all linear combinations of local terms $\mathit{l}_i$ (which can occur in diagram $3.3$) can be written gauge invariantly, if the relation $d_Q(\textbf{H}(\mathcal{M}')) \supseteq \mathcal{L}'$ can be proven.

\paragraph{The Subgroup \boldmath$\textbf{B}(\mathcal{M}')$}																				\label{BERECH2.5.1}

Obviously all elements of $\textbf{B}(\mathcal{M}')$ must include (according to the equations $\eqref{eich19a}$--$\eqref{eich19f}$) at least one derivation on an external field operator. Since the elements $\mathit{m}_{0,1}, \mathit{m}_{0,2}$ of $\mathcal{M}'$ cannot fulfill this restriction, they surely are not elements of $\textbf{B}(\mathcal{M}')$. Furthermore one realizes immediately, that under the remaining expressions only the term $\mathit{m}_{0,4}$ with the restriction $\p^{\mu}_{x_1}\p^{\nu}_{x_2}$ is in $\textbf{B}(\mathcal{M}')$, because
\begin{equation}	\label{berech2.58}
\begin{aligned}
	\mathit{m}_{0,4}\bigr|_{\p^{\mu}_{x_1}\p^{\nu}_{x_2}} 
			&= :A^{\mu}(x_1)A^{\nu}(x_2):\p^{\mu}_{x_1}\p^{\nu}_{x_2}\df	\\
			&= \n \,\, - :\p^{\mu}_{x_1}A^{\mu}(x_1)A^{\nu}(x_2): 																		\p^{\nu}_{x_2}\df		\\
			&= \n \,\, + :\p^{\mu}_{x_1}A^{\mu}(x_1)\p^{\nu}_{x_2}
													A^{\nu}(x_2):\df		\\
			&= \n + d_Q (:d_Q (\ti{u}) \, \ti{u}:)
\end{aligned}
\end{equation}
and no other term can be rewritten (with use of the above mentioned equations) in the form
\begin{align}	\label{berech2.59}
	d_Q(f) 	&  \equiv \mathit{m}_{0,j} \pmod \n 
			&& f \in \mathcal{L}\bigr|_{n_{u}(f)=-1} ,\; j \in \{3,4,5\}
\end{align}

\paragraph{The Subgroup \boldmath$\textbf{Z}(\mathcal{M}')$}																				\label{BERECH2.5.2}

According to definition $\eqref{algeb22}$, one has to show, whether gauge transformed elements $d_Q(\mathit{m}_{0,i})$ exist, which have the general form
\begin{align}	\label{berech2.60}
	d_Q(\mathit{m}_{0,i}) 	&	= 0 + \n + \mathit{\ti{l}}'_i
							&&	\mathit{\ti{l}}'_i \in \mathcal{L}
\end{align}
with $\mathit{\ti{l}}'_i$ \emph{one single} element out of $\mathcal{L}'$ only, or not.
To show if this meets any term, one has to calculate all gauge transformed expressions $d_Q(\mathit{m}_{i,j})$: 
\begin{align}
	d_Q(\mathit{m}_{1,1})																	&= d_Q(:A^{\mu}(x_1) A^{\mu}(x_2):\df)					\notag		\\
		&=(:\p^{\mu}u(x_1) A^{\mu}(x_2): + 
				:A^{\mu}(x_1) \p^{\mu}u(x_2):)\df		\label{berech2.61}
\\[1cm]
	d_Q(\mathit{m}_{0,2})																	&= d_Q(:u(x_1)\tilde{u}(x_2):\df)						\notag		\\
		&= (0 \,\, + :u(x_1)\p^{\mu}A^{\mu}(x_2):)\df	\label{berech2.62}	
\\[1cm]
	d_Q(\mathit{m}_{0,3})																	&= d_Q(:A^{\mu}(x_1)A^{\mu}(x_2):\p^{\nu}\p^{\nu}\df)	\notag		\\	
		&= (:\p^{\mu}u(x_1)A^{\mu}(x_2): + :A^{\mu}(x_1)\p^{\mu}u(x_2):)
				\p^{\nu}\p^{\nu}\df						\label{berech2.63}
\\[1cm]
	d_Q(\mathit{m}_{0,4})																	&= d_Q(:A^{\mu}(x_1)A^{\nu}(x_2):\p^{\mu}\p^{\nu}\df)	\notag		\\
		&= (:\p^{\mu}u(x_1)A^{\nu}(x_2): + :A^{\mu}(x_1)\p^{\nu}u(x_2):) 							\p^{\mu}\p^{\nu}\df						\label{berech2.64}
\\[1cm]
	d_Q(\mathit{m}_{0,5})																	&= d_Q(:u(x_1)\tilde{u}(x_2):\p^{\nu}\p^{\nu}\df)		\notag		\\
		&=(0 \,\, + :u(x_1)\p^{\mu}A^{\mu}(x_2): 
				\p^{\nu}\p^{\nu}\df						\label{berech2.65}
\end{align}
Before finishing this section, the following remark should be made
\begin{quote}
	Remembering definition $\eqref{algeb22}$, one immediately realizes that 			under $d_Q$ all terms $\mathit{m}_i$ are mapped onto $\mathcal{L}$. But 			there is no guarantee, that the image is a single element of $\mathcal{L}'$.
	
	Thus there remains to show in the last step, that the set of the gauge 				transformed elements of the group $\textbf{H}(\mathcal{M}')$ (comprising all 	transformed elements $d_Q(\mathit{m}_{i,j})$ which delivers a single element 	$\mathit{l}_i$ under the gauge transformation) covers the whole set 				$\mathcal{L}'$\;\footnote{Confer figure $1$.}.
\end{quote}
With this intermediate statement in mind we can go back to our task and note, that only the two images $d_Q(m_{0,2})$ and $d_Q(m_{0,5})$ fulfill the mentioned requirement. Thus the subgroup $\textbf{Z}$ simply comprises of
\begin{equation}	\label{berech2.66}
	\textbf{Z}(\mathcal{M}') = \{ m_{0,2} \, , \, m_{0,5} \}
\end{equation}

\paragraph{The Gauge--Factor Subgroup \boldmath$\textbf{H}(\mathcal{M}')$} 																	\label{BERECH2.5.3}

Without hesitation we can write down the gauge--factor group as
\begin{align}
	\textbf{H}(\mathcal{M}') 
	&= \, \raisebox{1ex}{$\textbf{Z}(\mathcal{M})$} \raisebox{0ex}{$\! \Big/\!$} 	\raisebox{-1ex}{$\textbf{B}(\mathcal{M})$}
	= \, \raisebox{1ex}{$\{ m_{0,2} \, , \, m_{0,5} \}$}
	\raisebox{0ex}{$\! \Big/\!$}
	\raisebox{-1ex}{$\{ m_{0,4}\bigr|_{\p^{\mu}_{x_1}\p^{\nu}_{x_2}}\}$}\notag\\
	&=\{ m_{0,2} \, , \, m_{0,5} \}							\label{berech2.67}
\end{align}
Comparing the set of gauge transformed elements of $\textbf{H}(\mathcal{M}')$ with $\mathcal{L}'$ one immediately finds, that $d_Q(\textbf{H}(\mathcal{M}'))$ contains $\mathcal{L}'$
\begin{equation}	\label{berech2.68}
	\mathcal{L}' \subseteq d_Q(\textbf{H}(\mathcal{M}'))
\end{equation}
With this result gauge invariance of all $2$--leg graphs is proven and we can direct our attention to the $3$--leg case.
\subsection{\boldmath $3$--Leg Calculations for Disjunct Arguments}					\label{BERECH3}

Similar to the preceeding section we will start discussing the $3$--leg calculations with a few basic restrictions which guide the calculations.
As mentioned in the $2$--leg case, the color structure plays an important r\^ole. Thus let us start with this topic first.

\begin{itemize}
\item
	If one takes into consideration the explanations given in the appendix of 			\cite{YMIV} as well as the fundamental works of Dittner \cite{dittner1}, 			\cite{dittner2} one immediately realizes, that all $3$--leg graphs (as well 		as all \emph{local} $3$--leg graphs) of any $T_n$  have the color tensor 			structure
	\begin{equation}	\label{berech3.1}
		f_{abc}\mathcal{D}^{\alpha}\!\!:\Lambda_a \Gamma_b \Delta_c:
			\mathcal{D}^{\beta}\df
	\end{equation}
	This result is mainly based on the even--odd theorem proven in appendix 			\textbf{A} of \cite{YMIV}, since the latter shows that odd numbers of 				$d_{abc}$\footnote{For more details about $d_{abc}$ and related items confer 	with the introduction to the $4$--leg calculations and the above mentioned 			literature.} cannot occur in any term of $T_n$\footnote{In \cite{dittner1}, 		\cite{dittner2} it is demonstrated that the only possible color tensors, 			which can occur in the three leg case, are the $d_{abc}$ and the $f_{abc}$. 		Together with the even--odd theorem, mentioned above, only the $f_{abc}$'s 			survive.}.

	Because all local terms have the same color--tensor structure, one can 				(except if noted explicitely the converse) omit the color tensors.
\item
	Additionally, equation $\eqref{berech1.2}$ has in the $3$--leg case the 			special form
	\begin{equation} 	\label{berech3.2}
		\sum_{j=1}^n \p_{x_j}^{\mu} \df = \p_{x_1}^{\mu} \df + 									\p_{x_2}^{\mu} \df + \p_{x_3}^{\mu} \df +
			\sum_{\trm{inner}} \p^{\mu}\df = 0
	\end{equation}
	whereas the inner derivatives simply yield to divergences again.
\end{itemize}
Similar to the calculations in the $2$--leg case, we start with the determination of the elements of $\mathcal{L}$ followed  by those of $\mathcal{M}$. In both cases, the terms are listed with disjunct arguments for the three field operators. Then gauge invariance for all local $3$--leg terms (with disjunct arguments) will be demonstrated. The subsequent chapter is devoted to the proof of gauge invariance for local terms based on reducible graphs.

\subsubsection{Determination of the Elements of \boldmath$\mathcal{L}$} 																	\label{BERECH3.1}

Again the set $\mathcal{L}$ will be broken up into subsets $\mathcal{L}_d$ with the same number of external derivatives on the field operator parts.
Equation $\eqref{algeb10}$ results in
\begin{equation}	\label{berech3.3}
	\omega(3\text{--leg}) \leqslant 4 - 3 - d + 1 = 2 - d
\end{equation}
and thus $\mathcal{L}$ can be written as the union of the three subsets $\mathcal{L}_0$, $\mathcal{L}_1$, $\mathcal{L}_2$
\begin{equation}	\label{berech3.4}
	\mathcal{L} = \mathcal{L}_0 \cup \mathcal{L}_1 \cup\mathcal{L}_2 				\end{equation}
Due to the fact, that the ghost number of the elements of $\mathcal{L}$ is given by $n_u(\mathcal{L}) = 1$, only graphs with the field operator part $AAu$ and $u\ti{u}u$ can occur.

\paragraph{The Subset \boldmath$\mathcal{L}_0$}		\label{BERECH3.1.0}

According to the above mentioned equation, the singular order is given here as
\begin{equation}	\label{berech3.5}
	\omega(\mathcal{L}_0) \leqslant 4 - 3 - 0 + 1 = 2
\end{equation}
which results in the general form of $\mathcal{L}_0$
\begin{equation}	\label{berech3.6}
	\mathcal{L}_0 = \Big\{:\Lambda_a(x_1)\Gamma_b(x_2)\Delta_c(x_3): 					\mathcal{D}^{\beta}\df \Bigm| \Lambda\Gamma\Delta \in \left\{ A^{\kappa} 			A^{\xi}u ,u\ti{u}u \right\} \; , \; |\beta| \leqslant 2 \; , \; x_1 				\leftrightarrow x_2 \leftrightarrow x_3 \Big\}
\end{equation}
As usual, the arguments in the expression above can occur in exchanged order as indicated by $x_1 \leftrightarrow x_2 \leftrightarrow x_3$.
Since terms with $|\beta| = 1$ cannot be Lorentz scalars, only terms with $|\beta| = 0$ and $|\beta| = 2$ must be considered in detail. Additionally the expressions represent gauge transformed terms, so at least one derivation must appear\footnote{For related considerations confer with $\eqref{eich19a}$--$\eqref{eich19f}$ and the additional remarks in chapter $\ref{EICH}$.}. But $\mathcal{L}_0$ restricted to $|\beta| = 0$ has no derivation at all and thus $\mathcal{L}_0$ consists of the terms\\
$\mathit{l}_{0,i} \in \mathcal{L}_0 \bigr|_{|\beta| = 2}$
\begin{equation}	\label{berech3.7}
\begin{aligned}
	\mathit{l}_{0,1}&\overset{def}{=}\, 														:A^{\mu}_a(x_1) A^{\mu}_b(x_2) u_c(x_3):\p^{\nu}\p^{\nu}\df		&
	\qquad\qquad
	\mathit{l}_{0,2}&\overset{def}{=}\, 														:A^{\mu}_a(x_1) A^{\nu}_b(x_2) u_c(x_3):\p^{\mu}\p^{\nu}\df		 \\
	\mathit{l}_{0,3}&\overset{def}{=}\, 
			:u_a(x_1) \ti{u}_b(x_2) u_c(x_3): \p^{\mu}\p^{\mu}\df
\end{aligned}
\end{equation}
only. It should be remembered, that in the above expressions, the derivations on $\df$'s have to again be understood as defined in $\eqref{berech2.9}$.

\paragraph{The Subset \boldmath$\mathcal{L}_1$}		\label{BERECH3.1.1}

The singular order is given as
\begin{equation}	\label{berech3.8}
	\omega(\mathcal{L}_1) \leqslant 4 - 3 - 1 + 1 = 1
\end{equation}
which leads to the general set
\begin{equation}	\label{berech3.9}
	\mathcal{L}_1 =
	 \Big\{\mathcal{D}^{\alpha}:\Lambda_a(x_1)\Gamma_b(x_2)\Delta_c(x_3): 				\mathcal{D}^{\beta}\df \Bigm| \Lambda\Gamma\Delta \in \left\{ A^{\kappa} 			A^{\xi}u ,u\ti{u}u \right\} \; , \; |\alpha| = 1 \; , \; |\beta| \leqslant 1 	\; , \;  x_1 \leftrightarrow x_2 \leftrightarrow x_3 \Big\}
\end{equation}
whereas terms with $|\beta| = 0$ are not Lorentz scalars. Altogether the list of all elements of $\mathcal{L}_1$ is given by\\
$\mathit{l}_{1,i} \in \mathcal{L}_1 \bigr|_{|\beta| = 1}$
\begin{equation}	\label{berech3.10}
\begin{aligned}
	\mathit{l}_{1,1}&\overset{def}{=}\,
		:\p^{\mu}A^{\mu}_a(x_1) A^{\nu}_b(x_2) u_c(x_3):\p^{\nu}\df			&
	\qquad
	\mathit{l}_{1,2}&\overset{def}{=}\,
		:\p^{\mu}A^{\nu}_a(x_1) A^{\mu}_b(x_2) u_c(x_3):\p^{\nu}\df 		\\
	\mathit{l}_{1,3}&\overset{def}{=}\,
		:\p^{\mu}A^{\nu}_a(x_1) A^{\nu}_b(x_2) u_c(x_3):\p^{\mu}\df			&
	\mathit{l}_{1,4}&\overset{def}{=}\,
		:A^{\mu}_a(x_1) A^{\mu}_b(x_2) \p^{\nu}u_c(x_3):\p^{\nu}\df 		\\
	\mathit{l}_{1,5}&\overset{def}{=}\,
		:A^{\mu}_a(x_1) A^{\nu}_b(x_2) \p^{\nu}u_c(x_3):\p^{\mu}\df 		\\
	\mathit{l}_{1,6}&\overset{def}{=}\,
		:\p^{\mu}u_a(x_1) \ti{u}_b(x_2) u_c(x_3):\p^{\mu}\df				&
	\mathit{l}_{1,7}&\overset{def}{=}\,
		:u_a(x_1) \p^{\mu}\ti{u}_b(x_2) u_c(x_3):\p^{\mu}\df
\end{aligned}
\end{equation}
Since the expressions above also represent terms with exchanged arguments, the list covers all elements of the set $\mathcal{L}_1$.

\paragraph{The Subset \boldmath$\mathcal{L}_2$}		\label{BERECH3.1.2}

This last subset has singular order $\omega(\mathcal{L}_2) = 0$ and thus the most common form
\begin{equation}	\label{berech3.11}
	\mathcal{L}_2 =
	\Big\{\mathcal{D}^{\alpha}:\Lambda_a(x_1)\Gamma_b(x_2)\Delta_c(x_3): \df 			\Bigm| \Lambda\Gamma\Delta \in \left\{ A^{\kappa} A^{\xi}u ,u\ti{u}u 				\right\} \; , \; |\alpha| = 2 \;  , \;  x_1 \leftrightarrow x_2 					\leftrightarrow x_3 \Big\}
\end{equation}
which leads to the explicit list\\
$\mathit{l}_{2,i} \in \mathcal{L}_2$
\begin{equation}	\label{berech3.12}
\begin{aligned}
	\mathit{l}_{2,1}&\overset{def}{=}\,
		:\p^{\mu}\p^{\nu}A^{\mu}_a(x_1)A^{\nu}_b(x_2)u_c(x_3):\df			&
	\qquad
	\mathit{l}_{2,2}&\overset{def}{=}\,
		:A^{\mu}_a(x_1)A^{\nu}_b(x_2)\p^{\mu}\p^{\nu}u_c(x_3):\df			\\
	\mathit{l}_{2,3}&\overset{def}{=}\,
		:\p^{\mu}A^{\mu}_a(x_1)\p^{\nu}A^{\nu}_b(x_2)u_c(x_3):\df			&	
	\mathit{l}_{2,4}&\overset{def}{=}\,
		:\p^{\mu}A^{\nu}_a(x_1)\p^{\mu}A^{\nu}_b(x_2)u_c(x_3):\df			\\
	\mathit{l}_{2,5}&\overset{def}{=}\,
		:\p^{\mu}A^{\nu}_a(x_1)\p^{\nu}A^{\mu}_b(x_2)u_c(x_3):\df			&	
	\mathit{l}_{2,6}&\overset{def}{=}\,
		:\p^{\mu}A^{\mu}_a(x_1)A^{\nu}_b(x_2)\p^{\nu}u_c(x_3):\df			\\
	\mathit{l}_{2,7}&\overset{def}{=}\,
		:\p^{\mu}A^{\nu}_a(x_1)A^{\mu}_b(x_2)\p^{\nu}u_c(x_3):\df			&	
	\mathit{l}_{2,8}&\overset{def}{=}\,
		:\p^{\mu}A^{\nu}_a(x_1)A^{\nu}_b(x_2)\p^{\mu}u_c(x_3):\df			\\
	\mathit{l}_{2,9}&\overset{def}{=}\,
		:\p^{\mu}u_a(x_1)\p^{\mu}\ti{u}_b(x_2)u_c(x_3):\df					&	
	\mathit{l}_{2,10}&\overset{def}{=}\,
		:\p^{\mu}u_a(x_1)\ti{u}_b(x_2)\p^{\mu}u_c(x_3):\df
\end{aligned}
\end{equation}
All other terms which can be achieved by the remaining allocations of the derivations among the field operators equal zero due to the wave equations $\eqref{eich10a}$ and $\eqref{eich10d}$.
In the next step we will determine the factor set $\mathcal{L}' :=\,\raisebox{1ex} {$\mathcal{L}$}\raisebox{0ex}{$\! \Big/\!$}\raisebox{-1ex}{$\n$}$.

\subsubsection{Determination of Equivalent Elements in \boldmath$\mathcal{L}$} 																	\label{BERECH3.2}

The same procedure as in the $2$--leg case also applies here, and so we can start the calculations without a lot of explanatory remarks. Only the transformation of the first term is executed in a slightly extended form
\begin{equation}	\label{berech3.13}
\begin{aligned}
	\mathit{l}_{0,1}\bigr|_{\p^{\nu}_{x_1}}
		&= :A^{\mu}_a(x_1) A^{\mu}_b(x_2) u_c(x_3): 												\p^{\nu}_{x_1}\p^{\nu}\df 									\\
		&= \p^{\nu}_{x_1}\left(:A^{\mu}_a(x_1)A^{\mu}_b(x_2) u_c(x_3): 								\p^{\nu}\df \right) - : \p^{\nu}_{x_1} A^{\mu}_a(x_1) 								A^{\mu}_b(x_2) u_c(x_3): \p^{\nu}\df					 	\\
		&= \n - :\p^{\nu}_{x_1}A^{\mu}_a(x_1) A^{\mu}_b(x_2)u_c(x_3): 								\p^{\nu}\df												 	\\
		& \equiv -	\mathit{l}_{1,3} \pmod{\n}
\end{aligned}
\end{equation}
All the remaining ones appear in abbreviated form
\begin{equation}	\label{berech3.14}
\begin{aligned}
	\mathit{l}_{0,1}\bigr|_{\p^{\nu}_{x_2}}
		&= :A^{\mu}_a(x_1) A^{\mu}_b(x_2) u_c(x_3):
										\p^{\nu}_{x_2}\p^{\nu}\df			
	\qquad\qquad															&
	\mathit{l}_{0,1}\bigr|_{\p^{\nu}_{x_3}}
		&= :A^{\mu}_a(x_1) A^{\mu}_b(x_2) u_c(x_3):
										\p^{\nu}_{x_3}\p^{\nu}\df			\\
		&= \n - :A^{\mu}_a(x_1)\p^{\nu}_{x_2}A^{\mu}_b(x_2)u_c(x_3):
										\p^{\nu}\df							&
		&= \n - :A^{\mu}_a(x_1)A^{\mu}_b(x_2)\p^{\nu}_{x_3}u_c(x_3): 														\p^{\nu}\df							\\
		& \equiv - \mathit{l}_{1,3} \pmod{\n}								&
		& \equiv - \mathit{l}_{1,4} \pmod{\n}
\end{aligned}
\end{equation}
\\
\begin{equation}	\label{berech3.15}
\begin{aligned}
	\mathit{l}_{0,2}\bigr|_{\p^{\mu}_{x_1}}
	 	&= :A^{\mu}_a(x_1) A^{\nu}_b(x_2) u_c(x_3): 																		\p^{\mu}_{x_1}\p^{\nu}\df			\\
		&= \n - :\p^{\mu}_{x_1}A^{\mu}_a(x_1) A^{\nu}_b(x_2)u_c(x_3):
										\p^{\nu}\df							\\
		& \equiv -\mathit{l}_{1,1} \pmod{\n}
\\[.6cm]
	\mathit{l}_{0,2}\bigr|_{\p^{\mu}_{x_2}}
		&= :A^{\mu}_a(x_1) A^{\nu}_b(x_2) u_c(x_3): 																		\p^{\mu}_{x_2}\p^{\nu}\df			
	\qquad\qquad															&
	\mathit{l}_{0,2}\bigr|_{\p^{\mu}_{x_3}}
	 	&= :A^{\mu}_a(x_1) A^{\nu}_b(x_2) u_c(x_3): 																		\p^{\mu}_{x_3}\p^{\nu}\df			\\ 
		&= \n - :A^{\mu}_a(x_1) \p^{\mu}_{x_2}A^{\nu}_b(x_2)u_c(x_3): 														\p^{\nu}\df					 		&
		&= \n - :A^{\mu}_a(x_1) A^{\nu}_b(x_2) \p^{\mu}_{x_3}u_c(x_3): 														\p^{\nu}\df							\\
		& \equiv - \mathit{l}_{1,2} \pmod{\n}								&
		& \equiv - \mathit{l}_{1,5} \pmod{\n}
\end{aligned}
\end{equation}
\\
\begin{equation}	\label{berech3.16}
\begin{aligned}
	\mathit{l}_{0,3}\bigr|_{\p^{\mu}_{x_1}}
	 	&= :u_a(x_1) \ti{u}_b(x_2) u_c(x_3):\p^{\mu}_{x_1}\p^{\mu}\df		\\
		&= \n - :\p^{\mu}_{x_1}u_a(x_1) \ti{u}_b(x_2)u_c(x_3):\p^{\mu}\df	\\
		& \equiv - \mathit{l}_{1,6} \pmod{\n}
\\[.6cm]
	\mathit{l}_{0,3}\bigr|_{\p^{\mu}_{x_2}}
	 	&= :u_a(x_1) \ti{u}_b(x_2) u_c(x_3):\p^{\mu}_{x_2}\p^{\mu}\df		
	\qquad\qquad															&
	\mathit{l}_{0,3}\bigr|_{\p^{\mu}_{x_3}}
	 	&= :u_a(x_1) \ti{u}_b(x_2) u_c(x_3):\p^{\mu}_{x_3}\p^{\mu}\df		\\
		&= \n - :u_a(x_1) \p^{\mu}_{x_2}\ti{u}_b(x_2)u_c(x_3):\p^{\mu}\df	&
		&= \n - :u_a(x_1)\ti{u}_b(x_2) \p^{\mu}_{x_3}u_c(x_3):\p^{\mu}\df	\\
		& \equiv - \mathit{l}_{1,7} \pmod{\n}								&
		& \equiv - \mathit{l}_{1,6} \pmod{\n}
\end{aligned}
\end{equation}
In summary, the above results state the fact, that $\mathcal{L}_0$ and $\mathcal{L}_1$ are equivalent
\begin{equation}	\label{berech3.17}
	\mathcal{L}_0 \sim \mathcal{L}_1
\end{equation}
It remains to show, if equivalences between elements of $\mathcal{L}_0$ or $\mathcal{L}_1$ with those of $\mathcal{L}_2$ exist. This is most easily realized by transforming all elements of $\mathcal{L}_1$ and demonstrating, that the achieved terms cover the elements of $\mathcal{L}_2$.
If this can be done, the following relations hold
\begin{align}	\label{berech3.18}
	\mathcal{L}_0 \sim \mathcal{L}_1 && \mathcal{L}_1 \sim \mathcal{L}_2
\end{align}
and the whole set $\mathcal{L}$ can be represented by the single subset $\mathcal{L}_0$ alone.
The first few transformations of $\mathcal{L}_1$ explicitely read
\begin{equation}	\label{berech3.19}
\begin{aligned}
	\mathit{l}_{1,1}\bigr|_{\p^{\nu}_{x_1}}	
		&= :\p^{\mu}_{x_1}A_a^{\mu}(x_1) A^{\nu}_b(x_2)u_c(x_3): 											\p^{\nu}_{x_1}\df									\\
		&= \n - :\p^{\nu}_{x_1}\p^{\mu}_{x_1}
						A_a^{\mu}(x_1)A^{\nu}_b(x_2) u_c(x_3)\df			\\
		& \equiv - \mathit{l}_{2,1} \pmod{\n}
\\[.6cm]
	\mathit{l}_{1,1}\bigr|_{\p^{\nu}_{x_2}} 	
		&= :\p^{\mu}_{x_1}A_a^{\mu}(x_1) A^{\nu}_b(x_2)u_c(x_3): 											\p^{\nu}_{x_2}\df									
	\qquad\qquad															&
	\mathit{l}_{1,1}\bigr|_{\p^{\nu}_{x_3}}
	 	&= :\p^{\mu}_{x_1}A_a^{\mu}(x_1) A^{\nu}_b(x_2)u_c(x_3): 											\p^{\nu}_{x_3}\df									\\
		&= \n - :\p^{\mu}_{x_1}A_a^{\mu}(x_1)\p^{\nu}_{x_2}A^{\nu}_b(x_2)
						u_c(x_3): \df										&
		&= \n - :\p^{\mu}_{x_1}A_a^{\mu}(x_1) A^{\nu}_b(x_2) 												\p^{\nu}_{x_3}u_c(x_3):\df							\\
		& \equiv - \mathit{l}_{2,3} \pmod{\n}								&
		& \equiv - \mathit{l}_{2,6} \pmod{\n} 
\end{aligned}
\end{equation}
\\
\begin{equation}	\label{berech3.20}
\begin{aligned}
	\mathit{l}_{1,2}\bigr|_{\p^{\nu}_{x_1}} 
		&= :\p^{\mu}_{x_1}A_a^{\nu}(x_1) A^{\mu}_b(x_2)u_c(x_3): 											\p^{\nu}_{x_1}\df									\\
		&= \n - :\p^{\mu}_{x_1}\p^{\nu}_{x_1}A_a^{\nu}(x_1) 												A^{\mu}_b(x_2) u_c(x_3): \df						\\
		& \equiv - \mathit{l}_{2,1} \pmod{\n}
\\[.6cm]
	\mathit{l}_{1,2}\bigr|_{\p^{\nu}_{x_2}}
	 	&= :\p^{\mu}_{x_1}A_a^{\nu}(x_1) A^{\mu}_b(x_2)u_c(x_3): 											\p^{\nu}_{x_2}\df									
	\qquad\qquad															&
	\mathit{l}_{1,2}\bigr|_{\p^{\nu}_{x_3}}
	 	&= :\p^{\mu}_{x_1}A_a^{\nu}(x_1) A^{\mu}_b(x_2)u_c(x_3): 											\p^{\nu}_{x_3}\df									\\
		&= \n - :\p^{\mu}_{x_1}A_a^{\nu}(x_1)\p^{\nu}_{x_2} 												A^{\mu}_b(x_2) u_c(x_3): \df						&
		&= \n - :\p^{\mu}_{x_1}A_a^{\nu}(x_1) A^{\mu}_b(x_2) 												\p^{\nu}_{x_3}u_c(x_3): \df							\\
		& \equiv - \mathit{l}_{2,5} \pmod{\n}								&
		& \equiv - \mathit{l}_{2,7} \pmod{\n} 
\end{aligned}
\end{equation}
\\
\begin{equation}	\label{berech3.21}
\begin{aligned}
	\mathit{l}_{1,3}\bigr|_{\p^{\mu}_{x_1}}	
		&= : \p^{\mu}_{x_1}A_a^{\nu}(x_1) A^{\nu}_b(x_2)u_c(x_3): 											\p^{\mu}_{x_1}\df									\\
		&= \n - :\p^{\mu}_{x_1}\p^{\mu}_{x_1}A_a^{\nu}(x_1) 												A^{\nu}_b(x_2)u_c(x_3): \df							\\
		& \equiv 0\
\\[.6cm]
	\mathit{l}_{1,3}\bigr|_{\p^{\mu}_{x_2}}
	 	&= : \p^{\mu}_{x_1}A_a^{\nu}(x_1) A^{\nu}_b(x_2)u_c(x_3): 											\p^{\mu}_{x_2}\df									
	\qquad\qquad															&
	\mathit{l}_{1,3}\bigr|_{\p^{\mu}_{x_3}}
	 	&= : \p^{\mu}_{x_1}A_a^{\nu}(x_1) A^{\nu}_b(x_2)u_c(x_3): 											\p^{\mu}_{x_3}\df									\\
		&= \n - :\p^{\mu}_{x_1}A_a^{\nu}(x_1)\p^{\mu}_{x_2} 												A^{\nu}_b(x_2) u_c(x_3):\df							&
		&= \n - : \p^{\mu}_{x_1}A_a^{\nu}(x_1) A^{\nu}_b(x_2) 												\p^{\mu}_{x_3}u_c(x_3):\df							\\
		& \equiv - \mathit{l}_{2,4} \pmod{\n}								&
		& \equiv - \mathit{l}_{2,8} \pmod{\n} 
\end{aligned}
\end{equation}
The terms $\mathit{l}_{1,4}$ and $\mathit{l}_{1,6}$ do not lead to new equivalences with elements of $\mathcal{L}_2$. Thus the calculations for these expressions are dropped. The remaining elements yield
\begin{equation}	\label{berech3.22}
\begin{aligned}
	\mathit{l}_{1,5}\bigr|_{\p^{\mu}_{x_1}}
	 	&= : A_a^{\mu}(x_1) A^{\nu}_b(x_2)\p^{\mu}_{x_3}u_c(x_3): 											\p^{\nu}_{x_1}\df									\\
		&= \n - : \p^{\nu}_{x_1}A_a^{\mu}(x_1) A^{\nu}_b(x_2) 												\p^{\mu}_{x_3} u_c(x_3):\df							\\
		& \equiv - \mathit{l}_{2,7} \pmod{\n} 
\\[.6cm]
	\mathit{l}_{1,5}\bigr|_{\p^{\mu}_{x_2}}
	 	&= : A_a^{\mu}(x_1)A^{\nu}_b(x_2)\p^{\mu}_{x_3}u_c(x_3): 											\p^{\nu}_{x_2}\df									
	\qquad\qquad															&
	\mathit{l}_{1,5}\bigr|_{\p^{\mu}_{x_3}}
	 	&= : A_a^{\mu}(x_1)A^{\nu}_b(x_2)\p^{\mu}_{x_3} 													u_c(x_3):\p^{\nu}_{x_3}\df							\\
		&= \n - : A_a^{\mu}(x_1)\p^{\nu}_{x_2}A^{\nu}_b(x_2) 												\p^{\mu}_{x_3} u_c(x_3):\df							&
		&= \n - : A_a^{\mu}(x_1)A^{\nu}_b(x_2)\p^{\mu}_{x_3} 												\p^{\nu}_{x_3}u_c(x_3):\df							\\
		& \equiv - \mathit{l}_{2,6} \pmod{\n}								&
		& \equiv - \mathit{l}_{2,2} \pmod{\n} 
\end{aligned}
\end{equation}
\\
\begin{equation}	\label{berech3.23}
\begin{aligned}
	\mathit{l}_{1,7}\bigr|_{\p^{\mu}_{x_1}}
		&= : \p^{\mu}_{x_1}u_a(x_1) \ti{u}_b(x_2)u_c(x_3):
						\p^{\mu}_{x_1}\df									\\
		&= \n - : \p^{\mu}_{x_1}\p^{\mu}_{x_1}u_a(x_1) 													\ti{u}_b(x_2) u_c(x_3): \df								\\
		& \equiv 0 
\\[.6cm]
	\mathit{l}_{1,7}\bigr|_{\p^{\mu}_{x_2}}
	 	&= :\p^{\mu}_{x_1}u_a(x_1) \ti{u}_b(x_2)u_c(x_3):\p^{\mu}_{x_2}\df	
	\qquad\qquad															&
	\mathit{l}_{1,7}\bigr|_{\p^{\mu}_{x_3}}
	 	&= :\p^{\mu}_{x_1}u_a(x_1) \ti{u}_b(x_2)u_c(x_3):\p^{\mu}_{x_3}\df	\\
		&= \n - : \p^{\mu}_{x_1}u_a(x_1) \p^{\mu}_{x_2} 													\ti{u}_b(x_2) u_c(x_3): \df							&
		&= \n - : \p^{\mu}_{x_1}u_a(x_1) \ti{u}_b(x_2) 														\p^{\mu}_{x_3} u_c(x_3): \df						\\
		& \equiv - \mathit{l}_{2,9} \pmod{\n}								&
		& \equiv - \mathit{l}_{2,10} \pmod{\n} 
\end{aligned}
\end{equation}
Collecting the equivalences shows, that all elements of $\mathcal{L}_2$ can be represented by those of $\mathcal{L}_1$. Thus the factor set $\mathcal{L}'$ comprises the three elements
\begin{align}	\label{berech3.24}
	\mathcal{L}' 
	:= \,\raisebox{1ex}{$\mathcal{L}$}\raisebox{0ex}
				{$\! \Big/\!$}\raisebox{-1ex}{$\n$}
	= \big\{\mathit{l}_{0,1}\,,\,\mathit{l}_{0,2}\,,\,\mathit{l}_{0,3}\big\}
\end{align}
only. Up until now we made no use of the additional restriction that all singular terms must have a permutation invariant form in there arguments. Taking this into consideration for the factor group above, one finds that not all found terms are different from zero.
Thus we next determine the permutation invariant forms of the elements of $\mathcal{L}'$.

\subsubsection{The Permutation Invariant Elements of \boldmath$\mathcal{L}'$}																	\label{BERECH3.3}

\paragraph{The Elements \boldmath$\mathit{l}_{0,1}$}	\label{BERECH3.3.1}

First we remember, that the derivatives $\p^{\nu}\p^{\nu}$ occurring in the considered expressions represent the set of all possible combinations of derivations for different arguments\footnote{Confer with definition $\eqref{berech2.9}$.}. So, in the following one has to distinguish between the different cases\footnote{The derivative--pairs with $\sum_{\trm{inner}}\df$ must not be treated further, since they are equivalent to zero.}
\begin{equation}	\label{berech3.25}
	\bigg\{ \p^{\nu}_{x_1}\p^{\nu}_{x_1}, \p^{\nu}_{x_1}\p^{\nu}_{x_2},
			\p^{\nu}_{x_1}\p^{\nu}_{x_3}, 														\ldots
			\p^{\nu}_{x_3}\p^{\nu}_{x_3} \bigg\}
\end{equation}	
Then, the general permutation invariant form of $\mathit{l}_{0,1}$ is given by
\begin{align}	\label{berech3.26}
	\mathit{l}_{0,1}\big|_{\trm{P}} &= \sum_{ \trm{P} \{ x_1,x_2,x_3 \} } 					:A_a^{\mu}(x_1) A^{\mu}_b(x_2) u_c(x_3):\p^{\nu}_{x_l} \p^{\nu}_{x_m} 				\df	&&\forall (x_l,x_m) \in \{ x_1,x_2,x_3 \}
\end{align}
whereas the sum runs over all permutations of the elements $\{ x_i,x_j,x_k \}$. In full length this is written as (again for all pairs $(x_l,x_m)$ of elements out of $\{ x_i,x_j,x_k \}$)
\begin{subequations}
\begin{align}
	\mathit{l}_{0,1}\big|_{\trm{P}}
	&=\, :A_a^{\mu}(x_1) A^{\mu}_b(x_2) u_c(x_3): 										\p^{\nu}_{\trm{P}(x_l)} \p^{\nu}_{\trm{P}(x_m)}\df	\label{berech3.27a}	\\
	&\hphantom{= }+	:A_a^{\mu}(x_1) A^{\mu}_b(x_3) u_c(x_2): 							\p^{\nu}_{\trm{P}(x_l)} \p^{\nu}_{\trm{P}(x_m)}\df	\label{berech3.27b} \\
	&\hphantom{= }+ :A_a^{\mu}(x_2) A^{\mu}_b(x_1) u_c(x_3): 							\p^{\nu}_{\trm{P}(x_l)} \p^{\nu}_{\trm{P}(x_m)}\df	\label{berech3.27c}	\\
	&\hphantom{= }+	:A_a^{\mu}(x_2) A^{\mu}_b(x_3) u_c(x_1): 							\p^{\nu}_{\trm{P}(x_l)} \p^{\nu}_{\trm{P}(x_m)}\df	\label{berech3.27d}	\\
	&\hphantom{= }+	:A_a^{\mu}(x_3) A^{\mu}_b(x_2) u_c(x_1): 							\p^{\nu}_{\trm{P}(x_l)} \p^{\nu}_{\trm{P}(x_m)}\df	\label{berech3.27e}	\\
	&\hphantom{= }+	:A_a^{\mu}(x_3) A^{\mu}_b(x_1) u_c(x_2): 							\p^{\nu}_{\trm{P}(x_l)} \p^{\nu}_{\trm{P}(x_m)}\df	\label{berech3.27f}					\end{align}
\end{subequations}
For the sum above, the various cases of derivative--pairs lead to different results. One obtains for

\subparagraph{The Case \boldmath$\p^{\nu}_{x_i}\p^{\nu}_{x_i}$}
												\label{BERECH3.3.1.0}

The wave equations make the terms equal to zero.

\subparagraph{The Case \boldmath$\p^{\nu}_{x_1}\p^{\nu}_{x_2}$}																	\label{BERECH3.3.1.1} 

With this restriction (the derivations for the various terms in $\eqref{berech3.27a}$ -- $\eqref{berech3.27f}$ must be written with appropriate arguments) the term $\mathit{l}_{0,1}\big|_{\trm{P}}$ is
\begin{subequations}
\begin{align}
	\mathit{l}_{0,1}\big|_{\trm{P},\p^{\nu}_{x_1}\p^{\nu}_{x_2}}
		&=\, :A_a^{\mu}(x_1) A^{\mu}_b(x_2) u_c(x_3): 												\p^{\nu}_{x_1}\p^{\nu}_{x_2}\df 	\label{berech3.28a}		\\
		&\hphantom{= }+	:A_a^{\mu}(x_1) A^{\mu}_b(x_3) u_c(x_2): 									\p^{\nu}_{x_1}\p^{\nu}_{x_3}\df		\label{berech3.28b} 	\\
		&\hphantom{= }+ :A_a^{\mu}(x_2) A^{\mu}_b(x_1) u_c(x_3): 									\p^{\nu}_{x_2}\p^{\nu}_{x_1}\df		\label{berech3.28c}		\\
		&\hphantom{= }+	:A_a^{\mu}(x_2) A^{\mu}_b(x_3) u_c(x_1): 									\p^{\nu}_{x_2}\p^{\nu}_{x_3}\df		\label{berech3.28d}		\\
		&\hphantom{= }+	:A_a^{\mu}(x_3) A^{\mu}_b(x_2) u_c(x_1): 									\p^{\nu}_{x_3}\p^{\nu}_{x_2}\df		\label{berech3.28e}		\\
		&\hphantom{= }+	:A_a^{\mu}(x_3) A^{\mu}_b(x_1) u_c(x_2): 									\p^{\nu}_{x_3}\p^{\nu}_{x_1}\df		\label{berech3.28f}					\end{align}
\end{subequations}
Obviously the sub-sums $\eqref{berech3.28a}+\eqref{berech3.28c}$ , $\eqref{berech3.28b}+\eqref{berech3.28f}$ and $\eqref{berech3.28d}+\eqref{berech3.28e}$ add under the antisymmetric color contraction with $f_{abc}$ up to zero.

\subparagraph{The Similar Pairs \boldmath $\p^{\nu}_{x_1} \p^{\nu}_{x_3}$ and $\p^{\nu}_{x_2}\p^{\nu}_{x_3}$}																						\label{BERECH3.3.1.2} 
 
With the aid of equation $\eqref{berech3.2}$ both cases can be rewritten in the form $\p^{\nu}_{x_1}\p^{\nu}_{x_2}$. As an example, the latter is shown for $\mathit{l}_{0,1}\big|_{\trm{P},\p^{\nu}_{x_1}\p^{\nu}_{x_3}}$ $\eqref{berech3.13}$ in detail
\begin{align}
	\mathit{l}_{0,1}\big|_{\trm{P},\p^{\nu}_{x_1}\p^{\nu}_{x_3}} 
		= : \ldots : \p^{\nu}_{x_1}\p^{\nu}_{x_3} \df
			&= -: \ldots : \p^{\nu}_{x_1} \big( \p^{\nu}_{x_1} +\p^{\nu}_{x_2} 						+\sum_{\trm{inner}}\p^{\nu} \big)	
				&&\big|\text{ with }\eqref{berech3.2}			\notag \\
			&= -: \ldots :
					 \big( \underbrace{\p^{\nu}_{x_1}\p^{\nu}_{x_1}\df}_{0} +
					\p^{\nu}_{x_1}\p^{\nu}_{x_2}\df +
				\underbrace{\p^{\nu}_{x_1}\sum
				\p^{\nu}\df}_{\n} \big)								\notag \\
			& \equiv - : \ldots : \p^{\nu}_{x_1}\p^{\nu}_{x_2}\df \pmod{\n}
														\label{berech3.29}
\end{align}
And again, after summation with $f_{abc}$ the terms cancel in pairs. Thus, in conclusion, the term $\mathit{l}_{0,1}$ equals zero (for any derivative pair) and so the set $\mathcal{L}'$ comprises, at most, the elements $\mathit{l}_{0,2}$ and $\mathit{l}_{0,3}$ only.

\paragraph{The Elements \boldmath$\mathit{l}_{0,2}$}	\label{BERECH3.3.2}

The permutation invariant form of $\mathit{l}_{0,2}$ is written as\footnote{Again, as instated in the remark preceding equation $\eqref{berech3.27a}$, the derivation pair $(\p^{\mu}_{x_l} \p^{\nu}_{x_m})$ represent all elements of the set $\eqref{berech3.25}$}
\begin{subequations}
\begin{align}
	\mathit{l}_{0,2}\big|_{\trm{P}}
	&=\, :A^{\mu}_a(x_1) A^{\nu}_b(x_2) u_c(x_3): 										\p^{\mu}_{\trm{P}(x_l)} \p^{\nu}_{\trm{P}(x_m)}\df	\label{berech3.30a}	\\
	&\hphantom{= }+	:A_a^{\mu}(x_1) A^{\nu}_b(x_3) u_c(x_2): 							\p^{\mu}_{\trm{P}(x_l)} \p^{\nu}_{\trm{P}x_m}\df	\label{berech3.30b} \\
	&\hphantom{= }+ :A_a^{\mu}(x_2) A^{\nu}_b(x_1) u_c(x_3): 							\p^{\mu}_{\trm{P}(x_l)} \p^{\nu}_{\trm{P}(x_m)}\df	\label{berech3.30c}	\\
	&\hphantom{= }+	:A_a^{\mu}(x_2) A^{\nu}_b(x_3) u_c(x_1): 							\p^{\mu}_{\trm{P}(x_l)} \p^{\nu}_{\trm{P}(x_m)}\df	\label{berech3.30d}	\\
	&\hphantom{= }+	:A_a^{\mu}(x_3) A^{\nu}_b(x_2) u_c(x_1): 							\p^{\mu}_{\trm{P}(x_l)} \p^{\nu}_{\trm{P}(x_m)}\df	\label{berech3.30e}	\\
	&\hphantom{= }+	:A_a^{\mu}(x_3) A^{\nu}_b(x_1) u_c(x_2): 							\p^{\mu}_{\trm{P}(x_l)} \p^{\nu}_{\trm{P}(x_m)}\df	\label{berech3.30f}					\end{align}
\end{subequations}
As above, one has to consider the different restrictions on the derivatives .

\subparagraph{The Cases \boldmath $\p^{\mu}_{x_1}\p^{\nu}_{x_2}$, $\p^{\mu}_{x_2}\p^{\nu}_{x_1}$ and $\p^{\mu}_{x_3}\p^{\nu}_{x_3}$}																		\label{BERECH3.3.2.0}

Analogous to the preceeding terms $\mathit{l}_{0,1}\big|_{\trm{P}, \p^{\nu}_{x_i} \p^{\nu}_{x_j}}$, each sum is zero under the appropriate antisymmetric summation.

\subparagraph{The Remaining Cases}				\label{BERECH3.3.2.1}

There are no further terms which are equal zero, thus one can restrict oneself to find existing equivalences among the remaining expressions. Equation $\eqref{berech3.2}$ results in
\small
\\[.5em]
\begin{align}
	\mathit{l}_{0,2}\big|_{\trm{P},\p^{\mu}_{x_3}\p^{\nu}_{x_1}} 
		= : \ldots : \p^{\mu}_{x_3}\p^{\nu}_{x_1} \df
			&= -: \ldots : \p^{\mu}_{x_3} \big( \p^{\nu}_{x_2} +\p^{\nu}_{x_3} 						+\sum_{\trm{inner}}\p^{\nu} \big)	
				&&\big|\text{ with }\eqref{berech3.2}			\notag \\
			&= -: \ldots :
				\big( \p^{\mu}_{x_3}\p^{\nu}_{x_2}\df +
				\underbrace{\p^{\mu}_{x_3}\p^{\nu}_{x_3}\df}_{0} +
				\underbrace{\p^{\mu}_{x_3}\sum\p^{\nu}\df}_{\n} \big) 
				&&\big|\text{ due to case $\p^{\mu}_{x_3}\p^{\nu}_{x_3}$ 																above }	\notag \\
			& \equiv - : \ldots : \p^{\mu}_{x_3}\p^{\nu}_{x_2}\df \pmod{\n}
														\label{berech3.31}
\\[1.5em]
	\mathit{l}_{0,2}\big|_{\trm{P},\p^{\mu}_{x_3}\p^{\nu}_{x_1}} 
		= : \ldots : \p^{\mu}_{x_3}\p^{\nu}_{x_1} \df
			&= -: \ldots : \big( \p^{\mu}_{x_1} +  \p^{\mu}_{x_2}  									+\sum_{\trm{inner}}\p^{\mu} \big) \p^{\nu}_{x_1} 				
				&&\big|\text{ with }\eqref{berech3.2}			\notag \\
			&= -: \ldots :
				\big( \p^{\mu}_{x_1}\p^{\nu}_{x_1}\df +
				\underbrace{\p^{\mu}_{x_2}\p^{\nu}_{x_1}\df}_{0} +
				\underbrace{\sum\p^{\mu}\p^{\nu}_{x_1}\df}_{\n} \big) 
				&&\big|\text{ due to case $\p^{\mu}_{x_2}\p^{\nu}_{x_1}$ 																above }\notag \\
			& \equiv - : \ldots : \p^{\mu}_{x_1}\p^{\nu}_{x_1}\df \pmod{\n}
														\label{berech3.32}
\end{align}
\begin{align}
	\mathit{l}_{0,2}\big|_{\trm{P},\p^{\mu}_{x_2}\p^{\nu}_{x_3}} 
		= : \ldots : \p^{\mu}_{x_2}\p^{\nu}_{x_3} \df
			&= -: \ldots : \p^{\mu}_{x_2} \big( \p^{\nu}_{x_1} +\p^{\nu}_{x_2} 						+\sum_{\trm{inner}}\p^{\nu} \big)	
				&&\big|\text{ with }\eqref{berech3.2}			\notag \\
			&= -: \ldots :
				\big( \underbrace{\p^{\mu}_{x_2}\p^{\nu}_{x_1}\df}_{0} +
				\p^{\mu}_{x_2}\p^{\nu}_{x_2}\df +
				\underbrace{\p^{\mu}_{x_2}\sum\p^{\nu}\df}_{\n} \big) 
				&&\big|\text{ due to case $\p^{\mu}_{x_2}\p^{\nu}_{x_2}$ 																above }\notag \\
			& \equiv - : \ldots : \p^{\mu}_{x_2}\p^{\nu}_{x_2}\df \pmod{\n}
														\label{berech3.33}
\\[1.5em]
	\mathit{l}_{0,2}\big|_{\trm{P},\p^{\mu}_{x_2}\p^{\nu}_{x_3}} 
		= : \ldots : \p^{\mu}_{x_2}\p^{\nu}_{x_3} \df
			&= -: \ldots : \big( \p^{\mu}_{x_1} +  \p^{\mu}_{x_3}  									+\sum_{\trm{inner}}\p^{\mu} \big) \p^{\nu}_{x_3} 				
				&&\big|\text{ with }\eqref{berech3.2}			\notag \\
			&= -: \ldots :
				\big( \p^{\mu}_{x_1}\p^{\nu}_{x_3}\df +
				\underbrace{\p^{\mu}_{x_3}\p^{\nu}_{x_3}\df}_{0} +
				\underbrace{\sum\p^{\mu}\p^{\nu}_{x_3}\df}_{\n} \big) 
				&&\big|\text{ due to case $\p^{\mu}_{x_3}\p^{\nu}_{x_3}$ 																above }\notag \\
			& \equiv - : \ldots : \p^{\mu}_{x_1}\p^{\nu}_{x_3}\df \pmod{\n}
														\label{berech3.34}
\end{align}
\normalsize
Whereas the two remaining terms (the terms on the left hand side of $\eqref{berech3.31}$ -- $\eqref{berech3.34}$)
\begin{equation}	\label{berech3.35}
\begin{aligned}
	\mathit{l}_{0,2}&\big|_{\trm{P},\p^{\mu}_{x_3}\p^{\nu}_{x_1}}
		= \sum_{ \trm{P} \{ x_1,x_2,x_3 \} }:A^{\mu}_a(x_1) A^{\nu}_b(x_2) 							u_c(x_3): \p^{\mu}_{x_3}\p^{\nu}_{x_1}\df					\\
	\mathit{l}_{0,2}&\big|_{\trm{P},\p^{\mu}_{x_2}\p^{\nu}_{x_3}}
		= \sum_{ \trm{P} \{ x_1,x_2,x_3 \} }:A^{\mu}_a(x_1) A^{\nu}_b(x_2) 							u_c(x_3): \p^{\mu}_{x_2}\p^{\nu}_{x_3}\df
\end{aligned}
\end{equation}
represent equivalent permutation invariant sums, if one takes the totally antisymmetric contraction over the color indices into account.
Additionally, the terms $\eqref{berech3.35}$
\begin{align}
	\mathit{l}_{0,2}&\big|_{\trm{P},\p^{\mu}\p^{\nu}}
		= \sum_{ \trm{P} \{ x_1,x_2,x_3 \} }:\p^{\nu}A^{\mu}_a(x_1) 							A^{\nu}_b(x_2) \p^{\mu}u_c(x_3): \df	\pmod{\n}\label{berech3.36}
\intertext{can be written in the form}
	\mathit{l}_{0,2}&\big|_{\trm{P},\p^{\mu}\p^{\nu}}
		= \sum_{ \trm{P} \{ x_1,x_2,x_3 \} }:F^{\nu\mu}_a(x_1) A^{\nu}_b(x_2) 					\p^{\mu}u_c(x_3): \df	\pmod{\n}				\label{berech3.37}
\end{align}
since the local term $:\p^{\mu}A^{\nu}_a A^{\nu}_b \p^{\mu}u_c: \df$ which is added in the transition from $\eqref{berech3.36}$ to $\eqref{berech3.37}$ equals zero according to the conclusive result for the term $\mathit{l}_{0,1}$ in this section.
Finally one has to investigate the elements of $\mathit{l}_{0,3}$.

\paragraph{The Elements \boldmath$\mathit{l}_{0,3}$}	\label{BERECH3.3.3}

The term $\mathit{l}_{0,3}$
\begin{equation}	\label{berech3.38}
	\mathit{l}_{0,3} = :u_a(x_1) \ti{u}_b(x_2)u_c(x_3):\p^{\mu}\p^{\mu}\df
\end{equation}
obviously equals zero due to the wave equation, if both derivatives act on the same argument. Thus only the pairs
\begin{equation}	\label{berech3.39}
	\big\{ \p^{\mu}_{x_1}\p^{\mu}_{x_2}\, ,\, \p^{\mu}_{x_1}\p^{\mu}_{x_3}\, 				,\,\p^{\mu}_{x_2}\p^{\mu}_{x_3} 
	\big\}
\end{equation}
must be discussed in detail. With the help of $\eqref{berech3.2}$ one immediately sees the equivalence
\small
\begin{align}	\label{berech3.40}
	\mathit{l}_{0,3}\big|_{\trm{P},\p^{\mu}_{x_1}\p^{\mu}_{x_2}}
		&= : \ldots : \p^{\mu}_{x_1}\p^{\mu}_{x_2} \df
		&& \big|\text{ since }\eqref{berech3.2}				\notag \\
		&= - : \ldots : \big( 
			\underbrace{\p^{\mu}_{x_1}\p^{\mu}_{x_1}}_{0} + 
			\p^{\mu}_{x_1}\p^{\mu}_{x_3} + 
			\underbrace{\p^{\mu}_{x_1}\sum\p^{\mu}}_{\n}
			\big) \df											\notag \\
		&\equiv - : \ldots : \p^{\mu}_{x_1}\p^{\mu}_{x_3} \df 	\pmod{\n}
\end{align}
\normalsize
of the first and the second derivative--pair. Additionally, $\p^{\mu}_{x_1}\p^{\mu}_{x_2}$ and $\p^{\mu}_{x_2}\p^{\mu}_{x_3}$ lead to the same permutation invariant sum. Conclusively the set $\mathcal{L}'$ only comprises the elements
\begin{align}	\label{berech3.41}
	\mathcal{L}'_{\trm{P}}
		&= \{ \mathit{l}_{0,2}\big|_{\trm{P},\p^{\mu}_{x_3}\p^{\nu}_{x_1}}\,,\,
			  \mathit{l}_{0,3}\big|_{\trm{P},\p^{\mu}_{x_1}\p^{\mu}_{x_2}}
			\} 	 													\notag 	\\
		&=\big\{ \sum_{\trm{P} \{ x_1,x_2,x_3 \} }:F^{\nu\mu}_a(x_1) 								A^{\nu}_b(x_2) \p^{\mu}u_c(x_3): \df  \, , \,
				 	\sum_{\trm{P} \{ x_1,x_2,x_3 \} }\p^{\mu}_{x_1}u_a(x_1) 								\p^{\mu}_{x_2} \ti{u}_b(x_2) u_c(x_3):\df \big\}
\end{align}
Next we investigate the set $\mathcal{M}$.

\subsubsection{Determination of the Elements of \boldmath$\mathcal{M}$}																			\label{BERECH3.4}

As in the preceeding sections, each subset $\mathcal{M}_i$ of $\mathcal{M}$
\begin{equation}	\label{berech3.42}
	\mathcal{M} = \mathcal{M}_0 \cup \mathcal{M}_1
\end{equation}
with the same number of external derivatives will be considered separately\footnote{The following equation immediately shows that $\mathcal{M}$ can be divided into two subsets only.}. According to the inequality $\eqref{algeb8}$, the singular order of the subsets can be estimated by
\begin{align}	\label{berech3.43}
	\omega(2\text{--leg}) \leqslant 4 - 3 - d = 1 - d
\end{align}
Furthermore, all elements of $\mathcal{M}$ must have ghost number $n_u(\mathcal{M}) = 0$ which can only be satisfied by the following two field operator parts
\begin{align}	 \label{berech3.44}
	&A^{\kappa}(x_1)A^{\xi}(x_2)A^{\zeta}(x_3) 											&A^{\kappa}(x_1)u(x_2)\ti{u}(x_3)
\end{align}
The subsets can easily be found by the previous technique.

\paragraph{The Subset \boldmath$\mathcal{M}_0$}		\label{BERECH3.4.0}

This subset features the singular order
\begin{equation}	\label{berech3.45}
	\omega(\mathcal{M}_0) \leqslant 4 - 3 - 0 = 1
\end{equation}
Thus, the general set for $\mathcal{M}_0$ comprises the elements
\begin{equation}	\label{berech3.46}
	\mathcal{M}_0 =
	\Bigl\{ :\Lambda_a(x_1)\Gamma_b(x_2)\Delta_c(x_3): \, \mathcal{D}^{\beta} 			\df \Bigm| \left( \Lambda \Gamma \Delta \right) \in \left\{ 						A^{\kappa}A^{\xi}A^{\zeta} ,A^{\kappa}u \ti{u} \right\} \,, \,|\beta| 				\leqslant 1 \, , \,\, x_1 \leftrightarrow x_2 \leftrightarrow x_3 \Bigr\}
\end{equation}
But terms with $|\beta| = 0$ are not Lorentz scalars, and so $\mathcal{M}_0$ consists of\\
$\mathit{m}_{0,i} \in \mathcal{M}_0$
\begin{align}	\label{berech3.47}	
	\mathit{m}_{0,1}&\overset{def}{=}\,
		:A^{\mu}_a(x_1) A^{\mu}_b(x_2)A^{\nu}_c(x_3):\p^{\nu}\df			&
	\mathit{m}_{0,2}&\overset{def}{=}\,
		:A^{\mu}_a(x_1) u_b(x_2)\ti{u}_c(x_3):\p^{\mu}\df
\end{align}
only.

\paragraph{The Subset \boldmath$\mathcal{M}_1$}		\label{BERECH3.4.1}

The same considerations as above lead to
\begin{equation}	\label{berech3.48}
	\omega(\mathcal{M}_1) \leqslant 4 - 3 - 1 = 0
\end{equation}
and
\begin{equation}	\label{berech3.49}
	\mathcal{M}_1 =
	\Bigl\{ \mathcal{D}^{\alpha}:\Lambda_a(x_1)\Gamma_b(x_2)\Delta_c(x_3): \, 			\df \Bigm| \left( \Lambda \Gamma \Delta \right) \in \left\{ 						A^{\kappa}A^{\xi}A^{\zeta} ,A^{\kappa}u \ti{u} \right\} \,, \,|\alpha| = 1 			\, , \,\, x_1 \leftrightarrow x_2 \leftrightarrow x_3 \Bigr\}
\end{equation}
which result in the list of elements\\
$\mathit{m}_{1,i} \in \mathcal{M}_1$
\begin{equation}	\label{berech3.50}
\begin{aligned}
	\mathit{m}_{1,1}&\overset{def}{=}\,
		:\p^{\mu}A^{\mu}_a(x_1) A^{\nu}_b(x_2) A^{\nu}_c(x_3): \df			&
	\qquad\qquad
	\mathit{m}_{1,2}&\overset{def}{=}\,
		:\p^{\mu}A^{\nu}_a(x_1) A^{\mu}_b(x_2) A^{\nu}_c(x_3): \df			\\
	\mathit{m}_{1,3}&\overset{def}{=}\,
		:\p^{\mu}A^{\mu}_a(x_1) u_b(x_2) \ti{u}_c(x_3):\df	 				&
	\mathit{m}_{1,4}&\overset{def}{=}\,
		:A^{\mu}_a(x_1) \p^{\mu}u_b(x_2) \ti{u}_c(x_3):\df					\\
	\mathit{m}_{1,5}&\overset{def}{=}\,
		:A^{\mu}_a(x_1) u_b(x_2) \p^{\mu}\ti{u}_c(x_3):\df
\end{aligned}
\end{equation}
for the set $\mathcal{M}_1$.

\subsubsection{Determination of Equivalent Elements in \boldmath$\mathcal{M}$} 																\label{BERECH3.5}

Obviously the set $\mathcal{M}_1$ can be equivalently represented by the elements of $\mathcal{M}_0$. This can be seen by the transformations
\begin{equation}	\label{berech3.51}
\begin{aligned}
	\mathit{m}_{0,1}\bigr|_{\p^{\nu}_{x_1}}	
		&= :A^{\mu}_a(x_1) A^{\mu}_b(x_2)A^{\nu}_c(x_3):\p^{\nu}_{x_1}\df 	\\
		&= \n - 
			:\p^{\nu}_{x_1}A^{\mu}_a(x_1) A^{\mu}_b(x_2)A^{\nu}_c(x_3):\df	\\
		& \equiv - \mathit{m}_{1,2} \pmod{\n}
\\[.6cm]
	\mathit{m}_{0,1}\bigr|_{\p^{\nu}_{x_2}} 	
		&= :A^{\mu}_a(x_1) A^{\mu}_b(x_2)A^{\nu}_c(x_3):\p^{\nu}_{x_2}\df 
		\qquad\qquad														& 
	\mathit{m}_{0,1}\bigr|_{\p^{\nu}_{x_3}}
	 	&= :A^{\mu}_a(x_1) A^{\mu}_b(x_2)A^{\nu}_c(x_3):\p^{\nu}_{x_3}\df 	\\
		&= \n - 
			:A^{\mu}_a(x_1)\p^{\nu}_{x_2} A^{\mu}_b(x_2)A^{\nu}_c(x_3):\df	&
		&= \n - 
			:A^{\mu}_a(x_1) A^{\mu}_b(x_2)\p^{\nu}_{x_3}A^{\nu}_c(x_3):\df	\\
		& \equiv - \mathit{m}_{1,2} \pmod{\n}								&
		& \equiv - \mathit{m}_{1,1} \pmod{\n} 
\end{aligned}
\end{equation}
\\
\begin{equation}	\label{berech3.52}
\begin{aligned}
	\mathit{m}_{0,2}\bigr|_{\p^{\mu}_{x_1}} 
		&= :A^{\mu}_a(x_1) u_b(x_2)\ti{u}_c(x_3):\p^{\mu}_{x_1}\df 			\\
		&= \n - :\p^{\mu}_{x_1}A^{\mu}_a(x_1) u_b(x_2)\ti{u}_c(x_3):\df		\\
		& \equiv - \mathit{m}_{1,3} \pmod{\n}
\\[.6cm]
	\mathit{m}_{1,2}\bigr|_{\p^{\mu}_{x_2}}
	 	&= :A^{\mu}_a(x_1) u_b(x_2)\ti{u}_c(x_3):\p^{\mu}_{x_2}\df 						\qquad\qquad															&
	\mathit{m}_{1,2}\bigr|_{\p^{\mu}_{x_3}}
	 	&= :A^{\mu}_a(x_1) u_b(x_2)\ti{u}_c(x_3):\p^{\mu}_{x_3}\df 			\\
		&= \n - :A^{\mu}_a(x_1) \p^{\mu}_{x_2}u_b(x_2)\ti{u}_c(x_3):\df		&
		&= \n - :A^{\mu}_a(x_1) u_b(x_2)\p^{\mu}_{x_3}\ti{u}_c(x_3):\df		\\
		& \equiv - \mathit{m}_{1,4} \pmod{\n}								&
		& \equiv - \mathit{m}_{1,5} \pmod{\n} 
\end{aligned}
\end{equation}
The shorthand notation for $\mathcal{M}'$ thus reads as
\begin{align}	\label{berech3.53}
	\mathcal{M}' 
	:= \,\raisebox{1ex}{$\mathcal{M}$}\raisebox{0ex}
				{$\! \Big/\!$}\raisebox{-1ex}{$\n$}
	= \big\{ \mathit{m}_{0,1}, \mathit{m}_{0,2} \big\}
\end{align}

\subsubsection{The Subgroups \boldmath$\textbf{B}(\mathcal{M}')$, $\textbf{Z} (\mathcal{M}')$ and $\textbf{H}(\mathcal{M}')$}			\label{BERECH3.6}

As in the $2$--leg case, we start with the calculations for the subgroup $\textbf{B}(\mathcal{M}')$.

\subparagraph{The Subgroup \boldmath$\textbf{B}(\mathcal{M}')$}																\label{BERECH3.6.1}

None of the elements
\begin{equation}	\label{berech3.54}
\begin{aligned}
	\mathcal{M}' 
		&= \big\{ \mathit{m}_{0,1}, \mathit{m}_{0,2} \big\}					\\
		&= \big\{:A^{\mu}_a(x_1) A^{\mu}_b(x_2)A^{\nu}_c(x_3):\p^{\nu}\df \,,\,
			:A^{\mu}_a(x_1) u_b(x_2)\ti{u}_c(x_3):\p^{\mu}\df
		   \big\}
\end{aligned}
\end{equation}
can be written, with the aid of $\eqref{eich19a}-\eqref{eich19f}$, as the image of a term $f$ according to
\begin{align}	\label{berech3.55}
	d_Q(f) 	&  \equiv \mathit{m}_{0,j} \pmod{\n}
			&& f \in \mathcal{L}\bigr|_{n_{u}(f)=-1} ,\; j \in \{1,2\}
\end{align}
Thus the set $\textbf{B}$
\begin{equation} 	\label{berech3.56}
	\textbf{B}(\mathcal{M}') = \emptyset
\end{equation}
is empty.

\subparagraph{The Subgroup \boldmath$\textbf{Z}(\mathcal{M}')$}																		\label{BERECH3.6.2}

Let us treat the different restrictions $\p^{\nu}_{x_1}$, $\p^{\nu}_{x_2}$ separately. According to equation $\eqref{berech3.2}$, the third restriction $\p^{\nu}_{x_3}$ can be disregarded. This is true, since $\p^{\nu}_{x_3}$ is equivalent to the linear combination $\p^{\nu}_{x_1} + \p^{\nu}_{x_2}$ up to a divergence.
Similar to the remark given for $\mathit{l}_{0,2}$ on page $\pageref{berech3.37}$, the element
\begin{equation}	\label{berech3.57}
\begin{aligned}
	\mathit{m}_{0,1}\big|_{\p^{\nu}_{x_1}}
		&= :A^{\mu}_a(x_1) A^{\mu}_b(x_2)A^{\nu}_c(x_3):
											\p^{\nu}_{x_1}\df		 	\\
		&= :\p^{\nu}_{x_1}A^{\mu}_a(x_1) A^{\mu}_b(x_2)A^{\nu}_c(x_3):\df
																\pmod{\n}
\end{aligned}
\end{equation}
can only appear in the form
\begin{equation}	\label{berech3.58}
	\mathit{m}_{0,1}\big|_{\trm{P},\p^{\nu}_{x_1}} = :F^{\nu\mu}_a(x_1) 						A^{\mu}_b(x_2) A^{\nu}_c(x_3):\df
\end{equation}
which leads under the gauge transformation $d_Q$ to
\begin{align}	\label{berech3.59}	
	d_Q\big( f_{abc} \mathit{m}_{0,1}\big|_{\trm{P},\p^{\nu}_{x_1}}\big)
		&= f_{abc} \sum_{\trm{P} \{ x_1,x_2,x_3 \} }d_Q :F^{\nu\mu}_a(x_1) 						A^{\mu}_b(x_2) A^{\nu}_c(x_3):\df					\notag 	\\
		&= 0 + 2 f_{abc} \sum_{\trm{P} \{ x_1,x_2,x_3 \} }:F^{\nu\mu}_a(x_1) 						A^{\mu}_b(x_2) \p^{\nu}_{x_3}u_c(x_3):\df
\end{align}
The term $\mathit{m}_{0,1}\big|_{\trm{P},\p^{\nu}_{x_2}}$ obviously leads to the same expression $\eqref{berech3.59}$. Thus one can focus on term $\mathit{m}_{0,2}$.
The restriction $\p^{\mu}_{x_2}$ leads to
\begin{align}	\label{berech3.60}
	&d_Q\big(\mathit{m}_{0,2}\big|_{\trm{P},\p^{\mu}_{x_2}}\big)
		= 	\sum_{\trm{P} \{ x_1,x_2,x_3 \} }:\p^{\mu}u_a(x_1) 										\p^{\mu}u_b(x_2)\ti{u}_c(x_3):\df + \notag \\
	&\phantom{d_Q\big(\mathit{m}_{0,2}\big|_{\trm{P},\p^{\mu}_{x_2}}\big)
		= 	\sum_{\trm{P} \{ x_1,x_2,x_3 \} }:\p^{\mu}u_a(x_1)}	
			\sum_{\trm{P} \{ x_1,x_2,x_3 \}}:A^{\mu}_a(x_1)\p^{\mu}u_b(x_2) 						\p^{\nu}A^{\nu}_a(x_3):\df \pmod{\n}
\end{align}
wherein the second term on the right hand side equals $\mathit{l}_{0,2}$. This can be easily found with the aid of equation $\eqref{berech3.2}$.
In summary it is demonstrated, that the subgroup $\textbf{Z}(\mathcal{M}')$ at least comprises the two elements
\begin{equation}	\label{berech3.61}
	\textbf{Z}(\mathcal{M}')=
		\big\{\mathit{m}_{0,1}\big|_{\trm{P},\p^{\nu}_{x_1}}\,,\, 
			\mathit{m}_{0,2}\big|_{\trm{P},\p^{\mu}_{x_2}} - \frac{1}{2}										\mathit{m}_{0,1}\big|_{\trm{P},\p^{\nu}_{x_1}}
		\big\}
\end{equation}
This is true, since in the above line the second expression has the $d_Q$--transformed
\begin{equation}	\label{berech3.62}
\begin{aligned}	
	d_Q\big( \mathit{m}_{0,2} - \frac{1}{2}\mathit{m}_{0,1} \big)
			&= :\p^{\mu}u_a(x_1)\p^{\mu}u_b(x_2)\ti{u}_c(x_3):\df
\end{aligned}
\end{equation}
as $\eqref{berech3.60}$ shows.

\subparagraph{The Gauge--Factor Subgroup \boldmath$\textbf{H}(\mathcal{M}')$}															\label{BERECH3.6.3}

The results above show, that the factor group at least contains the elements
\begin{align}	\label{berech3.63}
	\textbf{H}(\mathcal{M}')
		= \, \raisebox{1ex}{$\textbf{Z}(\mathcal{M})$} \raisebox{0ex}{$\! 					\Big/\!$} 	\raisebox{-1ex}{$\textbf{B}(\mathcal{M})$}
		&= \big\{\mathit{m}_{0,1}\,,\, \mathit{m}_{0,2} - \mathit{m}_{0,1}
			\big\}													
\end{align}
Comparing the set $\mathcal{L}'$ with the elements of $d_Q \big(\textbf{H}(\mathcal{M}')\big)$ shows, that
\begin{equation}	\label{berech3.64}
	d_Q \big(\textbf{H}(\mathcal{M}')\big) \supseteq \mathcal{L}'
\end{equation}
and thus all elements of $\mathcal{L}'$ are gauge invariant writable.
This statement finishes the calculations for the $3$--leg case with disjunct arguments and one can consider in the next part the reducible $3$--leg problem.
\subsection{\boldmath $3$--Leg Calculations for Terms Based on Reducible Graphs}																\label{BERECH3a.1}
Trivially, the elements $\mathit{l}_{0,1}\,,\,\mathit{l}_{0,2}$ and $\mathit{l}_{0,3}$ in $\eqref{berech3.24}$ represent in the sum over all permuted arguments and under the transition to two identical arguments all elements of $\mathcal{L}(\text{$3$--leg})\big|_{\trm{P},x_1=x_2}$. This is best seen by the following construction of reducible terms out of the original ones with disjunct arguments:\\
In the permutation invariant expressions $\eqref{berech3.27a}$, $\eqref{berech3.30a}$ and $\eqref{berech3.38}$ the derivatives $\p^{\kappa}\p^{\xi}$ represent all possible derivation--pairs
\begin{equation}		\label{berech3a.1}
	\big\{ \p^{\kappa}_{x_1}\p^{\xi}_{x_1} \,,\, \p^{\kappa}_{x_1}\p^{\xi}_{x_2} 	\,, \, \ldots \,, \, \p^{\kappa}_{x_n}\p^{\xi}_{x_n} \big\}
\end{equation}
on the $\df$'s, which must be investigated separately. Once, one has been decided for one pair of derivatives, in a first step the derivatives must be shifted with the help of Leibnitz' rule. Due to the disjunct arguments, every such derivative pair leads up to a divergence to one expression with derivatives acting on the field operators only. Now --- with all derivatives acting on the field operators only --- all dispersions of derivatives onto field operators occur --- the transition to two identical arguments will be executed. Obviously this action leads again to permutation invariant sums for each term. This shows, that according to the above construction, all elements of $\mathcal{L} (\text{$3$--leg}) \big|_{\trm{P},x_1=x_2}$ are represented by the three terms $\mathit{l}_{0,1}\,, \,\mathit{l}_{0,2}$ and $\mathit{l}_{0,3}$ under the restriction $x_1=x_2$.

Furthermore, after the remarks concerning the proofs for reducible graphs given in the introduction to the calculations, only equivalences which are based on identity $\eqref{berech3.2}$ or the Leibnitz' rule must be recalculated after transition to identical arguments. All identities following $\eqref{berech3.25}$ will be investigated separately under this point of view. Immediately one is lead to the insight, that the statements for $\p^{\nu}_{x_i}\p^{\nu}_{x_i}$ and $\p^{\nu}_{x_1}\p^{\nu}_{x_2}$ on page $\pageref{BERECH3.3.1.1}$ hold true under the restriction to two identical arguments\footnote{Since no use of equation $\eqref{berech3.2}$ nor of the Leibnitz' rule was made.}. Also the equivalence of $\mathit{l}_{0,1}\big|_{\trm{P},\p^{\nu}_{x_1} \p^{\nu}_{x_3}}$ and $\mathit{l}_{0,1}\big|_{\trm{P},\p^{\nu}_{x_2}\p^{\nu}_{x_3}}$ stated after $\eqref{berech3.28f}$ remain correct. Since in both permutation invariant sums occur up to interchanged color states $a \longleftrightarrow b$ the same terms. Moreover one realizes immediately, that the terms equal zero respectively. This is best seen, if one rewrites $\mathit{l}_{0,1}\big|_{\trm{P},\p^{\nu}_{x_1} \p^{\nu}_{x_3},x_1=x_2}$ in full length and then makes use of $\eqref{berech3.2}$ and the wave equations.
One requires: 
\begin{align}
	\mathit{l}_{0,1}\big|_{\trm{P},\p^{\nu}_{x_1}\p^{\nu}_{x_3},x_1=x_2}
				&=\, 	f_{abc}:\p^{\nu}A_a^{\mu}(x_1) A^{\mu}_b(x_2)\p^{\nu} 										u_c(x_3): \df \big|_{x_1=x_2} 	\notag		\\
	&\hphantom{= }+		f_{abc}:\p^{\nu}A_a^{\mu}(x_1) A^{\mu}_b(x_3)\p^{\nu} 										u_c(x_2): \df \big|_{x_1=x_2}	\notag		\\
	&\hphantom{= }+ 	f_{abc}:\p^{\nu}A_a^{\mu}(x_2) A^{\mu}_b(x_1) \p^{\nu} 										u_c(x_3):  \df \big|_{x_1=x_2}	\notag		\\
	&\hphantom{= }+		f_{abc}:\p^{\nu}A_a^{\mu}(x_2) A^{\mu}_b(x_3) \p^{\nu} 										u_c(x_1): \df \big|_{x_1=x_2}	\notag		\\
	&\hphantom{= }+		f_{abc}:\p^{\nu}A_a^{\mu}(x_3) A^{\mu}_b(x_2) \p^{\nu} 										u_c(x_1): \df \big|_{x_1=x_2}	\notag		\\
	&\hphantom{= }+		f_{abc}:\p^{\nu}A_a^{\mu}(x_3) A^{\mu}_b(x_1) \p^{\nu} 										u_c(x_2):\df \big|_{x_1=x_2}	\notag		\\
	\notag	\\[.1ex]
	&\begin{aligned}		\label{berech3a.2}
	&= 2\, f_{abc}:\p^{\nu}A_a^{\mu}(x_1) A^{\mu}_b(x_1)\p^{\nu} u_c(x_3):\df \\
	&\hphantom{= }+
	   2\, f_{abc}:\p^{\nu}A_a^{\mu}(x_1) A^{\mu}_b(x_3)\p^{\nu} u_c(x_1):\df \\
	&\hphantom{= }+
	   2\, f_{abc}:\p^{\nu}A_a^{\mu}(x_3) A^{\mu}_b(x_1)\p^{\nu} u_c(x_1):\df \\
	\end{aligned}
\end{align}
whereas the first term in $\eqref{berech3a.2}$ above leads to (without the leading factor $2$)
\begin{align}
	f_{abc}:\p^{\nu}A_a^{\mu}(x_1) A^{\mu}_b(x_1)\p^{\nu} u_c(x_3):\df
	&= \n_{x_3} 														\notag
	- f_{abc}:\p^{\nu}A_a^{\mu}(x_1) A^{\mu}_b(x_1)u_c(x_3):\p^{\nu}_{x_3} \df
	\\[2ex]
	&= \n_{x_3} + \n_{x_1}												\notag
	- f_{abc}:\underbrace{\Box A_a^{\mu}(x_1)}_{=0} A^{\mu}_b(x_1)u_c(x_3):\df\\
	&\hphantom{= \n_{x_3} + \n_{x_1}- f_{abc}}							\notag
	- \underbrace{f_{abc}:\p^{\nu}A_a^{\mu}(x_1) \p^{\nu}A^{\mu}_b(x_1) 								u_c(x_3):}_{=0 \text{ (antisymmetry of $f_{abc}$)}} \df\\
	&= 0 \pmod{\n}											\label{berech3a.3}
\end{align}
the second term gives
\begin{align}
	f_{abc}:\p^{\nu}A_a^{\mu}(x_1) A^{\mu}_b(x_3)\p^{\nu} u_c(x_1):\df
	&= \n_{x_1} 														\notag
	- f_{abc}:\underbrace{\Box A_a^{\mu}(x_1)}_{=0} A^{\mu}_b(x_3)u_c(x_1):\df\\
	&\hphantom{=\n_{x_1} - f_{abc}:\p}									\notag
	- f_{abc}:\p^{\nu} A_a^{\mu}(x_1) A^{\mu}_b(x_3)u_c(x_1):\p^{\nu}_{x_1}\df
	\\[2ex]
	&= \n_{x_1} + \n_{x_3}									\label{berech3a.4}
	- f_{abc}:\p^{\nu}A_a^{\mu}(x_1) \p^{\nu}A^{\mu}_b(x_3)u_c(x_1): \df			\end{align}
and the third term can be written as
\begin{align}
	f_{abc}:\p^{\nu}A_a^{\mu}(x_3) A^{\mu}_b(x_1)\p^{\nu} u_c(x_1):\df
	&= \n_{x_1} 														\notag
	- f_{abc}:\p^{\nu}A_a^{\mu}(x_3) \p^{\nu}A^{\mu}_b(x_1) u_c(x_1):\df	\\
	&\hphantom{= \n_{x_1} - f_{abc}:\p}									\notag
	- f_{abc}:\p^{\nu}A_a^{\mu}(x_3) A^{\mu}_b(x_1) u_c(x_1):\p^{\nu}_{x_1}\df
	\\[2ex]
	&= \n_{x_1}															\notag
	- f_{abc}:\p^{\nu}A_a^{\mu}(x_3) \p^{\nu}A^{\mu}_b(x_1) u_c(x_1):\df	\\
	&\hphantom{= \n_{x_1} - f_{abc}:\p}						\label{berech3a.5}
	+ f_{abc}:\underbrace{\Box A_a^{\mu}(x_3)}_{=0} A^{\mu}_b(x_1) u_c(x_1):\df
\end{align}
Thus, the transformed second and third terms also add up to zero under the permutation of $a \longleftrightarrow b$ and the antisymmetry of $f_{abc}$. In conclusion, this states the fact, that $\mathit{l}_{0,1}\big|_{\trm{P},x_1=x_2}$ equals zero for all derivation pairs.

Next one considers the term $\mathit{l}_{0,2}$ in detail. Again one establishes, that for the derivative restrictions $\p^{\nu}_{x_1}\p^{\mu}_{x_2} \,,\, \p^{\mu}_{x_1}\p^{\nu}_{x_2}$ and $\p^{\mu}_{x_3}\p^{\nu}_{x_3}$ the permutation invariant sum figures up to zero absolutely analogous to $\eqref{berech3.28a}$. So there remains only to show, that the equivalences $\eqref{berech3.31}$ -- $\eqref{berech3.34}$ hold correct. At first one shows correctness of  $\eqref{berech3.31}$
\begin{equation}		\label{berech3a.6}
	\mathit{l}_{0,2}\big|_{\trm{P},\p^{\mu}_{x_3} \p^{\nu}_{x_1},x_1=x_2}
		\equiv
	\mathit{l}_{0,2}\big|_{\trm{P},\p^{\mu}_{x_3} \p^{\nu}_{x_2},x_1=x_2}
		\pmod{\n}
\end{equation}
for restricted arguments, which is demonstrated, if the following sum equals zero:
\begin{equation}		\label{berech3a.7}
\begin{aligned}
	\frac{1}{2} \big( \mathit{l}_{0,2}\big|_{\trm{P},\p^{\mu}_{x_3} 						\p^{\nu}_{x_1},x_1=x_2} + \mathit{l}_{0,2}\big|_{\trm{P},\p^{\mu}_{x_3} 				\p^{\nu}_{x_2},x_1=x_2} \big)
	&= 	f_{abc}:\p^{\nu}A_a^{\mu}(x_1) A^{\nu}_b(x_1) \p^{\mu}u_c(x_3):\df	\\
	&\hphantom{= }+
		f_{abc}:\p^{\nu}A_a^{\mu}(x_1) A^{\nu}_b(x_3) \p^{\mu}u_c(x_1):\df	\\
	&\hphantom{= }+
		f_{abc}:\p^{\nu}A_a^{\mu}(x_3) A^{\nu}_b(x_1) \p^{\mu}u_c(x_1):\df	
	\\[2ex]
	&\hphantom{= +f_{abc}:\p^{\nu} }+
		f_{abc}:A_a^{\mu}(x_1) \p^{\nu}A^{\nu}_b(x_1) \p^{\mu}u_c(x_3):\df	\\
	&\hphantom{= +f_{abc}:\p^{\nu} }+
		f_{abc}:A_a^{\mu}(x_1) \p^{\nu}A^{\nu}_b(x_3) \p^{\mu}u_c(x_1):\df	\\
	&\hphantom{= +f_{abc}:\p^{\nu} }+
		f_{abc}:A_a^{\mu}(x_3) \p^{\nu}A^{\nu}_b(x_1) \p^{\mu}u_c(x_1):\df	\\
\end{aligned}
\end{equation}
With help of the Leibnitz' rule and the wave equations the first term on the right hand side of $\eqref{berech3a.7}$ gives
\begin{align}
	f_{abc}:\p^{\nu}A_a^{\mu}(x_1) A^{\nu}_b(x_1) \p^{\mu}u_c(x_3):\df
	&=  \n_{x_1}												\notag
	- f_{abc}:A_a^{\mu}(x_1) \p^{\nu}A^{\nu}_b(x_1) \p^{\mu} u_c(x_3):\df	\\
	&\hphantom{=\n_{x_1}- f_{abc}:\p}							\notag
	- f_{abc}:A_a^{\mu}(x_1) A^{\nu}_b(x_1) \p^{\mu}u_c(x_3):\p^{\nu}_{x_1}\df
	\\[2ex]
	&= \n_{x_1} + \n_{x_3}										\notag
	- f_{abc}:A_a^{\mu}(x_1) \p^{\nu}A^{\nu}_b(x_1) \p^{\mu} u_c(x_3):\df	\\
	&\hphantom{=\n_{x_1} + \n_{x_3}- f_{abc}:\p}			\label{berech3a.8}
	- \underbrace{f_{abc}:A_a^{\mu}(x_1)A^{\nu}_b(x_1)\p^{\nu}\p^{\mu} 							u_c(x_3):}_{=0 \text{ (antisymmetry of $f_{abc}$)}} \df
\end{align}
which cancels against the fourth term. The second term leads to
\begin{align}
	f_{abc}:\p^{\nu}A_a^{\mu}(x_1) A^{\nu}_b(x_3) \p^{\mu}u_c(x_1):\df
	&=  \n_{x_1}												\notag
	- \underbrace{f_{abc}:A_a^{\mu}(x_1) A^{\nu}_b(x_3) \p^{\nu}\p^{\mu} 							u_c(x_1):}_{=0 \text{ (antisymmetry of $f_{abc}$)}}\df		\\
	&\hphantom{ \n_{x_1}f_{abc}:\p}								\notag
	- f_{abc}:A_a^{\mu}(x_1) A^{\nu}_b(x_3) \p^{\mu}u_c(x_1):\p^{\nu}_{x_1}\df
	\\[2ex]
	&= \n_{x_1} + \n_{x_3}									\label{berech3a.9}
	- f_{abc}:A_a^{\mu}(x_1) \p^{\nu}A^{\nu}_b(x_3) \p^{\mu}u_c(x_1):\df
\end{align}
which cancels against the fifth term. The third term gives
\begin{align}
	f_{abc}:\p^{\nu}A_a^{\mu}(x_3) A^{\nu}_b(x_1) \p^{\mu}u_c(x_1):\df
	&=  \n_{x_3}												\notag
	- f_{abc}:A_a^{\mu}(x_3) A^{\nu}_b(x_1)\p^{\mu}u_c(x_1): \p^{\nu}_{x_3}\df
	\\[2ex]
	&=	\n_{x_3} + \n_{x_1}										\notag
	- f_{abc}:A_a^{\mu}(x_3) \p^{\nu}A^{\nu}_b(x_1) \p^{\mu}u_c(x_1):\df	\\			&\hphantom{\n_{x_3} + \n_{x_1}	- f_{abc}:\p}			\label{berech3a.10}
	- \underbrace{f_{abc}:A_a^{\mu}(x_3) A^{\nu}_b(x_1) 									\p^{\nu}\p^{\mu}u_c(x_1):}_{=0 \text{ (antisymmetry of $f_{abc}$)}}\df
\end{align}
which cancels against the sixth term.
Altogether, the established results deliver the identity
\begin{equation}		\label{berech3a.11} 											\mathit{l}_{0,2}\big|_{\trm{P},\p^{\mu}_{x_3} \p^{\nu}_{x_1},x_1=x_2} +
	\mathit{l}_{0,2}\big|_{\trm{P},\p^{\mu}_{x_3} \p^{\nu}_{x_2},x_1=x_2} = 
	0 \pmod{\n}
\end{equation}
and so $\eqref{berech3.31}$ also holds for reducible local terms. With this result there remains only to show the validity of equivalence $\eqref{berech3.32}$ for local terms with identical arguments, since $\mathit{l}_{0,2}\big|_{\trm{P},\p^{\mu}_{x_3} \p^{\nu}_{x_1}}$ and $\mathit{l}_{0,2}\big|_{\trm{P},\p^{\mu}_{x_2} \p^{\nu}_{x_3}}$ on the left hand side of the equations $\eqref{berech3.31}$ -- $\eqref{berech3.34}$ are obviously equivalent under the permutations $a \longleftrightarrow b$ , $\mu \longleftrightarrow \nu$ and the restriction $x_1=x_2$.
To establish the equivalence $\eqref{berech3.32}$ under the restriction $x_1=x_2$, one demonstrates again, that the sum of the two terms in question equals zero:
\begin{equation}		\label{berech3a.12}
\begin{aligned}
	\frac{1}{2} \big( \mathit{l}_{0,2}\big|_{\trm{P},\p^{\mu}_{x_3} 						\p^{\nu}_{x_1},x_1=x_2} + \mathit{l}_{0,2}\big|_{\trm{P},\p^{\mu}_{x_1} 				\p^{\nu}_{x_1},x_1=x_2} \big)
	&= 	f_{abc}:\p^{\nu}A_a^{\mu}(x_1) A^{\nu}_b(x_1) \p^{\mu}u_c(x_3):\df	\\
	&\hphantom{= }+
		f_{abc}:\p^{\nu}A_a^{\mu}(x_1) A^{\nu}_b(x_3) \p^{\mu}u_c(x_1):\df	\\
	&\hphantom{= }+
		f_{abc}:\p^{\nu}A_a^{\mu}(x_3) A^{\nu}_b(x_1) \p^{\mu}u_c(x_1):\df	
	\\[2ex]
	&\hphantom{= +f_{abc}:\p^{\nu} }+
		f_{abc}:\p^{\mu}\p^{\nu}A_a^{\mu}(x_1) A^{\nu}_b(x_1) u_c(x_3):\df	\\
	&\hphantom{= +f_{abc}:\p^{\nu} }+
		f_{abc}:\p^{\mu}\p^{\nu}A_a^{\mu}(x_1) A^{\nu}_b(x_3) u_c(x_1):\df	\\
	&\hphantom{= +f_{abc}:\p^{\nu} }+
		f_{abc}:\p^{\mu}\p^{\nu}A_a^{\mu}(x_3) A^{\nu}_b(x_1) u_c(x_1):\df	\\
\end{aligned}
\end{equation}
Rewriting the fourth, fifth and sixth term leads again to the insight, that the sum equals zero. Explicitely the fourth term in $\eqref{berech3a.12}$ gives
\begin{align}
	f_{abc}:\p^{\mu}\p^{\nu}A_a^{\mu}(x_1) A^{\nu}_b(x_1) u_c(x_3):\df
	&=  \n_{x_1}												\notag
	- \underbrace{f_{abc}:\p^{\nu}A_a^{\mu}(x_1) \p^{\mu}A^{\nu}_b(x_1) 						u_c(x_3):}_{=0 \text{ (antisymmetry of $f_{abc}$)}}\df			\\
	&\hphantom{\n_{x_1}f_{abc}:\p}								\notag
	- f_{abc}:\p^{\nu}A_a^{\mu}(x_1) A^{\nu}_b(x_1) u_c(x_3):\p^{\mu}_{x_1}\df
	\\[2ex]
	&= \n_{x_1} + \n_{x_3}									\label{berech3a.13}
	- f_{abc}:\p^{\nu}A_a^{\mu}(x_1)A^{\nu}_b(x_1)\p^{\mu}u_c(x_3): \df
\end{align}
which cancels against the first term. The fifth term leads to
\begin{align}
	f_{abc}:\p^{\mu}\p^{\nu}A_a^{\mu}(x_1) A^{\nu}_b(x_3) u_c(x_1):\df
	&=  \n_{x_1}												\notag
	- f_{abc}:\p^{\nu}A_a^{\mu}(x_1) A^{\nu}_b(x_3)\p^{\mu} u_c(x_1):\df	\\
	&\hphantom{\n_{x_1}- f_{abc}:\p}							\notag
	- f_{abc}:\p^{\nu}A_a^{\mu}(x_1) A^{\nu}_b(x_3) u_c(x_1):\p^{\mu}_{x_1}\df
	\\[2ex]
	&= \n_{x_1} + \n_{x_3}										\notag
	- f_{abc}:\p^{\nu}A_a^{\mu}(x_1) A^{\nu}_b(x_3)\p^{\mu} u_c(x_1):\df	\\
	&\hphantom{\n_{x_1} + \n_{x_3}- f_{abc}:\p}				\label{berech3a.14}
	- f_{abc}:\p^{\nu}A_a^{\mu}(x_1) \p^{\mu}A^{\nu}_b(x_3)u_c(x_1):\df
\end{align}
whereas the first term on the right hand side of $\eqref{berech3a.14}$ cancels against the second term in $\eqref{berech3a.12}$. The sixth term gives
\begin{align}
	f_{abc}:\p^{\mu}\p^{\nu}A_a^{\mu}(x_3) A^{\nu}_b(x_1) u_c(x_1):\df
	&=  \n_{x_3}												\notag
	- f_{abc}:\p^{\nu}A_a^{\mu}(x_3) A^{\nu}_b(x_1)u_c(x_1): \p^{\mu}_{x_3}\df
	\\[2ex]
	&=	\n_{x_3} + \n_{x_1}										\notag
	- f_{abc}:\p^{\nu}A_a^{\mu}(x_3) \p^{\mu}A^{\nu}_b(x_1) u_c(x_1):\df	\\
	&\hphantom{\n_{x_3} + \n_{x_1}	- f_{abc}:\p}			\label{berech3a.15}
	- f_{abc}:\p^{\nu}A_a^{\mu}(x_3) A^{\nu}_b(x_1)\p^{\mu}u_c(x_1):\df
\end{align}
whereas the second term on the right hand side of $\eqref{berech3a.15}$ cancels against the third term in $\eqref{berech3a.12}$ and the remaining terms in $\eqref{berech3a.14}$ and $\eqref{berech3a.15}$ figure up to zero separately. Thus the whole sum equals zero as expected.
All results above lead to the conclusive result, that the reducible local terms $\mathit{l}_{0,1}\big|_{\trm{P},x_1=x_2}$ and $\mathit{l}_{0,2}\big|_{\trm{P},x_1=x_2}$ fulfill the same equivalences as for disjunct arguments and so the term $\eqref{berech3.37}$ under the restriction $x_1=x_2$ of the arguments represents the whole set $\big\{ \mathit{l}_{0,1}\big|_{\trm{P},x_1=x_2}\,,\, \mathit{l}_{0,2}\big|_{\trm{P},x_1=x_2} \big\}$ again.
Finally it remains to show, whether the equivalence $\eqref{berech3.40}$ also holds for reducible local terms $\mathit{l}_{0,3}\big|_{\trm{P},x_1=x_2}$.
Easily one finds equivalent expressions for $\mathit{l}_{0,3}\big|_{\trm{P},\p^{\mu}_{x_1} \p^{\nu}_{x_2}, x_1=x_2}$ as follows:
\begin{align}
	\mathit{l}_{0,3}\big|_{\trm{P},\p^{\mu}_{x_1} \p^{\nu}_{x_2}, x_1=x_2}
	&= 	:\p^{\mu}u_a(x_1) \p^{\mu}\ti{u}_b(x_2)u_c(x_3):\df \big|_{x_1=x_2}	
																\notag		\\
	&\hphantom{= }+												\notag
		:\p^{\mu}u_a(x_1) \p^{\mu}\ti{u}_b(x_3)u_c(x_2):\df \big|_{x_1=x_2}	\\
	&\hphantom{= }+												\notag
		:\p^{\mu}u_a(x_3) \p^{\mu}\ti{u}_b(x_2)u_c(x_1):\df \big|_{x_1=x_2}	
	\\[2ex]
	&= \n_{x_1}													\notag
	- :\underbrace{\Box u_a(x_1)}_{=0} \ti{u}_b(x_1)u_c(x_3):\df
	- :\p^{\mu} u_a(x_1) \ti{u}_b(x_1)u_c(x_3):\p^{\mu}_{x_1}\df			\\
	&\hphantom{= } + \n_{x_3}									\notag
	- :\p^{\mu}u_a(x_1) \ti{u}_b(x_3)u_c(x_1):\p^{\mu}_{x_3}\df				\\
	&\hphantom{= } + \n_{x_1}									\notag
	- :\p^{\mu}u_a(x_3) \ti{u}_b(x_1)\p^{\mu}u_c(x_1):\df
	- :\p^{\mu}u_a(x_3) \ti{u}_b(x_1)u_c(x_1):\p^{\mu}_{x_1}\df				
	\\[2ex]
	&= \n_{x_1} + \n_{x_3}										\notag
	- :\p^{\mu} u_a(x_1) \ti{u}_b(x_1)\p^{\mu}u_c(x_3):\df					\\
	&\hphantom{= } + \n_{x_3} + \n_{x_1}						\notag
	- :\underbrace{\Box u_a(x_1)}_{=0} \ti{u}_b(x_3)u_c(x_1):\df
	- :\p^{\mu} u_a(x_1) \ti{u}_b(x_3)\p^{\mu}u_c(x_1):\df					\\
	&\hphantom{= } + \n_{x_1} + \n_{x_3}						\notag
	- :\p^{\mu}u_a(x_3) \ti{u}_b(x_1)\p^{\mu}u_c(x_1):\df
	- :\underbrace{\Box u_a(x_3)}_{=0} \ti{u}_b(x_1)u_c(x_1):\df			\\
	&= \mathit{l}_{0,3}\big|_{\trm{P},\p^{\mu}_{x_1} \p^{\nu}_{x_3}, x_1=x_2}
	  \pmod{\n}											\label{berech3a.16}
\end{align}
The above equivalence and the trivial statement (for the permutation invariant sum)
\begin{equation}		\label{berech3a.17}
	\mathit{l}_{0,3}\big|_{\trm{P},\p^{\mu}_{x_1} \p^{\nu}_{x_2}, x_1=x_2}
		\equiv
	\mathit{l}_{0,3}\big|_{\trm{P},\p^{\mu}_{x_2} \p^{\nu}_{x_3}, x_1=x_2}
		\pmod{\n}
\end{equation}
finishes the gauge invariance proof for the reducible local $3$--leg terms, since with the above results the set $\mathcal{L}'(\text{$3$--leg})\big|_{\trm{P},x_1=x_2}$ can be written as $\eqref{berech3.41}$ restricted to two identical arguments
\begin{equation}		\label{berech3a.18}
	\mathcal{L}'\big|_{\trm{P},x_1=x_2} =
		\mathcal{L}'_{\trm{P}}(\ref{berech3.41})\big|_{x_1=x_2}
\end{equation}
and all further steps in the gauge invariance proof remain unchanged under the restriction to two identical arguments. In detail, the transition to two identical arguments obviously does not destroy the remaining equivalences $\eqref{berech3.42}$ -- $\eqref{berech3.64}$ \footnote{Since neither the Leibnitz' rule nor equation $\eqref{berech3.2}$ were used in that connection.}, and so the rest of the $3$--leg proof for disjunct arguments apply without change to the case of reducible $3$--leg terms under the restriction to two equal arguments. 
\subsection{\boldmath $4$--Leg Calculations for Disjunct Arguments}																			\label{BERECH4}

As in the preceeding sections, we start with general remarks introducing the calculations.
\begin{itemize}
\item
	All $4$--leg graphs have the general structure
	\begin{equation}	\label{berech4.1}
		\mathcal{P}_{abcd} \mathcal{D}^{\alpha}:\!\!:\Lambda_a \Gamma_b 					\Delta_c \Pi_d : \mathcal{D}^{\beta}\df
	\end{equation}
	with $\mathcal{P}_{abcd}$ any color tensor of the below introduced list.

\item
	According to \cite{dittner1}, \cite{dittner2}, each color tensor can be 			written as the trace
	\begin{equation}	\label{berech4.2}
		\mathcal{P}_{abcd} = \mathbf{Tr}(\lambda_a\lambda_b\lambda_c\lambda_d)
	\end{equation}
	of different Lie algebra generators $\lambda_a , \ldots , \lambda_d$ of 			$su(N)$. Furthermore it is shown in \cite{YMIV}, that for all $N \geqslant 			2$ the non trivial color tensors comprise of linear combinations of the 			following nine base tensors:
	\begin{align}
		& \df_{ab}\df_{cd}	&& \df_{ac}\df_{bd}	&& \df_{ad}\df_{bc}
	 													\label{berech4.3}	\\
		& d_{abr}d_{cdr}	&& d_{acr}d_{bdr}	&& d_{adr}d_{bcr}
	 													\label{berech4.4}	\\
		& d_{abr}f_{cdr}	&& d_{acr}f_{bdr} 	&& d_{adr}f_{bcr}
	 													\label{berech4.5}
	\end{align}
	As usual, the $f_{abc}$'s represent totally antisymmetric connection 				coefficients defined by
	\begin{equation}	\label{berech4.6}
		[\lambda_a,\lambda_b]= 2 i f_{abc} \lambda_c
	\end{equation}
	and the $d_{abc}$'s stand for the totally symmetric tensors belonging to the 	universal covering algebra of $su(N)$. Finally, the $\df$'s are the usual 			Kronecker deltas.
	
	Remembering the even--odd theorem mentioned in the introductory part of the 		$3$--leg calculations, one immediately realizes that the terms with a color 		tensor of $\eqref{berech4.5}$ must equal zero. Thus the color structure is 			given by the six tensors of $\eqref{berech4.3}$ and $\eqref{berech4.4}$ 			only.
	
\item
	The two remaining triples of color base--tensors can be investigated 				separately. To minimize work further we mention, that
	\begin{itemize}
	\item
		terms are not equal to zero only, if both field operators and color 				tensors have the same (anti)symmetry in there color 								arguments\footnote{Symmetry or antisymmetry under exchange of two color 			indices respectively.}. Furthermore, one can build by means of linear 				combinations for \emph{each} (anti)symmetrical \emph{pair} of field 				operators a new color tensor basis which consist of symmetrical and 				anti--symmetrical color tensors in the considered \emph{color pair} 				only.
		
		As an \emph{example}, the color basis for any (anti)symmetrical field 				operator pair in the arguments $(a,b)$ for the elements 							$\eqref{berech4.3}$ could be given by
		\begin{equation}	\label{berech4.7}
			\df_{ab}\df_{cd}\, , \, 
			\df_{ac}\df_{bd} + \df_{ad}\df_{bc} \, , \, 										\df_{ac}\df_{bd} - \df_{ad}\df_{bc}
		\end{equation}
		were the first two base elements are symmetrical and the last one is 				anti--symmetrical under the exchange of $a \leftrightarrow b$ 						respectively. The same considerations obviously apply to the color 					basis elements $\eqref{berech4.4}$.

	\item
		Immediately one realizes, that each anti--symmetrical field operator 				pair occurs in connection with only \emph{one single}\footnote{The last 			tensor in the above list.} anti--symmetrical color tensor in a adapted 				base similar to $\eqref{berech4.7}$ above.
		Similarly, any symmetrical field operator pair merely arises in 					connection with two symmetrical tensors\footnote{The first two tensors 				in the list above.} only.
	\end{itemize}

\item
	Equation $\eqref{berech1.2}$ has in the $4$--leg case the special form
	\begin{equation} 	\label{berech4.8}
		\sum_{j=1}^n \p_{x_j}^{\mu} \df = \p_{x_1}^{\mu} \df + \p_{x_2}^{\mu} 				\df + \p_{x_3}^{\mu} \df + \p_{x_4}^{\mu} \df + \sum_{\trm{inner}} 					\p^{\mu}\df = 0
	\end{equation}
\end{itemize}

\subsubsection{Determination of the Elements of \boldmath$\mathcal{L}$} 																	\label{BERECH4.1}

As in the preceeding sections, the determination of local field expressions will be done for subsets with the same number of external derivatives separately. 
Equation $\eqref{algeb10}$ shows
\begin{equation}	\label{berech4.9}
	\omega(4\text{--leg}) \leqslant 4 - 4 - d + 1 = 1 - d
\end{equation}
that local terms arise only if non, or at most one external derivative in the expressions under consideration occurs. Thus, one merely has to investigate the graphs of the two subsets
\begin{equation}	\label{berech4.10}
	\mathcal{L} = \mathcal{L}_0 \cup \mathcal{L}_1
\end{equation}
with respect to the restriction $n_u(\mathcal{L}) = 1$. The latter can be fulfilled by the two field operator combinations $AAAu$ and $Au\ti{u}u$ only.

\paragraph{The Subset \boldmath$\mathcal{L}_0$}		\label{BERECH4.1.0}

This set has singular order
\begin{equation}	\label{berech4.11}
	\omega(\mathcal{L}_0) \leqslant  1 - 0 = 1
\end{equation}
which leads to the general set
\begin{equation}	\label{berech4.12}
\begin{aligned}
	\mathcal{L}_0 &= \Big\{\mathcal{P}_{abcd}:\Lambda_a(x_1)\Gamma_b(x_2) 					\Delta_c(x_3) \Pi_d(x_4): \mathcal{D}^{\beta}\df \Bigm| 							\Lambda\Gamma\Delta\Pi \in \left\{ A^{\kappa} A^{\xi}A^{\sigma}u 					,A^{\kappa}u\ti{u}u \right\}\; , \\ & \phantom{= 									\Big\{:\Lambda_a(x_1)\Gamma_b(x_2)\Delta_c(x_3)\Pi_d(x_4):
		\mathcal{D}^{\beta}\df\Bigm| \Lambda\Gamma\Delta}
		|\beta| \leqslant 1 \;,\; x_1 \leftrightarrow x_2 									\leftrightarrow x_3 \leftrightarrow x_4 \Big\}
\end{aligned}
\end{equation}
But terms with $|\beta| = 0$ are not Lorentz scalars and thus $\mathcal{L}_0$ comprises the local expressions\\
$\mathit{l}_{0,i} \in \mathcal{L}_0$
\begin{equation}	\label{berech4.13}
\begin{aligned}
	\mathit{l}_{0,1}&\overset{def}{=}\, 													\mathcal{P}_{abcd}:A^{\mu}_a(x_1)A^{\mu}_b(x_2)A^{\nu}_c(x_3)u_d(x_4):
	 												\p^{\nu}\df				\\
	\mathit{l}_{0,2}&\overset{def}{=}\, 
		\mathcal{P}_{abcd}:A^{\mu}_a(x_1)u_b(x_2) \ti{u}_c(x_3) u_d(x_4): 																\p^{\mu}\df
\end{aligned}
\end{equation}
only. According to the color symmetry of $\mathit{l}_{0,1}$ under exchange of $a \leftrightarrow b$ \footnote{In the permutation invariant sum over all arguments.}, the set of adapted color tensors which don't lead to terms equal to zero consist of
\begin{equation} 	\label{berech4.14}
	\mathcal{P}_{abcd}(\mathit{l}_{0,1}) \in \big\{ \df_{ab}\df_{cd}, 						\df_{ac}\df_{bd}+ \df_{ad}\df_{bc}, d_{abr}d_{cdr}, d_{acr}d_{bdr} + 				d_{adr}d_{bcr} \big\}
\end{equation}
Under the restriction $\p^{\nu}_{x_1}$\footnote{All the following considerations also apply one--to--one to the case $\p^{\nu}_{x_2}$.} the element $\mathit{l}_{0,1}$ can be written as
\begin{align}		\label{berech4.15}
	\mathit{l}_{0,1}\big|_{\p^{\nu}_{x_1}}
		&= \mathcal{P}_{abcd}: \p^{\nu}A^{\mu}_a(x_1)A^{\mu}_b(x_2) 							A^{\nu}_c(x_3) u_d(x_4): \df \pmod \n 
		&&\big| \text{ with }\mathcal{P}_{abcd} = \mathcal{P}_{bacd}
\end{align}
We will demonstrate in the following, that the latter term must equal zero due to the special color structure.
But before proving, we start with a few principal remarks. As mentioned in the introduction to this section, the two sets of three color base tensors can be treated separately\footnote{Only the calculations for the elements of $\eqref{berech4.3}$ are explicitely executed. The same considerations can be applied one--to--one to the color tensors $\eqref{berech4.4}$.}. Furthermore we will use the fact, that all local terms occurring in the splitting procedure are only determined up to a constant $c_q$ as stated in $\eqref{cqft27}$, $\eqref{cqft30}$ and $\eqref{cqft31}$. Additionally the two color bases adapted to the exchange of $a \leftrightarrow b$ and $b \leftrightarrow c$ respectively
\begin{align}
	a \leftrightarrow b: \qquad	&A_{abcd} := \df_{ab}\df_{cd}	
							&&B_{abcd}:= \df_{ac}\df_{bd} + \df_{ad}\df_{bc}	
							&&C_{abcd}:= \df_{ac}\df_{bd} - \df_{ad}\df_{bc}
							\label{berech4.17}								\\
	b \leftrightarrow c: \qquad	&A_{abcd}' := \df_{ad}\df_{bc}	
							&&B_{abcd}':= \df_{ab}\df_{cd} + \df_{ac}\df_{bd}	
							&&C_{abcd}':= \df_{ab}\df_{cd} - \df_{ac}\df_{bd}
							\label{berech4.18}
\end{align}
will play the r\^ole of prime importance.
Obviously, the tensors $\big\{A_{abcd} ,\, B_{abcd} ,\, C_{abcd} \big\}$ can be written in the basis $\big\{A_{abcd}' ,\, B_{abcd}' ,\, C_{abcd}' \big\}$ as follows
\begin{subequations}		\label{berech4.19}
\begin{align}
	A_{abcd} &= \frac{1}{2} \big[ B_{abcd}' + C_{abcd}' \big]	\\
	B_{abcd} &= \frac{1}{2} \big[ B_{abcd}' - C_{abcd}' \big] + A_{abcd}'	\\
	C_{abcd} &= \frac{1}{2} \big[ B_{abcd}' - C_{abcd}' \big] - A_{abcd}'
\end{align}
\end{subequations}
With all the preliminary notes in mind, one can easily prove the above statement that $\mathit{l}_{0,1}\big|_{\p^{\nu}_{x_1}}$ equals zero, by contradiction:
Supposing the term $\mathit{l}_{0,1}\big|_{\p^{\nu}_{x_1}}$ $\big( \mathit{l}_{0,1}\big|_{\p^{\nu}_{x_2}} \big)$ of $\eqref{berech4.15}$ does not equal zero. Since $\mathit{l}_{0,1}$ features a symmetry under exchange of $a \leftrightarrow b$, only the color tensors $\big\{A_{abcd} ,\, B_{abcd} \big\}$\footnote{In the color tensor base $\big\{A_{abcd} ,\, B_{abcd} ,\, C_{abcd} \big\}$.} lead, under the color contraction, to expressions not equal to zero.
Rewriting the latter terms with the aid of $\eqref{berech4.19}$ one acquires for $i \in \{1 ,\, 2 \}$ (remember there is a constant $c_q$ in the expressions of local terms which can be freely chosen)
\begin{align}
	&\mathit{l}_{0,1}\big|_{\p^{\nu}_{x_i}, \, A_{abcd}}
		= c_q\,A_{abcd}: \p^{\nu}A^{\mu}_a(x_1)A^{\mu}_b(x_2) A^{\nu}_c(x_3) 						u_d(x_4): \df									\notag		\\ 
		&\phantom{\mathit{l}_{0,1}\big|_{\p^{\nu}_{x_i}, \, A_{abcd}}}
		= c_q\,A_{abcd}: F^{\nu\mu}_a(x_1)A^{\mu}_b(x_2)A^{\nu}_c(x_3) 
			u_d(x_4): \df +\notag\\
		&\phantom{\mathit{l}_{0,1}\big|_{\p^{\nu}_{x_i}, \, 									A_{abcd}}\p^{\nu}_{x_i}, \, A_{abcd}A_{abcd}\p^{\nu}_{x_i}, \,}
				c_q\,A_{abcd}: \p^{\mu}A^{\nu}_a(x_1)A^{\mu}_b(x_2) 								A^{\nu}_c(x_3) u_d(x_4): \df					\notag		\\ 
		&\phantom{\mathit{l}_{0,1}\big|_{\p^{\nu}_{x_i}, \, A_{abcd}}}
		= c_q\,\frac{1}{2} \big[ B_{abcd}' + C_{abcd}' \big]: F^{\nu\mu}_a(x_1) 					A^{\mu}_b(x_2) A^{\nu}_c(x_3) u_d(x_4): \df +\notag\\
		&\phantom{\mathit{l}_{0,1}\big|_{\p^{\nu}_{x_i}, \, 									A_{abcd}}\p^{\nu}_{x_i}, \, A_{abcd}A_{abcd}\p^{\nu}_{x_i}, \,}
				c_q\,A_{abcd}: \p^{\mu}A^{\nu}_a(x_1)A^{\mu}_b(x_2) 								A^{\nu}_c(x_3) u_d(x_4): \df					\notag		\\
		&\phantom{\mathit{l}_{0,1}\big|_{\p^{\nu}_{x_i}, \, A_{abcd}}}
		= c_q\,\frac{1}{2} C_{abcd}': F^{\nu\mu}_a(x_1)A^{\mu}_b(x_2) 									A^{\nu}_c(x_3) u_d(x_4): \df +\notag\\
		&\phantom{\mathit{l}_{0,1}\big|_{\p^{\nu}_{x_i}, \, 									A_{abcd}}\p^{\nu}_{x_i}, \, A_{abcd}A_{abcd}\p^{\nu}_{x_i}, \,}
				c_q\,A_{abcd}: \p^{\mu}A^{\nu}_a(x_1)A^{\mu}_b(x_2) 								A^{\nu}_c(x_3) u_d(x_4): \df					\notag
\end{align}
since the symmetric color tensor $B_{abcd}'$ sums up to zero under the color contraction with the given Wick monomial part. The last line states the equality
\begin{align}		\label{berech4.20}
	&c_q\,A_{abcd}\big(: \p^{\nu}A^{\mu}_a(x_1)A^{\mu}_b(x_2) 								A^{\nu}_c(x_3) u_d(x_4) -  \p^{\mu}A^{\nu}_a(x_1)A^{\mu}_b(x_2) 					A^{\nu}_c(x_3) u_d(x_4):\big) \df =						\notag		\\
	 	&\phantom{\Longleftrightarrow \quad A_{abcd}: \p^{\nu} A^{\mu}_a(x_1) 				A^{\mu}_b(x_2) A^{\nu}_c(x_3)A_{abcd}:}
		c_q\,\frac{1}{2} C_{abcd}': F^{\nu\mu}_a(x_1) A^{\mu}_b(x_2) 						A^{\nu}_c(x_3) u_d(x_4): \df
\end{align}
A similar calculation leads for $\mathit{l}_{0,1}\big|_{\p^{\nu}_{x_i}, \, B_{abcd}}$ to
\begin{align}
	&\mathit{l}_{0,1}\big|_{\p^{\nu}_{x_i}, \, B_{abcd}}
		= c_{q'}\,B_{abcd}: \p^{\nu}A^{\mu}_a(x_1)A^{\mu}_b(x_2) A^{\nu}_c(x_3) 					u_d(x_4): \df									\notag		\\ 
		&\phantom{\mathit{l}_{0,1}\big|_{\p^{\nu}_{x_i}, \, B_{abcd}}}
		= c_{q'}\,B_{abcd}: F^{\nu\mu}_a(x_1)A^{\mu}_b(x_2) A^{\nu}_c(x_3)
			u_d(x_4): \df +\notag\\
		&\phantom{\mathit{l}_{0,1}\big|_{\p^{\nu}_{x_i}, \, 									A_{abcd}}\p^{\nu}_{x_i}, \, A_{abcd}A_{abcd}\p^{\nu}_{x_i}, \,}
			c_{q'}\,B_{abcd}: \p^{\mu}A^{\nu}_a(x_1)A^{\mu}_b(x_2) 									A^{\nu}_c(x_3) u_d(x_4): \df					\notag		\\ 
		&\phantom{\mathit{l}_{0,1}\big|_{\p^{\nu}_{x_i}, \, B_{abcd}}}
		= c_{q'}\,\big( \frac{1}{2} \big[ B_{abcd}' - C_{abcd}' \big] + 						{A}_{abcd}' \big) : F^{\nu\mu}_a(x_1) A^{\mu}_b(x_2) A^{\nu}_c(x_3)
				u_d(x_4): \df +\notag		\\
		&\phantom{\mathit{l}_{0,1}\big|_{\p^{\nu}_{x_i}, \, 									A_{abcd}}\p^{\nu}_{x_i}, \, A_{abcd}A_{abcd}\p^{\nu}_{x_i}, \,}
		c_{q'}\,B_{abcd}: \p^{\mu}A^{\nu}_a(x_1)A^{\mu}_b(x_2) 								A^{\nu}_c(x_3) u_d(x_4): \df							\notag		\\
		&\phantom{\mathit{l}_{0,1}\big|_{\p^{\nu}_{x_i}, \, B_{abcd}}}
		= -c_{q'}\,\frac{1}{2} C_{abcd}': F^{\nu\mu}_a(x_1)A^{\mu}_b(x_2) 						A^{\nu}_c(x_3) u_d(x_4): \df +\notag\\
		&\phantom{\mathit{l}_{0,1}\big|_{\p^{\nu}_{x_i}, \, 									A_{abcd}}\p^{\nu}_{x_i}, \, A_{abcd}A_{abcd}\p^{\nu}_{x_i}, \,}
			c_{q'}\,A_{abcd}: \p^{\mu}A^{\nu}_a(x_1)A^{\mu}_b(x_2) 									A^{\nu}_c(x_3) u_d(x_4): \df					\notag
\end{align}
which states the equality
\begin{align}		\label{berech4.21}
	&c_{q'}\,B_{abcd}\big(: \p^{\nu}A^{\mu}_a(x_1)A^{\mu}_b(x_2) A^{\nu}_c(x_3)
		u_d(x_4) - \p^{\mu}A^{\nu}_a(x_1)A^{\mu}_b(x_2) A^{\nu}_c(x_3) 
		u_d(x_4):\big) \df =										\notag	\\
		&\phantom{\Longleftrightarrow \quad A_{abcd}: \p^{\nu} A^{\mu}_a(x_1) 				A^{\mu}_b(x_2) A^{\nu}_c(x_3)A_{abcd}:}
		 -c_{q'}\,\frac{1}{2} C_{abcd}': F^{\nu\mu}_a(x_1)A^{\mu}_b(x_2) 
		A^{\nu}_c(x_3) u_d(x_4): \df
\end{align}
The equations $\eqref{berech4.20}$ and $\eqref{berech4.21}$ state anti--linear dependence of the two color tensor base elements $A_{abcd}$ and $B_{abcd}$. This is in contradiction to the supposition, and thus
\begin{align}		\label{berech4.22}
	\mathit{l}_{0,1}\big|_{\p^{\nu}_{x_i}} &= 0 && \qquad i \in \{ 1,2\}
\end{align}
must equal zero. In fact, the constants $c_{q}$ and $c_{q'}$ must equal zero.
This finishes the proof and one can next turn to the terms with one external derivative.

\paragraph{The Subset \boldmath$\mathcal{L}_1$}		\label{BERECH4.1.1}

The set $\mathcal{L}_1$ comprises the elements with singular order
\begin{equation}	\label{berech4.23}
	\omega(\mathcal{L}_1) \leqslant  1 - 1 = 0
\end{equation}
only and thus has the general form
\begin{equation}	\label{berech4.24}
\begin{aligned}
	\mathcal{L}_1 &= \Big\{\mathcal{P}_{abcd} \mathcal{D}^1:\Lambda_a(x_1) 					\Gamma_b(x_2) \Delta_c(x_3) \Pi_d(x_4): \df \Bigm| \Lambda\Gamma \Delta 			\Pi \in \left\{ A^{\kappa} A^{\xi}A^{\sigma}u ,A^{\kappa}u\ti{u}u 					\right\}\; ,\\ & \phantom{= \Big\{:\Lambda_a(x_1)\Gamma_b(x_2) 						\Delta_c(x_3) \Pi_d(x_4): \mathcal{D}^1\df\Bigm| 									\Lambda\Gamma\Delta|\beta| \leqslant 1 \;,:\Lambda_a(x_1)\Gamma}
			\; x_1 \leftrightarrow x_2 									\leftrightarrow x_3 \leftrightarrow x_4 \Big\}
\end{aligned}
\end{equation}
And so the full list reads as
\begin{equation}	\label{berech4.25}
\begin{aligned}
	\mathit{l}_{1,1} &\overset{def}{=}\, \mathcal{P}_{abcd}: 								F^{\mu\nu}_a(x_1)A^{\mu}_b(x_2)A^{\nu}_c(x_3)u_d(x_4): \df = 0 						\qquad\qquad
		&&\text{due to $\eqref{berech4.22}$}								\\
	\mathit{l}_{1,2} &\overset{def}{=}\, \mathcal{P}_{abcd}: 								A^{\mu}_a(x_1)F^{\mu\nu}_b(x_2) A^{\nu}_c(x_3)u_d(x_4):\df = 0 
		&&\text{due to $\eqref{berech4.22}$}								\\
	\mathit{l}_{1,3} &\overset{def}{=}\, \mathcal{P}_{abcd}: 								A^{\mu}_a(x_1)A^{\mu}_b(x_2)\p^{\nu}A^{\nu}_c(x_3)u_d(x_4): \df 
		&&\text{with $\mathcal{P}_{abcd}= \mathcal{P}_{bacd}$}				\\
	\mathit{l}_{1,4} &\overset{def}{=}\, \mathcal{P}_{abcd}: 								A^{\mu}_a(x_1)A^{\mu}_b(x_2)A^{\nu}_c(x_3)\p^{\nu}u_d(x_4): \df 
		&&\text{with $\mathcal{P}_{abcd}= \mathcal{P}_{bacd}$}				\\
	\mathit{l}_{1,5} &\overset{def}{=}\, \mathcal{P}_{abcd}: 								\p^{\nu}A^{\nu}_a(x_1)u_b(x_2) \ti{u}_c(x_3)u_d(x_4): \df 
		&&\text{with $\mathcal{P}_{abcd}= - \mathcal{P}_{adcb}$}				\\
	\mathit{l}_{1,6} &\overset{def}{=}\, \mathcal{P}_{abcd}: 								A^{\nu}_a(x_1)\p^{\nu}u_b(x_2) \ti{u}_c(x_3)u_d(x_4): \df 
		&&\text{with $\mathcal{P}_{abcd}= - \mathcal{P}_{adcb}$}				\\
	\mathit{l}_{1,7} &\overset{def}{=}\, \mathcal{P}_{abcd}: 								A^{\nu}_a(x_1)u_b(x_2) \p^{\nu}\ti{u}_c(x_3)u_d(x_4): \df 
		&&\text{with $\mathcal{P}_{abcd}= - \mathcal{P}_{adcb}$}				\\
	\mathit{l}_{1,8} &\overset{def}{=}\, \mathcal{P}_{abcd}: 								A^{\nu}_a(x_1)u_b(x_2) \ti{u}_c(x_3)\p^{\nu}u_d(x_4): \df 
		&&\text{with $\mathcal{P}_{abcd}= - \mathcal{P}_{adcb}$}
\end{aligned}
\end{equation}

\subsubsection{Determination of Equivalent Elements in \boldmath$\mathcal{L}$} 																	\label{BERECH4.2}

Equation $\eqref{berech4.8}$ leads in connection with $\mathit{l}_{0,1}$ immediately to the equivalence of $\mathit{l}_{1,3}$ and $\mathit{l}_{1,4}$, since
\begin{equation}	\label{berech4.26}
\begin{aligned}
	\mathit{l}_{0,1}\big|_{\sum_{\trm{inner}}\p^{\nu}}
		 &\equiv 0 \pmod{\n} 												\\
		 &= \n - \mathcal{P}_{abcd}:\p^{\nu}A^{\mu}_a(x_1)A^{\mu}_b 								(x_2)A^{\nu}_c(x_3) u_d(x_4):\df							\\
		 &\phantom{= \n - \mathcal{C}_{abcabc}} - 												 \mathcal{P}_{abcd}:A^{\mu}_a(x_1)\p^{\nu}A^{\mu}_b 									(x_2)A^{\nu}_c(x_3) u_d(x_4):\df   							\\
		 &\phantom{= \n - \mathcal{C}_{abcabc}}
		   - \mathcal{P}_{abcd}:A^{\mu}_a(x_1)A^{\mu}_b(x_2)\p^{\nu} 								A^{\nu}_c(x_3) u_d(x_4):\df									\\
		 &\phantom{= \n - \mathcal{C}_{}= \n - \mathcal{C}_{abcabcabcabc}}
	 	   - \mathcal{P}_{abcd}:A^{\mu}_a(x_1)A^{\mu}_b(x_2)A^{\nu}_c(x_3) 							\p^{\nu}u_d(x_4):\df
\end{aligned}
\end{equation}
whereas in the sum above the first and second term equals zero according to $\eqref{berech4.22}$. Thus the last equation states
\begin{equation}		\label{berech4.27}
	\mathcal{P}_{abcd}:A^{\mu}_a(x_1)A^{\mu}_b(x_2)\p^{\nu} 								A^{\nu}_c(x_3) u_d(x_4):\df = - 														\mathcal{P}_{abcd}:A^{\mu}_a(x_1)A^{\mu}_b(x_2)A^{\nu}_c(x_3) 							\p^{\nu}u_d(x_4):\df 	\pmod{\n}
\end{equation}
Altogether, $\mathit{l}_{1,4}$ represents the terms $\mathit{l}_{0,1}$ and $\mathit{l}_{1,3}$. Furthermore, the permutation invariant sum over all exchanged arguments in connection with the special form of the color tensor leads to the equivalence of the elements $\mathit{l}_{1,6}$ and $\mathit{l}_{1,8}$ which establishes
\begin{align}
	\mathit{l}_{1,7} 	&= \n - \mathit{l}_{1,5} - \mathit{l}_{1,6} - 									\mathit{l}_{1,8}						\label{berech4.28}
					&&\qquad\qquad \big|\mathit{l}_{1,6}\sim \mathit{l}_{1,8} \\
						&\equiv - \mathit{l}_{1,5} - 2\mathit{l}_{1,6} \pmod{\n}
															\label{berech4.28a}
\end{align}
Collecting all the above results one realizes, that $\mathcal{L}'$ comprises the elements
\begin{align}	\label{berech4.29}
	\mathcal{L}'
	:= \,\raisebox{1ex}{$\mathcal{L}$}\raisebox{0ex}
				{$\! \Big/\!$}\raisebox{-1ex}{$\n$}
	= \big\{ \mathit{l}_{1,4}\,,\, \mathit{l}_{1,5}\,,\, \mathit{l}_{1,6} \big\}
\end{align}
only.
That finishes this subsection and one can proceed to the set $\mathcal{M}$.

\subsubsection{Determination of Elements of \boldmath$\mathcal{M}$}	\label{BERECH4.3}

Since $\mathcal{M}$ has singular order $\omega(\mathcal{M}) = 0$, no derivatives occur neither on the field operators nor on the $\df$--distributions. Thus the set comprises of\\
$\mathit{m}_{i} \in \mathcal{M}$
\begin{equation}	\label{berech4.30}
\begin{aligned}
	\mathit{m}_{1} &\overset{def}{=}\, \mathcal{P}_{abcd}: 									A^{\mu}_a(x_1)A^{\mu}_b(x_2)A^{\nu}_c(x_3)A^{\nu}_d(x_4): \df							\qquad\qquad
		&&\text{with $\mathcal{P}_{abcd}= \mathcal{P}_{bacd}$}				\\
	\mathit{m}_{2} &\overset{def}{=}\, \mathcal{P}_{abcd}: 									A^{\mu}_a(x_1)A^{\mu}_b(x_2) \ti{u}_c(x_3)u_d(x_4):\df 
		&&\text{with $\mathcal{P}_{abcd}= \mathcal{P}_{bacd}$} 				\\
	\mathit{m}_{3} &\overset{def}{=}\, \mathcal{P}_{abcd}: 									\ti{u}_a(x_1)u_b(x_2) \ti{u}_c(x_3)u_d(x_4): \df 
		&&\text{with $\mathcal{P}_{abcd}= - \mathcal{P}_{cbad}$}
\end{aligned}
\end{equation}
with the explicit list of color tensors\footnote{Which lead to expressions not equal to zero.} for each term given as
\begin{subequations}	\label{berech4.31}
\begin{align}	
	\mathcal{P}_{abcd}(\mathit{m}_{1})
		& \in \big\{ \df_{ab}\df_{cd}, \df_{ac}\df_{bd}+ \df_{ad}\df_{bc}, 						d_{abr}d_{cdr}, d_{acr}d_{bdr} + d_{adr}d_{bcr} \big\} 			\\
	\mathcal{P}_{abcd}(\mathit{m}_{2})
		& \in \big\{ \df_{ab}\df_{cd}, \df_{ac}\df_{bd}+ \df_{ad}\df_{bc}, 						d_{abr}d_{cdr}, d_{acr}d_{bdr} + d_{adr}d_{bcr} \big\} 			\\
	\mathcal{P}_{abcd}(\mathit{m}_{3})
		& \in \big\{ \df_{ab}\df_{cd} - \df_{ad}\df_{bc}, d_{abr}d_{cdr} - 						d_{adr}d_{bcr} \big\}
\end{align}
\end{subequations}
Obviously the three elements above cannot be equivalent and so $\mathcal{M}'$ simply writes as
\begin{align}	\label{berech4.32}
	\mathcal{M}'
	:= 	\,\raisebox{1ex}{$\mathcal{M}$}
		\raisebox{0ex}
		{$\! \Big/\!$}\raisebox{-1ex}{$\n$} \;
	 = \; \big\{\mathit{m}_{1}\,,\,\mathit{m}_{2}\,,\,\mathit{m}_{3}  \big\} 		\end{align}

\subsubsection{The Subgroups \boldmath$\textbf{B}(\mathcal{M}')$, 					$\textbf{Z}(\mathcal{M}')$ and $\textbf{H}(\mathcal{M}')$}\label{BERECH4.4}

The preliminary results simplify the following calculations and one easily determines the related subgroup elements.

\paragraph{The Subgroup \boldmath$\textbf{B}(\mathcal{M}')$}																			\label{BERECH4.4.1}

The exact subgroup in the four leg case trivially is empty. This is true, since any $\mathit{m}_{i} \in \textbf{B}(\mathcal{M}')$ would be the image of a $\chi$ ($n_u(\chi) = -1$) under the mapping $d_Q$. But in the terms of $\mathcal{M}'$ not a single derivative does occur due to the singular order of the set. This contradicts the differential algebra property of $\mathcal{F}$, that each gauge transformation $d_Q$ leads to a derivative in the transformed field operator expression. Thus
\begin{equation}	\label{berech4.33}
	\textbf{B}(\mathcal{M}') = \emptyset
\end{equation}
$\textbf{B}(\mathcal{M}')$ is empty.
The latter also reveals, that the subgroup $\textbf{Z}(\mathcal{M}')$ and $\textbf{H}(\mathcal{M}')$ coincide\footnote{According to the definition of factor groups.}.

\paragraph{The Subgroup \boldmath$\textbf{Z}(\mathcal{M}')$ and the Gauge--Factor Subgroup \boldmath$\textbf{H}(\mathcal{M}')$}																			\label{BERECH4.4.2}

As a consequence of the last statement, the terms of $\textbf{Z}(\mathcal{M}')$ fully determine the group $\textbf{H}(\mathcal{M}')$. Similar to the earlier calculations, the elements of $\textbf{Z}(\mathcal{M}')$ are found by $d_Q$--transforming the elements of \nolinebreak[4]$\mathcal{M}'$
\begin{align}
&\begin{aligned}		\label{berech4.34}
	d_Q\big(\mathit{m}_{1}\big|_{\trm{P}}\big)
		&= \sum_{{\trm{P} \{ x_1,x_2,x_3,x_4 \} }}\mathcal{P}_{abcd}\, d_Q 						:A^{\mu}_a(x_1) A^{\mu}_b(x_2) A^{\nu}_c(x_3) A^{\nu}_d(x_4):\df\\
		&= \sum_{{\trm{P} \{ x_1,x_2,x_3,x_4 \} }}\mathcal{P}_{abcd} 							:\p^{\mu}u_a(x_1) A^{\mu}_b(x_2) A^{\nu}_c(x_3) A^{\nu}_d(x_4):\df\\
		& \equiv \mathit{l}_{1,4} \quad \pmod{\n} \qquad
			\big| \; \forall \; \mathcal{P}_{abcd}(\mathit{l}_{1,4}) 				 \end{aligned}
\\[1.5em]
&\begin{aligned}		\label{berech4.35}
	d_Q\big(\mathit{m}_{2}\big|_{\trm{P}}\big)
		&= \sum_{{\trm{P} \{ x_1,x_2,x_3,x_4 \} }} \mathcal{P}_{abcd}\, d_Q 					:A^{\mu}_a(x_1)A^{\mu}_b(x_2) \ti{u}_c(x_3)u_d(x_4):\df\\
		&=  \sum_{{\trm{P} \{ x_1,x_2,x_3,x_4 \} }}\mathcal{P}_{abcd}:\p^{\mu} 					u_a(x_1) A^{\mu}_b(x_2) \ti{u}_c(x_3)u_d(x_4): \df 	\,+\,\\
		&\phantom{\sum_{\trm{P}}\mathcal{P}_{abcd}:\p^{\mu}u_a(x_1)A^{\mu}}
			\sum_{{\trm{P} \{ x_1,x_2,x_3,x_4 \} }}\mathcal{P}_{abcd} 							:A^{\mu}_a(x_1)A^{\mu}_b(x_2) \p^{\nu} A^{\nu}_c(x_3)u_d(x_4):\df\\	
		& \equiv 2 \mathcal{P}_{abcd}(\mathit{m}_{2})\,\mathit{l}_{1,6} + 						\mathcal{P}_{abcd}(\mathit{l}_{1,3})\,\mathit{l}_{1,3} 
				\quad \pmod{\n} \quad
				\big| \text{ since }\mathcal{P}_{abcd}(\mathit{m}_2)=
				\mathcal{P}_{abcd}(\mathit{l}_{1,3}) 					 
 \end{aligned}
\\[1.5em]
 &\begin{aligned}		\label{berech4.36}
	d_Q\big(\mathit{m}_{3}\big|_{\trm{P}}\big)
		&= \sum_{{\trm{P} \{ x_1,x_2,x_3,x_4 \} }}\mathcal{P}_{abcd}\,d_Q 						:\ti{u}_a(x_1)u_b(x_2) \ti{u}_c(x_3)u_d(x_4):\df				\\
		&= \sum_{{\trm{P} \{ x_1,x_2,x_3,x_4 \} }}\mathcal{P}_{abcd} 							:\p^{\mu}A^{\mu}_a(x_1)u_b(x_2)\ti{u}_c(x_3)u_d(x_4):\df		\\
		& \equiv \mathit{l}_{1,5} \quad \pmod{\n} \qquad
			\big| \; \forall \; \mathcal{P}_{abcd}(\mathit{l}_{1,5})
 \end{aligned}
\end{align}
Considering equivalence $\eqref{berech4.27}$, the $d_Q$--transformed $m_2$ simply rewrites as
\begin{equation}	\label{berech4.37}
	d_Q\big(\mathit{m}_{2}\big|_{\trm{P}}\big)
		= \mathcal{P}_{abcd}(\mathit{m}_{2})\,\mathit{l}_{1,6} +
			d_Q(\mathit{m}_{1}\big|_{\trm{P}}\big)
\end{equation}
since $\eqref{berech4.27}$ states $\mathit{l}_{1,3} \sim \mathit{l}_{1,4}$ and $\mathit{l}_{1,4}$ equals $d_Q\big(\mathit{m}_{1}\big|_{\trm{P}}\big)$ as shown in $\eqref{berech4.34}$\footnote{The color tensors $\mathcal{P}_{abcd}$ $(\mathit{l}_{1,3})$ $= \mathcal{P}_{abcd}(\mathit{l}_{1,4})$ are equivalent.}.
Now there only remains to show, that one can construct the needed $\mathcal{P}_{abcd}(\mathit{l}_{1,6}) \; \big( = \mathcal{P}_{abcd} (\mathit{m}_{3}) \big)$ with the aid of the color tensors $\mathcal{P}_{abcd} (\mathit{m}_{2})$ of expression $\eqref{berech4.37}$.
This can be shown immediately, if one takes into account, that due to the anti--symmetry of $\big\{ \p^{\mu}u_a(x_1) , \, u_d(x_4) \big\}$\footnote{In the permutation invariant sum.} all terms with symmetrical color tensors\footnote{Under the exchange of $a \leftrightarrow d$.} equal zero\footnote{Symmetrical color tensors under $a \leftrightarrow d$ arise in the second terms of the sums in $\mathcal{P}_{abcd}(\mathit{m}_{2})$.}. 
Consequently, only the color tensors
\begin{equation}		\label{berech4.38}
	\mathcal{P}_{abcd}'(\mathit{m}_{2}) \in
		\big\{ \df_{ab}\df_{cd}\,,\, \df_{ac}\df_{bd}\,,\, 												d_{abr}d_{cdr}\,,\, d_{acr}d_{bdr} \big\}
\end{equation}
arise in the first term of expression $\eqref{berech4.37}$. Then, simple linear combinations of the latter tensors can express the required color tensors
\begin{equation}	\label{berech4.39}
	\mathcal{P}_{abcd}(\mathit{l}_{1,6}) \in
		\big\{ \df_{ab}\df_{cd} - \df_{ac}\df_{bd}\,,\, 												d_{abr}d_{cdr} - d_{acr}d_{bdr} \big\}
\end{equation}
and thus $\mathcal{P}_{abcd}(\mathit{l}_{1,6}) \mathit{l}_{1,6}$ is writable as
\begin{equation}	\label{berech4.40}
	d_Q\bigg(\big( \mathit{m}_{2}\big|_{\trm{P}} - \mathit{m}_{1} 							\big|_{\trm{P}}\big)
	\Big|_{\begin{subarray}{l}
		\mathcal{P}_{abcd}=\df_{ab}\df_{cd}\\
		\mathcal{P}_{abcd}=d_{abr}d_{cdr}
			\end{subarray}}\bigg)
	-d_Q\bigg(\big( \mathit{m}_{2}\big|_{\trm{P}} - \mathit{m}_{1} 							\big|_{\trm{P}}\big)
	\Big|_{\begin{subarray}{l}
		\mathcal{P}_{abcd}=\df_{ac}\df_{bd} - \df_{ad}\df_{cb}\\
		\mathcal{P}_{abcd}= d_{acr}d_{bdr} - d_{adr}d_{cbr}
			\end{subarray}}\bigg)
	\equiv \mathcal{P}_{abcd}(\mathit{l}_{1,6}) \mathit{l}_{1,6}
\end{equation}
whereas $\mathcal{P}_{abcd}(\mathit{l}_{1,6})$ on the right hand side of $\eqref{berech4.40}$ represents the two cases indicated by the two subscript--terms on the left hand side.

This finishes the proof in the $4$--leg case for disjunct arguments, since with this result all terms of $\mathcal{L}'$ can be written as $d_Q$--transformed elements of 
\begin{align}	\label{berech4.41}
	\textbf{H}(\mathcal{M}')
		&= \, \raisebox{1ex}{$\textbf{Z}(\mathcal{M})$} \raisebox{0ex}{$\! 					\Big/\!$} 	\raisebox{-1ex}{$\textbf{B}(\mathcal{M})$}		\notag	\\	
		&= \big\{\mathit{m}_{1} \,,\, \mathit{m}_{3} \,,\, \big(\mathit{m}_{2} - 				\mathit{m}_{1}\big)
					\Big|_{\begin{subarray}{l}
					\mathcal{P}_{abcd}=\df_{ab}\df_{cd}\\
					\mathcal{P}_{abcd}=d_{abr}d_{cdr}
					\end{subarray}} - 
			\big(\mathit{m}_{2} - \mathit{m}_{1}\big)													\Big|_{\begin{subarray}{l}
					\mathcal{P}_{abcd}=\df_{ac}\df_{bd} - \df_{ad}\df_{cb}\\
					\mathcal{P}_{abcd}= d_{acr}d_{bdr} - d_{adr}d_{cbr}
					\end{subarray}}
			\big\}													
\end{align}
\subsection{\boldmath $4$--Leg Calculations for Terms Based on Reducible Graphs}																\label{BERECH4a.1}

Up to the listing $\eqref{berech4.25}$ no use was made of the Leibnitz' rule nor of equation $\eqref{berech4.8}$, and thus all equivalences, up to this point, hold true under the transition to equal arguments\footnote{In detail, all equivalences $\eqref{berech4.9}$ -- $\eqref{berech4.25}$ simply can be written, without further changes, for the terms of reducible graphs as the original expressions restricted by $\big|_{x_1=x_2}$ or $\big|_{x_1=x_2 ,x_3=x_4}$ respectively.}.
In the next transformation $\eqref{berech4.26}$ there is made use of equation $\eqref{berech4.8}$. Thus one has to show that $\eqref{berech4.27}$ holds true under the restriction of equal arguments. Without further work one can state, that $\eqref{berech4.27}$ remains unchanged, since the Leibnitz' rule leads to exactly the same sum of four terms on the right hand side of $\eqref{berech4.26}$ under the restrictions to equal arguments, and the first two terms again equal zero according to the (trivial adapted) restricted\footnote{Restricted to terms of reducible graphs.} calculations $\eqref{berech4.20}$ and $\eqref{berech4.21}$.

To finish the proof for the reducible $4$--leg terms there remains only to show the validity of equivalence $\eqref{berech4.28a}$ for the cases of equal arguments. This is true, since all further proof steps can be simply achieved by the restrictions to equal arguments without changing the calculations nor the results, because no use of the Leibnitz' rule nor of identity $\eqref{berech4.8}$ was made.

So one focuses on $\eqref{berech4.28a}$ next.
First, one notices, that in the permutation invariant sum for disjunct arguments the terms $\mathit{l}_{1,6}$ and $\mathit{l}_{1,8}$ are identical. This equality obviously is not destroyed by restricting the arguments to $(x_1=x_2)$ or $(x_1=x_2 ,x_3=x_4)$ of reducible terms in the permutation invariant sum. Furthermore, the identity $\eqref{berech4.28}$ also remains correct for the reducible terms, since the Leibnitz' rule --- as for identity $\eqref{berech4.26}$ --- leads to the same sum of terms (in $\eqref{berech4.28}$) on the right hand side even for equal arguments. This states in fact, that the equivalence $\eqref{berech4.28a}$ also holds for reducible terms.
This finishes the proof of gauge invariance for reducible $4$--leg terms and thus the possibility of a gauge invariant normalization is shown for all $T_n$'s.
\newpage
	
\end{document}